{}

\documentclass[11pt,a4paper]{article}

\newif\ifpublic\publictrue
\newif\iffancy\fancytrue

\usepackage{jheppub}
\usepackage{setspace}
\usepackage{rotating}

\usepackage{amsmath,amssymb, amsfonts}
\usepackage{dsfont}

\usepackage[fonts]{mpostinl}

\usepackage[usenames,dvipsnames,table]{xcolor}
\definecolor{mygreen}{rgb}{0,0.4,0}
\definecolor{myblue}{rgb}{0,0.0,0.4}
\definecolor{refrcolor}{rgb}{0,0.4,0}
\definecolor{cgreen}{rgb}{0,0.7,0}
\definecolor{ecolor}{rgb}{.52,.03,.06}
\definecolor{bgcolor}{rgb}{.96,.95,.80}
\definecolor{bgcolordark}{rgb}{.80,.80,.67}
\definecolor{faint}{rgb}{.80,.80,.80}

\newcommand{\ph}[1]{\phantom{#1}}

\def\beq{\begin{equation}}
\def\eeq{\end{equation}}

\usepackage{enumitem}

\makeatletter
\providecommand*{\shuffle}{%
  \mathbin{\mathpalette\shuffle@{}}%
}
\newcommand*{\shuffle@}[2]{%
  \sbox0{$#1\vcenter{}$}%
  \kern .15\ht0 
  \rlap{\vrule height .25\ht0 depth 0pt width 2.5\ht0}%
  \raise.1\ht0\hbox to 2.5\ht0{%
    \vrule height 1.75\ht0 depth -.1\ht0 width .17\ht0 %
    \hfill
    \vrule height 1.75\ht0 depth -.1\ht0 width .17\ht0 %
    \hfill
    \vrule height 1.75\ht0 depth -.1\ht0 width .17\ht0 %
  }%
  \kern .15\ht0 
}
\makeatother

\makeatletter
\providecommand*{\boldshuffle}{%
  \mathbin{\mathpalette\boldshuffle@{}}%
}
\newcommand*{\boldshuffle@}[2]{%
  \sbox0{$#1\vcenter{}$}%
  \kern .15\ht0 
  \rlap{\vrule height .25\ht0 depth 0.25pt width 2.5\ht0}%
  \raise.1\ht0\hbox to 2.5\ht0{%
    \vrule height 1.75\ht0 depth -.1\ht0 width .31\ht0 %
    \hfill
    \vrule height 1.75\ht0 depth -.1\ht0 width .31\ht0 %
    \hfill
    \vrule height 1.75\ht0 depth -.1\ht0 width .31\ht0 %
  }%
  \kern .15\ht0 
}
\makeatother

\makeatletter
\g@addto@macro\bfseries{\boldmath}
\makeatother

\usepackage{xparse}

\ExplSyntaxOn
\NewDocumentCommand{\Gtargz}{m m}
{
 \Gt\left(\begin{smallmatrix}
 \Gtargz_print:n {#1} \\
 \Gtargz_print:n {#2}
 \end{smallmatrix};z\right)
}
\seq_new:N \l_Gtargz_list_seq
\cs_new_protected:Npn \Gtargz_print:n #1
{
  \seq_set_split:Nnn \l_Gtargz_list_seq { , } { #1 }
  \seq_use:Nn \l_Gtargz_list_seq { , & }
}
\ExplSyntaxOff

\ExplSyntaxOn
\NewDocumentCommand{\Gtargzt}{m m}
{
 \Gt\left(\begin{smallmatrix}
 \Gtargzt_print:n {#1} \\
 \Gtargzt_print:n {#2}
 \end{smallmatrix};z,\tau\right)
}
\seq_new:N \l_Gtargzt_list_seq
\cs_new_protected:Npn \Gtargzt_print:n #1
{
  \seq_set_split:Nnn \l_Gtargzt_list_seq { , } { #1 }
  \seq_use:Nn \l_Gtargzt_list_seq { , & }
}
\ExplSyntaxOff

\newcommand{\SI}[1]{\Sel[#1]}

\ExplSyntaxOn
\NewDocumentCommand{\SIE}{m m}
{
\SelEn\!\Big[\begin{smallmatrix}
 \SI_print:n {#1} \\
 \SI_print:n {#2}
 \end{smallmatrix}\Big]
}
\seq_new:N \l_SI_list_seq
\cs_new_protected:Npn \SI_print:n #1
{
  \seq_set_split:Nnn \l_SI_list_seq { , } { #1 }
  \seq_use:Nn \l_SI_list_seq { , & }
}
\ExplSyntaxOff

\ExplSyntaxOn
\NewDocumentCommand{\SIEzwei}{m m}
{
\SelEz\!\Big[\begin{smallmatrix}
 \SI_print:n {#1} \\
 \SI_print:n {#2}
 \end{smallmatrix}\Big]
}
\seq_new:N \l_SIz_list_seq
\cs_new_protected:Npn \SIz_print:n #1
{
  \seq_set_split:Nnn \l_SIz_list_seq { , } { #1 }
  \seq_use:Nn \l_SIz_list_seq { , & }
}
\ExplSyntaxOff

\PassOptionsToPackage{unicode}{hyperref}
\usepackage{graphbox}
\usepackage{float}
\usepackage{lmodern}
\usepackage{amsbsy}
\usepackage[utf8]{inputenc}
\usepackage[font=small,labelfont=bf]{caption}
\usepackage{tikz}
\usepackage{braids}

\iffancy

\def\showkeysrefformat#1{{\normalfont\tiny\ttfamily#1}}
\makeatletter
\def\SK@@ref#1>#2\SK@{%
 {\@inlabelfalse\leavevmode\vbox to\z@{%
 \vss\SK@refcolor\rlap{\vrule\raise .75em%
  \hbox{\showkeysrefformat{#2}}}}}}
\makeatother
\fi

\numberwithin{equation}{section}

\newcommand{\eqn}[1]{eq.~\eqref{#1}}
\newcommand{\Eqn}[1]{Equation~\eqref{#1}}
\newcommand{\eqns}[2]{eqs.~\eqref{#1} and~\eqref{#2}}

\newcommand{\rcite}[1]{ref.~\cite{#1}}
\newcommand{\rcites}[1]{refs.~\cite{#1}}

\providecommand{\href}[2]{#2}

\makeatletter
\def\mr@ignsp#1 {\ifx\:#1\@empty\else #1\expandafter\mr@ignsp\fi}%
\newcommand{\multiref}[1]{\begingroup
\xdef\mr@no@sparg{\expandafter\mr@ignsp#1 \: }%
\def\mr@comma{}%
\@for\mr@refs:=\mr@no@sparg\do{\mr@comma\def\mr@comma{,}\ref{\mr@refs}}%
\endgroup}
\renewcommand{\eqref}[1]{(\multiref{#1})}
\makeatother

\makeatletter
\newcommand{\namedref}[2]{#1~\hyperref[#2]{\ref*{#2}}}
\newcommand{\secref}{\@ifstar{\namedref{Section}}{\namedref{section}}}
\newcommand{\subsecref}{\@ifstar{\namedref{Subsection}}{\namedref{subsection}}}
\newcommand{\appref}{\@ifstar{\namedref{Appendix}}{\namedref{appendix}}}
\newcommand{\tabref}{\@ifstar{\namedref{Table}}{\namedref{table}}}
\newcommand{\figref}{\@ifstar{\namedref{Figure}}{\namedref{figure}}}
\makeatother

\providecommand{\hypersetup}[1]{}

\hypersetup{plainpages=false}
\hypersetup{pdfpagemode=UseNone}
\hypersetup{bookmarksnumbered=true}
\hypersetup{pdfstartview=FitH}
\hypersetup{colorlinks=false}
\hypersetup{citebordercolor={0.7 0.7 1}}
\hypersetup{urlbordercolor={.4 .8 1}}
\hypersetup{linkbordercolor={1 .8 .6}}
\hypersetup{colorlinks=true, urlcolor=[rgb]{0.13,0.30,0.45}, linkcolor=[rgb]{0.13,0.30,0.45}, citecolor=[rgb]{0.55,0.0,0.05}}
\makeatletter
\let\@keywords\@empty
\let\@subject\@empty
\providecommand{\keywords}[1]{\gdef\@keywords{#1}}
\providecommand{\subject}[1]{\gdef\@subject{#1}}
\def\thetitle{\@title}
\def\theauthor{\@author}
\def\thesubject{\@subject}
\def\thedate{\@date}
\def\thekeywords{\@keywords}
\makeatother
\AtBeginDocument{
\hypersetup{pdftitle={\thetitle}}%
\hypersetup{pdfauthor={\theauthor}}%
\hypersetup{pdfsubject={\thesubject}}%
\hypersetup{pdfkeywords={\thekeywords}}%
}

\RequirePackage{verbatim}

\makeatletter
\newwrite\bibinl@out
{\immediate\closeout\bibinl@out\@esphack}
\makeatother

\newif\ifnote 
\notetrue

\allowdisplaybreaks

\newcommand{\ad} {\mathrm{ad}}

\let\Re\relax\DeclareMathOperator{\Re}{Re}
\let\Im\relax\DeclareMathOperator{\Im}{Im}

\newcommand{\pd}{\partial}

\newcommand{\s}{\sigma}

\newcommand{\SL}{\mathrm{SL}}

\newcommand{\dd}{\mathrm{d}}

\newcommand{\ap}{\alpha'}

\newcommand{\vecb}{\left(\begin{array}{c}}
\newcommand{\vece}{\end{array}\right)}
\newcommand{\ccb}{\left(\begin{array}{cc}}
\newcommand{\cce}{\end{array}\right)}
\newcommand{\cccb}{\left(\begin{array}{ccc}}
\newcommand{\ccce}{\end{array}\right)}
\newcommand{\ccccb}{\left(\begin{array}{cccc}}
\newcommand{\cccce}{\end{array}\right)}
\newcommand{\cccccb}{\left(\begin{array}{ccccc}}
\newcommand{\ccccce}{\end{array}\right)}
\newcommand{\ccccccb}{\left(\begin{array}{cccccc}}
\newcommand{\cccccce}{\end{array}\right)}

\newcommand{\ZN}{\mathbb N}

\newcommand{\ZZ}{\mathbb Z}

\newcommand{\CG}{\mathcal{G}} 

\def\tree{\text{tree}}

\let\emptyset\varnothing

\newcommand{\nnl}{\nonumber\\}

\DeclareMathOperator{\Gt}{\tilde{\Gamma}}
\newcommand{\KN}{\mathrm{KN}}

\DeclareMathOperator{\Sel}{S}

\DeclareMathOperator{\SelEn}{S^E_\mathit{n}}
\DeclareMathOperator{\SelEz}{S^E_2}

\usepackage{nicefrac}

\newcommand{\phiChain}{\tilde{\varphi}^\tau}

\title{Open-string integrals with multiple unintegrated punctures at genus one}
\author[a,b]{Andr\'e Kaderli,}
\author[c]{Carlos Rodriguez}
\affiliation[a]{Institut f\"ur Mathematik und Institut f\"ur Physik, Humboldt-Universit\"at zu Berlin\\
	IRIS Adlershof, Zum Gro\ss{}en Windkanal 6, 12489 Berlin, Germany}
\affiliation[b]{Max-Planck-Institut f\"ur Gravitationsphysik, Albert-Einstein-Institut\\
	Am M\"uhlenberg 1, 14476 Potsdam, Germany}
\affiliation[c]{Department of Physics and Astronomy, Uppsala University\\
	75108 Uppsala, Sweden}
\emailAdd{andre.kaderli@gmail.com}
\emailAdd{carlos.rodriguez@physics.uu.se}
\abstract{
We study integrals appearing in intermediate steps of one-loop open-string amplitudes, with multiple unintegrated punctures on the $A$-cycle of a torus. We construct a vector of such integrals which closes after taking a total differential with respect to the $N$ unintegrated punctures and the modular parameter $\tau$. These integrals are found to satisfy the elliptic Knizhnik-Zamolodchikov-Bernard (KZB) equations, and can be written as a power series in $\ap$ -- the string length squared-- in terms of elliptic multiple polylogarithms (eMPLs). In the $N$-puncture case, the KZB equation reveals a representation of $B_{1,N}$, the  braid group of $N$ strands on a torus, acting on its solutions. We write the simplest of these braid group elements -- the braiding one puncture around another -- and obtain generating functions of analytic continuations of eMPLs. The   KZB equations in the so-called universal case is written in terms of the genus-one Drinfeld-Kohno algebra $\mathfrak{t}_{1,N} \rtimes \mathfrak{d}$, a graded algebra. Our construction determines matrix representations of various dimensions for several generators of this algebra which respect its grading up to commuting terms.
 }
\preprint{ UUITP-14/22}
\begin{document}
	\maketitle

\vspace*{20pt}
\newpage

\section{Introduction}

In perturbative field theory and string theory, various integrals have to be computed in order to obtain scattering amplitudes. In the case of field theory, these are integrals over loop momenta of virtual particles. In string theory, these are integrals over moduli spaces of punctured (super) Riemann surfaces.  In recent years, physicists, number theorists, and algebraic geometers have gained a better understanding of the methods necessary to perform these integrals and of the properties of the resulting functions and numbers -- or periods, respectively.

In field theory, the simplest loop integrals yield polylogarithms evaluated at certain kinematic points or numbers, which can be computed in dimensional regularization using methods based on differential equations or direct integration \cite{Henn:2013pwa,hyperInt}. However, even simple quantum field theories, such as a $\phi^4$-theory, can eventually produce more complicated periods at high enough number of loops \cite{Brown:kthree}. In string theory, the simplest amplitudes describe the tree-level scattering of massless external states. Their low-energy expansion returns periods of the moduli space of Riemann spheres with $n$ marked points $\mathcal{M}_{0,n}$ \cite{Brown:0606}. The marked points or punctures, respectively, correspond to the vertex insertion points of the external string states. Interesting properties of these genus-zero string amplitudes are made apparent, for example, in the KLT relations, which relate tree-level closed-string integrals to sums of bilinears of open-string integrals \cite{Kawai:1985xq}. These relations have beautiful mathematical interpretations in either intersection theory of \rcite{Mizera:2019gea} or in a more  succinct level, in terms of the single-valued integration of Brown, Dupont and Schnetz \cite{Schnetz:2013hqa,Brown:2018omk}.

Next in difficulty, there are integrals involving genus-one, or elliptic, curves. They appear quite non-trivially\footnote{In particular, it is hard to determine whether a given integral expressible via elliptic functions, for example, does not admit a representation in terms of genus-zero quantities.} in field theory amplitudes, while their appearance is obvious in computing genus-one string theory amplitudes, since the latter involve integrals over the moduli space of $n$-punctured  genus-one curves $\mathcal{M}_{1,n}$. A multitude of methods have recently been developed in computing these integrals as iterated integrals over either the punctures or the modular parameter of the torus. There has been progress in the computation of  amplitudes beyond genus-one in both string theory and field theory amplitudes, but the systematics in these cases are not as developed as in the genus-zero or genus-one cases.

An important tool among the methods above is the use of differential equations. In the context of string integrals, these are differential equations with respect to the moduli, e.g. the modular parameter $\tau$ parametrising the torus at genus one, or the vertex insertion points. Starting form these differential equations, one can perform a series-expansion in the inverse string tension $\ap$ for these integrals. In order to use this method, though, one needs a sufficiently large vector of integrals such that, under integration-by-parts (IBP) and partial-fractioning, one can define an integrable connection acting on such a vector of integrals. These  methods have been used with great success for tree-level and one-loop string integrals, as can be seen in  \rcites{Terasoma,Broedel:2013aza,Boels:2013jua,Puhlfuerst:2015zqw,Mafra:2019xms,Mafra:2019ddf,Broedel:2019gba,Kaderli:2019dny,Broedel:2020tmd,Britto:2021prf}. 

In the current work, we expand on the method of \rcite{Broedel:2020tmd} to the setting of a genus-one integral with multiple unintegrated punctures. In doing this, we are also generalizing some of the results of  \rcite{Britto:2021prf} into a genus-one setting. In this last reference, the authors studied open- and closed-string integrals with multiple unintegrated punctures and wrote them in terms of generating functions of multiple polylogarithms. In this paper, we are able to describe functions closely related to genus-one open string integrals with unintegrated punctures via a generating function of the elliptic multiple polylogarithms (eMPLs) $\tilde{\Gamma}\left(\begin{smallmatrix}
	k_1&\dots&k_r\\
	z_{1}&\dots&z_{r}
\end{smallmatrix}; z,\tau\right)$ of \rcite{Broedel:2017kkb}, and study some of their properties.

The functions studied in this work satisfy a KZB\footnote{For  Knizhnik,Zamolodchikov and Bernard, who first studied these equations \cite{Knizhnik:1984nr,Bernard:1987df}}  differential equations in $n$ punctures. This means that the differential equation itself provides a representation of a subalgebra\footnote{That is, we have a matrix representation of several elements of this algebra, with a few missing generators.} of the genus-one Drinfeld-Kohno algebra $\bar{\mathfrak{t}}_{1,n}$: this is a bigraded algebra that ensures the integrability of the differential equation \cite{KZB}. We find matrices that satisfy the algebra relations of $\bar{\mathfrak{t}}_{1,n}$ and that obey their grading, up to a shift. Moreover, the theory of the KZB equation explicitly describes the analytic continuation of its solutions. Thus, we are able to provide explicit identities for the analytic continuation of eMPLs.

\subsection{Outline}

In section 2, we review some of the special functions required to describe string integrals at tree level and one loop.  In particular, we describe how the integrals we study in this paper differ from string integrals in that we utilize meromorphic but  not doubly periodic functions for the integrands. This is in contrast with one-loop string integrals (after integration of loop momentum in the chiral splitting formalism \cite{Verlinde:1987sd,DHoker:1988pdl,DHoker:1989cxq}) , in which the integrands are doubly periodic but not meromorphic functions. In particular, the integrals we study can be seen as generating functions of the genus-one Selberg integrals of \rcite{Broedel:2019gba}.

In Section 3, we describe $\boldsymbol{Z}^\tau_{n,p}$, the vector of genus-one integrals with $n$ punctures on the elliptic curve, of which $p$ punctures are integrated over. The integrands of $\boldsymbol{Z}^\tau_{n,p}$ are built from certain products of meromorphic Kronecker-Eisenstein series $F(z,\eta,\tau)$. A related non-meromorphic version of these integrands has been conjectured to form an integrand basis, under integration by parts (IBP), for one-loop integrals before in \rcites{Mafra:2019xms,Mafra:2019ddf,Broedel:2020tmd}. In this section we succeed in using IBP relations and the Fay identity, a genus-one analog of partial fractioning, satisfied by the $F(z,\eta, \tau)$, to obtain a differential system for $\boldsymbol{Z}^\tau_{n,p}$.

We are furthermore able to compare the differential system of $\boldsymbol{Z}^\tau_{n,p}$ to the universal KZB equation, and find that the Schwarz integrability conditions of $\boldsymbol{Z}^\tau_{n,p}$ imply several of the commutation relations known from the theory of the KZB equation. Thus, we claim to find matrix-and-operator-valued representations $\boldsymbol{x}^{(k)}_{i,j}$ of the algebra underlying the KZB equations: the genus-one Drinfeld Kohno algebra $\bar{\mathfrak{t}}_{1,n}$ \cite{KZB}.  After showing how the initial values of our differential system for $\boldsymbol{Z}^\tau_{n,p}$ degenerates into tree-level string integrals, we end Section 3 with a worked out example of $\boldsymbol{Z}^\tau_{4,1}$.

In Section 4, we argue from two different viewpoints how to obtain back $\boldsymbol{Z}^\tau_{n,p}$  from a boundary value. In particular, we show how to recover the dependence of $\boldsymbol{Z}^\tau_{n,p}$ on the $n-p$ unintegrated punctures $z_i$ from a regularized initial (i.e. $z_i \rightarrow 0$) value $\boldsymbol{v}_{n,p}$ via generating functions of eMPLs. We find that this organization of the $z_i-$dependence (a) simplifies the $\tau$-dependence of $\boldsymbol{v}_{n,p}$  and moreover (b) for our particular representation $\boldsymbol{x}^{(n)}_{i,j}$ only one generating series of eMPLs suffices to describe all the $z_i-$dependence of $\boldsymbol{Z}^\tau_{n,p}$ for the branch choice $0=z_1<z_{p+2}<z_{p+3}<\ldots<z_n$ of the unintegrated punctures. We furthermore compare the computation of $\boldsymbol{Z}^\tau_{4,1}$ via this method to a direct-integration calculation.

In Section 5, we study the analytic continuation of our integrals $\boldsymbol{Z}^\tau_{n,p}$  away from our original branch choice $0=z_1<z_{p+2}<z_{p+3}<\ldots<z_n$.  With this, we extend the applicability of our solution $\boldsymbol{Z}^\tau_{n,p}$ beyond our initial choice of branch, or equivalently, find initial conditions compatible with a different branch choice. Moreover, we found these equations to  imply some interesting identities for the change of fibration basis of  eMPLs $\tilde{\Gamma}\left(\begin{smallmatrix}
	k_1&\dots&k_r\\
	z_{1}&\dots&z_{r}
\end{smallmatrix}; z,\tau\right)$ that we have been able to verify numerically.

In Section 6, we take a closer look to the matrix representations ${\boldsymbol{x}^{(n)}_{i,j},\boldsymbol{x}^{(0)}_j, \boldsymbol{\epsilon}^{(k)}}$ we obtain for the differential system of $\boldsymbol{Z}^\tau_{n,p}$ .  After building a dictionary between our representation and the mathematics literature, we note that our  representation respects the grading of the genus-one Drinfeld-Kohno algebra, $\bar{\mathfrak{t}}_{1,n}$ \cite{KZB}. Moreover, we can deduce the relations of Tsunogai's special derivation algebra from some combinations of these matrices 
 \cite{Tsunogai,Pollack}. 

In the appendices, we describe the details of the IBP and Fay identity toolkit and calculations that allow us to obtain the results of Section 3 and 4, and explain our particular choice of integration cycle for the \textit{integrated} punctures. In  \appref{app:DictionaryGenusOne} we lay out a dictionary relating our integrals $\boldsymbol{Z}^\tau_{n,p}$  to similar integrals appearing in the string amplitudes literature. The detailed derivation of the differential system satisfied by $\boldsymbol{Z}^\tau_{n,p}$ is documented in  \appref{app:DerivationOfDifferentialSystem}.  In  \appref{app:evalEq} we report on the eigenvalue equations that allow us to prove the simplicity of our initial values for $\boldsymbol{Z}^\tau_{n,p}$. In \appref{app:PoleSubraction} we detail a pole subtraction scheme for the direct integration of $\boldsymbol{Z}^\tau_{4,1}$.  In  \appref{app:fibrationBasisChange} we give some more detail on the analytic continuation of eMPLs from their generating functions. Finally, in  \appref{app:LastAppendixGenerlaIntegrationDomain}, we showcase some partial results about the different orderings of integration cycles for the \textit{integrated} punctures. In this last appendix we can see that some simplifications for the initial value of $\boldsymbol{Z}^\tau_{n,p}$ do not occur for generic integration contours.

\section{Review} \label{sec:ReviewSection}
\subsection{Tree-level open-string integrals}
In the computation of open-string amplitudes at tree level, one performs integrals over (ordered) marked points on the boundary of the upper-half plane, parametrising the vertex insertion points on the boundary of the genus-zero worldsheet. 
In the case of $n$ massless states, we need to perform integrals over the positions of $n$ ordered punctures along the real line. However, fixing residual gauge symmetries is taken into account by dividing out the volume of the conformal Killing group\footnote{$\textrm{SL}_2(\mathbb{R})$ is the  group of transformations  $z_j\rightarrow \frac{a z_j +b}{c z_j +d}$, for $a,b,c,d \in \mathbb{R}$ that keep the boundary of the upper-half plane fixed. The rational function $ \Phi(\{z_i\},\{s_{ij}\})$ is  $\textrm{SL}_2(\mathbb{R})-$covariant such that the whole integrand is $\textrm{SL}_2(\mathbb{R})-$invariant} $\textrm{SL}_2(\mathbb{R})$, which effectively fixes three punctures. The remaining $(n-3)-$fold string integral takes the form

\begin{align}\label{eqn:treeNpoint}
Z^\textrm{tree}_n= \int_{z_i<z_{i+1}}  \frac{ \dd z_1 \dd z_2 \dd z_3 \ldots \dd z_{n-1} \dd z_n}{\textrm{vol} \, \textrm{SL}_2(\mathbb{R})} \KN_{12\ldots n}  \,  \Phi(\{z_i\},\{s_{ij}\})   \, .
\end{align}
The factor $\Phi(\{z_i\},\{s_{ij}\})$ is a $\textrm{SL}_2(\mathbb{R})$-covariant, rational function of the punctures $\{z_i\}_{i=1}^{n}$ and other kinematic data such as polarizations $\{\epsilon_{i}\}_{i=1}^{n}$ of the external states and dimensionless Mandelstam variables
\begin{align}
s_{ij} = 2\,\alpha' k_i \cdot k_j\,,
\end{align}
where $k_i$ is the momentum of the $i$-th external state associated to the puncture $z_i$ and $\ap$ is a parameter proportional to the inverse of the string tension (and proportional to the square of the string length).
The other factor $\KN_{12\ldots n}$ is the Koba--Nielsen factor, which is a universal factor in $n$-point string integrals. For massless states, it takes the form
\begin{align}\label{eqn:treeKN}
	\KN_{12\ldots n} =  \prod_{1 \leq i<j\leq n} (z_j-z_i)^{ s_{ij}}\,.
\end{align}
When computing string integrals, we are usually interested in their $\ap$-expansion, that is, a series expansions in factors of $s_{ij}$.

Contemporary methods to compute tree-level string integrals rely on first rewriting the integrands $\Phi(\{z_i\},\{s_{ij}\}) $ in terms of a \textit{basis} of integrands whose $\ap$-expansion is known. This reduction into a basis of integrals is performed with integration by parts (IBP) and partial fractioning.  We are interested in IBP in the presence of a  Koba--Nielsen factor $\KN_{12\ldots n}$, amounting to the relation

\begin{align}\label{eqn:treeKNIBP}
	\int_\gamma  \frac{ \dd z_1 \dd z_2 \dd z_3 \ldots \dd z_{n-1} \dd z_n}{\textrm{vol} \, \textrm{SL}_2(\mathbb{R})}  \partial_{z_i} \big( \KN_{12\ldots n}  \Phi(\{z_i\},\{s_{ij}\})    \big) =  0 \, \quad \forall i  \, .
\end{align}
The integral vanishes because the Koba--Nielsen factor $\KN_{12\ldots n}$ vanishes at the endpoints (or boundary) of the integration contour $\partial \gamma$. This process of finding a basis of integrands modulo IBP-with-a-Koba--Nielsen and partial fraction identities is known to be well-defined in the genus-zero case \cite{Kaderli:2019dny}. These integrands define the basis of  an $(n-3)!$-dimensional vector space known as $(n-3)$th twisted cohomology group $H^{n-3}(\mathcal{M}_{0,n},\dd+\dd \log (\KN_{12\ldots n}))$. We will not make further use of this notation or the concepts of twisted cohomology and homology groups, but remark that these give a mathematically precise description to what we mean by a \textit{basis} of integrands under an ever-present Koba--Nielsen factor. For definitions of these twisted homology and cohomology groups, the reader is referred to  \rcites{aomoto1987,AomotoKita}, also see \cite{Mizera:2019gea,Mizera:2017cqs} for their introduction into the string-theory literature. 

A good candidate for this basis of integrands, originally found by Aomoto  \cite{aomoto1987}, and rediscovered by physicists \cite{Broedel:2013tta,Cachazo:2013gna}, is the basis of so-called Parke-Taylor factors\footnote{These are named for their similarity to the Parke-Taylor formula \cite{ParkeTaylor86} written in 4-dimensional spinor-helicity variables.}:

\begin{align}\label{eqn:parkeTaylorIntegrand}
\textrm{PT}(1,2, \ldots,n-1,n) = \frac{1}{(z_1-z_2)} \frac{1}{(z_2-z_3)} \ldots \frac{1}{(z_{n-1}-z_n)}\frac{1}{(z_n-z_1)}  \, .
\end{align}
One can form a $(n-3)!$-dimensional basis of integrands  by permuting the labels $(2,3,\ldots,n-2)$. With this basis in mind, the integral in \eqn{eqn:treeNpoint} can be written as

\begin{align}\label{eqn:treeNpointInPTBasis}
	Z^\textrm{tree}_n= \sum_{\rho \in S_{n-3}} b_\rho(\{s_{ij}\}) \int_{z_i<z_{i+1}}  \frac{ \dd z_1 \dd z_2 \dd z_3 \ldots \dd z_{n-1} \dd z_n}{\textrm{vol} \, \textrm{SL}_2(\mathbb{R})} \KN_{12\ldots n}  \, \textrm{PT}\left(1,\rho\left(2,\ldots,n-2\right),n-1,n\right)   \, ,
\end{align}
where the coefficients $b_\rho(\{s_{ij}\})$ are rational functions of the kinematic data, and can be found via IBP and partial fractioning, say with the methods of  \rcites{Cardona2016,SchlottererEinsteinYM2016,HeTengZhang2019}. 

A last step towards computing the $\ap$-expansion of the string integral of \eqn{eqn:treeNpointInPTBasis} is to compute each of the basis integrals. The integrals with a Parke-Taylor basis -- also known as $Z$-theory amplitudes -- have an $\ap$-expansion detailed in \rcite{Mafra:2016mcc}. We note that other alternative bases of integrals and their $\ap$-expansion have also been studied in the literature\cite{Oprisa:2005wu,Broedel:2013aza,Boels:2013jua,Kaderli:2019dny}.   

\subsection{Multiple polylogarithms and multiple zeta values}\label{subsec:g0}

We proceed to define a family of genus-zero iterated integrals closely related to the string integrals in \eqn{eqn:treeNpointInPTBasis}.
These iterated integrals are called multiple polylogarithms (MPLs) and recursively defined as follows \cite{Goncharov:2001iea}:
\begin{align}\label{eqn:Gmpldef}
	G(a_1,a_2,\ldots,a_n;z) = \int_{0}^{z} \dd t  \frac{1}{t-a_1}  G(a_2,a_3,\ldots ,a_n ; t)\, ,
\end{align}
with a base case $G(;z)=1$. A key property of MPLs, by virtue of being iterated integrals, is that they satisfy the shuffle product. For $A=(a_1,a_2,\ldots,a_n)$ and $B=(b_1,b_2,\ldots,b_m)$, we have:
\begin{align}\label{eqn:GmplShuffle}
	G(A;z) G(B;z) = \sum_{C\in A \shuffle B }  G(C;z)\, ,
\end{align}
where $A \shuffle B $ is the set of all permutations of $(a_1,\ldots , a_n, b_1,\ldots, b_m)$ that preserve the original ordering of the $a_i \in A$ and $b_j \in B$. An example of this shuffle product is given by:
\begin{align}\label{eqn:GmplShuffleExample}
	G(a_1;z) G(b_1,b_2;z) =G(a_1,b_1,b_2;z ) + G(b_1,a_1,b_2;z )+G(b_1,b_2,a_1;z )  \, .
\end{align}

One of the simplest example of an MPL is given by $G(1;z)=\log(1-z)$. The  apparently simpler case $G(0;z)$ has a basepoint divergence, but this divergence can be regularized via tangential basepoint regularization \cite{Panzer:2015ida}. The regularized value assigned to this MPL then becomes
\begin{align}\label{eqn:Gregularized}
G(0;z) = \log(z) \, .
\end{align}
Other MPLs that have a similar basepoiont divergence are regularized via shuffling with this regularized MPL above.

Finally, when the letters of an MPL $G(a_1,a_2,\ldots,a_n;z)$ are either $0$ or $1$, i.e.\ $a_i\in \{0,1\}$, and we furthermore integrate along the unit interval setting $z=1$, the MPLs evaluate to multiple zeta values (MZVs) 
\begin{align}\label{eqn:zetaFromMPLs}
\zeta_{n_1,n_2,\ldots , n_r}&=\sum_{1 \leq m_1 < m_2 < \ldots m_r} \frac{1}{{m_1}^{n_1}{m_2}^{n_2} \ldots {m_r}^{n_r}} \nnl &= (-1)^r G(0^{n_r-1},1,0^{n_{r-1}-1},1, \ldots , 0^{n_1-1} ,1 ; 1)
 \,,
\end{align}
where  $0^{n}$ denotes a string with $n$ zeroes, and $\zeta_{n_1,n_2,\ldots , n_r}$. The sum in \eqn{eqn:zetaFromMPLs} only converges if $n_r \geq 2$. Similar to \eqn{eqn:Gregularized}, the divergent values associated to MZVs can be regularised using the definition
\begin{align}\label{eqn:MZVregularized}
G(1;1) = 0
\end{align}
and the shuffle algebra for the remaining integrals with $n_r = 1$.

\subsection{Elliptic multiple polylogarithms  and A-cycle elliptic multiple zeta values}\label{subsec:eMPL}

We will proceed to define to define an elliptic -- or genus-one -- analogue of the MPLs from \eqn{eqn:Gmpldef}. We will follow the definitions of \rcites{Broedel:2017kkb,Mafra:2018pll}. But first, we will clarify the setting in which these iterated integrals are defined. 

Take $\tau$ to be a  complex number in the upper-half plane $\tau \in \mathbb{H}$. Then, for $z \in \mathbb{C}$ we can define doubly periodic, but not necessarily meromorphic functions $f(z,\tau)$ to be complex-valued functions such that 
\begin{align}\label{eqn:ellipticFunctionFDefinition}
f(z,\tau)=f(z+1,\tau)=f(z+\tau,\tau) \, .
\end{align}
For convenience, let's define the lattice
\begin{align}
\Lambda_\tau&=\mathbb{Z}+\tau \mathbb{Z}\,.
\end{align}
We note that elliptic functions are unchanged under displacements of $z$ by lattice points $\Lambda_\tau$. Thus, for an elliptic function it is natural to define the domain of $z$ to be in the fundamental parallelogram, or torus $\mathbb{C}/\Lambda_\tau$. 

That being said, we will mostly not deal with doubly periodic functions in the present work, but rather with quasi-periodic functions, periodic under $z \rightarrow z+1$ but with a well-defined transformation rule for $z \rightarrow z+ \tau$. Whenever this transformation rule is multiplicative, i.e.
\begin{align}\label{eqn:QuasiellipticTwistedFunctionFDefinition}
	h(z+1,\tau)&=h(z,\tau)\, , \nnl  h(z+\tau,\tau)  &= a (z,\tau)h(z,\tau)
\end{align}
with nonzero factor $a(z,\tau)$, the corresponding quasiperiodic functions can be understood to live in the universal cover of the torus, i.e. $\mathbb{C}$.

A quasi-periodic function we will extensively use is the Kronecker-Eisenstein series \cite{Kronecker,Weil:1976,ZagierF}
\begin{align}\label{eqn:KroneckerEisensteinDefn}
	F(z,\eta,\tau)= \frac{\theta_1(z+\eta,\tau) \theta'_1(0,\tau) }{\theta_1(z,\theta)\theta_1(\eta,\theta) }    \,,
\end{align}
where $\theta_1$ is the odd Jacobi theta function
\begin{align}\label{eqn:oddJacobiDefn}
\theta_1(z,\tau) =   2 q^{1/8} \sin(\pi z) \prod_{n=1}^{\infty}(1-q^n)(1-q^n e^{2 \pi i z})(1-q^n e^{- 2 \pi i z})
\end{align}
with
\begin{align}
q &= e^{2 \pi i \tau}
\end{align}
and $'$ denotes differentiation with respect to the first argument. This Kronecker-Eisenstein series is symmetric under an exchange of its first two arguments $F(z,\eta,\tau)=F(\eta,z,\tau)$, and has the following quasi-periodicity properties:
\begin{align}\label{eqn:KroneckerEisensteinQuasiP}
	F(z+1,\eta,\tau)&=  	F(z,\eta,\tau)   \, ,\nnl   	F(z+\tau,\eta,\tau) &= 	e^{- 2 \pi i  \eta }F(z,\eta,\tau)    \, .
\end{align}
The function $F(z,\eta,\tau)$ has a residue of 1 around $z=0$. One can obtain a doubly periodic but non-meromorphic function from $	F(z,\eta,\tau) $ by multiplying the appropriate prefactor \cite{BrownLev}

\begin{align}\label{eqn:KroneckerEisensteinDoublyP}
	\Omega(z,\eta,\tau)=  	e^{2 \pi i  \eta \frac{\Im z}{\Im \tau}}F(z,\eta,\tau)   \, .
\end{align}
The meromorphic Kronecker-Eisenstein series satisfies the mixed heat equation\footnote{The doubly periodic Kronecker-Eisenstein series $\Omega(u \tau + v,\eta,\tau)$ satisfies this equation with $\partial_v$ replacing  the $\partial_z$.}
\begin{align}\label{eqn:KroneckerEisensteinMixedHeat}
	2 \pi i \partial_\tau F (z,\eta,\tau)= \partial_z \partial_\eta F(z,\eta,\tau)   \, ,
\end{align}
and the Fay identity\footnote{The Fay identity also holds for  $\Omega(z,\eta,\tau)$.}
\begin{align}\label{eqn:KroneckerEisensteinFayIdentity}
F(z_{13},\eta_1,\tau)F(z_{23},\eta_2,\tau) = F(z_{13},\eta_{12},\tau)F(z_{21},\eta_{2},\tau)+ F(z_{23},\eta_{12},\tau)F(z_{12},\eta_{1},\tau)     \, ,
\end{align}
where we introduce short-hand notations for differences of punctures
\begin{align}
z_{ij}&=z_i-z_j
\end{align}
and for sums of $\eta$-variables
\begin{align}
\eta_{12\ldots n}&=\eta_{1}+\eta_{2}+\ldots+\eta_{n}\,.
\end{align}
The Fay identity \eqref{eqn:KroneckerEisensteinFayIdentity} should be considered to be a genus-one version of the partial fractioning identities that the functions $1/z_{ij}$ satisfy. In fact, if we replace every instance of $F(z,\eta,\tau)$ in \eqn{eqn:KroneckerEisensteinFayIdentity} by $1/z$, we obtain the familiar partial fraction identity:

\begin{align}\label{eqn:PartialFractions}
	\frac{1}{z_{13}}\frac{1}{z_{23}}=\frac{1}{z_{13}}\frac{1}{z_{21}} + \frac{1}{z_{23}}\frac{1}{z_{12}}     \, .
\end{align}

Expanding as a series in the auxiliary variable $\eta$, the Kronecker-Eisenstein series $F(z,\eta,\tau)$ is a generating function for an infinite class of functions $g^{(k)}(z,\tau)$:
\begin{align}\label{eqn:KroneckerEisensteinGenSeries}
	\eta F(z,\eta,\tau)=  \sum_{k=0}^{\infty} \eta^k g^{(k)}(z,\tau)  \, .
\end{align}
The functions $g^{(k)}(z,\tau)$ satisfy the symmetry relation
\begin{align}
g^{(k)}(-z,\tau)&=(-1)^{k} g^{(k)}(z,\tau)\,.
\end{align}
The first two examples are
\begin{align}\label{eqn:gKernelsFirstFewTerms}
g^{(0)}(z,\tau)&=1  \,,  \nnl g^{(1)}(z,\tau) &= \partial_z \log(\theta_1(z,\tau)) \, . 
\end{align}
Crucially, the only function $g^{(k)}(z,\tau)$ that has a simple pole at $z=0$ is $g^{(1)}(z,\tau)$, while the remaining functions are regular at $z=0$ for  $k\neq 1$. For other positive values of $k$, these functions have simple poles at $z=n \tau$, for $n\in \mathbb{Z}\backslash\{0\}$, as can be seen from their quasi-periodicity following from \eqn{eqn:KroneckerEisensteinQuasiP}:
\begin{align}\label{eqn:gKernelsQuasiP}
	g^{(k)}(z+1,\tau)& = g^{(k)}(z,\tau) \,  ,   \nnl	g^{(k)}(z+\tau,\tau)&=  \sum_{n=0}^{k} \frac{(-2 \pi i)^{n} }{n!}g^{(k-n)}(z,\tau)     \, . 
\end{align}
Also the Fay identity \eqref{eqn:KroneckerEisensteinFayIdentity} has an echo at the level of the functions $g^{(k)}(z,\tau)$. Using the short-hand notation
\begin{align}
g^{(n)}_{ij} &= g^{(n)}(z_{ij},\tau)
\end{align}
these identities are:
\begin{align}\notag
	  g^{(n)}_{12} g^{(m)}_{23}   = -g^{(n+m)}_{13} &+\sum_{j=0}^{n}(-1)^j  {m-1+j \choose j} g^{(n-j)}_{13} g^{(m+j)}_{23}\\
	      &  + \sum_{j=0}^{m}(-1)^j  {n-1+j \choose j} g^{(m-j)}_{13} g^{(n+j)}_{12} \, .
\label{eqn:gKernelsFayIdentities} 
\end{align}
The functions $g^{(k)}(z,\tau)$ evaluate to Eisenstein series when $z\rightarrow 0$:

\begin{align}\label{eqn:gToZeroEisenstein}
g^{(k)}(0,\tau) = -G_{k}(\tau)  \,,  \quad \forall k \in \mathbb{Z}_{\geq 2},
\end{align}
where the  Eisenstein series $G_k(\tau)$ are defined\footnote{The Eisenstein series $G_2(\tau)$ is not absolutely convergent, and requires a summation prescription. This is given by $G_2(\tau)=\sum_{n\in \mathbb{Z}\backslash \{0\}}\frac{1}{n^2}+\sum_{m\in \mathbb{Z}\backslash \{0\} }\sum_{n\in\mathbb{Z}} \frac{1}{(n+m \tau )^2}$ \cite{Mafra:2018pll} .} by
\begin{align}
\label{eqn:GkDef}
G_k(\tau)=\begin{cases}
-1&\text{if }k=0\,,\\
\sum\limits_{\begin{smallmatrix}n,m\in \ZZ\\
	(n,m)\neq(0,0)\end{smallmatrix}}\frac{1}{(n+m\tau)^k}&\text{if }k\in 2\ZN\,,\\
0&\text{if }k\in 2\ZN-1\,.
\end{cases}
\end{align}

Similarly, one can consider the non-meromorphic function $\Omega(z,\eta,\tau)$ defined in \eqn{eqn:KroneckerEisensteinDoublyP} to be a generating function of doubly-periodic functions $ f^{(k)}(z,\tau)$:
\begin{align}\label{eqn:KroneckerEisensteinNonMeroGenSeries}
	\eta \Omega(z,\eta,\tau)=  \sum_{k=0}^{\infty} \eta^k f^{(k)}(z,\tau)  \, .
\end{align}
These doubly-periodic functions $ f^{(k)}(z,\tau)$ also satisfy Fay relations, are non-meromorphic and coincide with the $ g^{(k)}(z,\tau)$ when $z\in \mathbb{R}$. 

We now recursively define\footnote{One can define iterated integrals $\Gamma \left(\begin{smallmatrix}
	k_1&k_2&\dots&k_r\\
	z_{1}&z_{2}&\dots&z_{r}
	\end{smallmatrix}; z,\tau\right)$ by using the doubly-periodic functions $ f^{(k)}(z,\tau)$ instead of the functions $ g^{(k)}(z,\tau)$ \cite{Broedel:2014vla}. These iterated integrals of non-meromorphic kernel are not homotopy invariant -- however, they can be lifted to be homotopy invariant integrals \cite{BrownLev,Broedel:2018izr}.  } iterated integrals of the integration kernels $g^{(k)}(z,\tau)$ as follows  \cite{Broedel:2017kkb}: 
\begin{align}\label{eqn:eMPLdef}
\tilde{\Gamma}\left(\begin{smallmatrix}
	k_1&k_2&\dots&k_r\\
	z_{1}&z_{2}&\dots&z_{r}
\end{smallmatrix}; z,\tau\right) = \int_{0}^{z} \dd t \,g^{(k_1)}(t-z_1)  
\tilde{\Gamma}\left(\begin{smallmatrix}
	k_2&k_3&\dots&k_r\\
	z_{2}&z_{3}&\dots&z_{r}
\end{smallmatrix}; t,\tau\right) \, ,
\end{align}
where the base case is given by $\tilde{\Gamma}(;z,\tau)=1$. We will refer to these iterated integrals as elliptic multiple polylogarithms (eMPLs) in the present work, even if they are not elliptic in the usual sense of the word. For an eMPL $\tilde{\Gamma} \left(\begin{smallmatrix}
	k_1&k_2&\dots&k_r\\
	z_{1}&z_{2}&\dots&z_{r}
\end{smallmatrix}; z,\tau\right)$, we define its length by $r$ and its weight by $k_1+k_2+\ldots+k_r$. The eMPLs  obey the shuffle algebra \eqref{eqn:GmplShuffle}, by virtue of being iterated integrals, but now shuffling each entry $\left( \begin{smallmatrix}
	k_r\\
	z_{r}
\end{smallmatrix} \right)$ as an individual letter, e.g.:
\begin{align}\label{eqn:GammaEmplShuffleExample}
	\tilde{\Gamma}\left(\begin{smallmatrix}
		n_1\\
		a_{1}
	\end{smallmatrix}; z,\tau\right) \tilde{\Gamma}\left(\begin{smallmatrix}
	m_1 & m_2\\
	b_{1} & b_2
\end{smallmatrix}; z,\tau\right)  =
	\tilde{\Gamma}\left(\begin{smallmatrix}
	n_1 & m_1 & m_2\\
	a_{1} & b_1 & b_2
\end{smallmatrix}; z,\tau\right)
+
\tilde{\Gamma}\left(\begin{smallmatrix}
	m_1 &n_1  & m_2\\
	b_1 & a_{1}  & b_2
\end{smallmatrix}; z,\tau\right)
+
\tilde{\Gamma}\left(\begin{smallmatrix}
	m_1 &m_2  & n_1\\
	b_1 & b_{2}  & a_1
\end{smallmatrix}; z,\tau\right)
 \, .
\end{align}
Like their genus-zero counterpart, eMPLs sometimes require regularization. We employ   tangential basepoint regularization as in \rcite{Broedel:2019tlz}, leading to the definition
\begin{align}\label{eqn:Gamma01reg}
\tilde{\Gamma}\left(\begin{smallmatrix}
	1\\
	0
\end{smallmatrix}; z,\tau\right)   \,  = \log(1-e^{2\pi i z}) - \pi i z + 2 \sum_{k,l>0} \frac{1}{ k}\left(1-\cos(2 \pi k z)\right) q^{k l} \, .
\end{align}
Numerical implementations of eMPLs, e.g. \rcite{Weinzierl_numerical_elliptic}, use this regularization, which is compatible with the shuffle product of eMPLs.

Finally, we can obtain elliptic analogues of MZVs by evaluating the eMPLs in the limit $z\to 1$ along the unit interval. In particular, we have:
\begin{align}\label{eqn:AcycleEMZVdefn}
 \lim_{z \rightarrow 1} \tilde{\Gamma} \left(\begin{smallmatrix}
	k_1&k_2&\dots&k_r\\
	0&0&\dots&0
\end{smallmatrix}; z,\tau\right) = \omega_A(k_r,k_{r-1},\ldots,k_1 ; \tau) \, ,
\end{align}
where the $A$ subscript denotes that the elliptic multiple zeta value (eMZV) has been evaluated by performing an integral along the $A$-cycle of the torus (i.e. from $0$ to $1$ along the unit interval). These eMZVs were first defined and studied by Enriquez \cite{Enriquez:Emzv}. We note that they will require some non-trivial endpoint regularization if $k_1=1$ and/or $k_r=1$, culminating in the definition
\begin{align}\label{eqn:AcycleEMZVregdefn}
\omega_A(1)&=0\,.
\end{align}
The remaining (regularized) eMZVs can be deduced from the well-defined instances of \eqns{eqn:AcycleEMZVdefn}{eqn:AcycleEMZVregdefn} and the shuffle algebra \eqref{eqn:GmplShuffle}. In this work we will drop the $A$-subscript from now on, and sometimes omit the  dependence of the  last variable $\tau$. Because our notation of eMZVs has the last variable written after a semicolon, there should be no an ambiguity when we omit $\tau$.

\subsection{$Z_n^{\tau}$-integrals}
\label{subsec:ZnTauIntegrals}
We will now define a genus-one analogue of the integrals in \eqn{eqn:treeNpoint}. For $\sigma \in S_{n-1}$ a permutation of $\{2,3,\ldots,n\}$, these will be of the form\footnote{We use $\tilde{\varphi}^\tau$ throughout this work to not clash with the $\varphi^\tau$ of \rcite{Gerken:2020xfv}. While $\varphi^\tau$ and $\tilde{\varphi}^\tau$ agree on the $A$-cycle, we note that the holomorphicity of the latter makes it easier to relate to the mathematical literature of KZB equations and twisted (co)homology. See \secref{sec:ConnectionMathLit} and \rcites{ManoWatanabe2012,FelderVarchenkoKZB95}. }:
\begin{align}\label{eqn:TauNpoint}
	Z_{n}^{\tau}(1,\sigma(2),\ldots , \sigma(n))= \int_{0<z_2<z_3<\ldots<z_n<1}  \dd z_2 \dd z_3 \ldots \dd z_n\, \KN^\tau_{12\ldots n}  \, \tilde{\varphi}^\tau(1,\sigma(2),\dots, \sigma(n))\,  ,
\end{align}
where $\KN^\tau_{12\ldots n}$ will be a genus-one version of the tree-level Koba--Nielsen factor from \eqn{eqn:treeKN}:

\begin{align}
	\KN^\tau_{12\ldots n}&=\exp\Big(-\sum\limits_{1\leq j<i\leq n} s_{ij}\CG^\tau_{ij}\Big)\, ,
	\label{eqn:KNloopSec2}
\end{align}
and where $\CG_{ij}^\tau$ is given by\footnote{This is a meromorphic version of the open-string Green's function. }
\begin{align}
	\CG^\tau_{ij}&=\tilde{\Gamma}\big(\begin{smallmatrix}1\\0\end{smallmatrix};z_{ij}\,,\tau\big)-\omega(1,0,\tau) 
	\, .
	\label{GFloopSec2} 
\end{align}
Furthermore, the integrands $\tilde{\varphi}^\tau(1,\sigma(2),\dots, \sigma(n))$ are written as chains of meromorphic Kronecker-Einsenstein series:
\begin{align}\label{eqn:EKchain}
	\tilde{\varphi}^\tau(a_1,a_2,\dots,a_m)&=\prod_{i=2}^m F(z_{a_{i-1}}-z_{a_i},\eta_{a_i,\dots,a_m},\tau)\,.\nonumber\\
	\tilde{\varphi}^\tau(a_1)&=1\,.
\end{align}
The auxiliary variables 
\begin{align}
	\eta_{a_i,\dots,a_m}&=\eta_{a_i}+\eta_{a_{i+1}}+\dots +\eta_{a_m} \, ,
\end{align}
in the Eisenstein--Kronecker series in \eqn{eqn:EKchain} are linear combinations of auxiliary variables $\eta_i$ associated\footnote{$\tilde{\varphi}^\tau$ is quasiperiodic $\tilde{\varphi}^\tau \rightarrow  \exp(2 \pi i \eta_j) \tilde{\varphi}^\tau  $ under $z_j \rightarrow z_j+\tau$. } to the punctures $z_i$, such that the chains $\tilde{\varphi}^\tau$ satisfy the following algebraic properties for two disjoint sequences $A$ and $B$ and a label $r\not \in A\cup B$:
\begin{align}
	\tilde{\varphi}^\tau(r,A)\tilde{\varphi}^\tau(r,B)&=\tilde{\varphi}^\tau(r,A\shuffle B)\,.
\end{align}
This identity is obtained by consecutive applications of the Fay identity \eqref{eqn:KroneckerEisensteinFayIdentity}. 

As defined here, these integrals coincide with the planar integrals of \rcites{Mafra:2019ddf,Mafra:2019xms}, but differ when considering non-planar integrals\footnote{\textit{Planar integrals} are integrals where  the punctures$z_i$ are integrated along the interval $z_i \in(0,1)$. \textit{Non-planar} integrals have a subset of the punctures integrated along the interval $z_i \in (\tau/2,\tau/2+1)$. This terminology is borrowed from \rcite{Green:1987mn}.}. We spell out the dictionary between the $Z_n^{\tau}$ integrals described in this section and the integrals of \rcites{Mafra:2019ddf,Mafra:2019xms} to  \appref{app:DictionaryGenusOne}.

The genus-one integrals of \eqn{eqn:TauNpoint} have two key properties. First, they can be seen as generating functions of (some of) the genus-one Selberg integrals of \rcite{Broedel:2019gba}, by looking at some of the components of their $\eta$-expansion. More importantly, the inclusion of all possible permutations $\sigma$ and forming the vector of integrals\footnote{We use bold face to refer to vectors.}
\begin{align}\label{eqn:TauNpointvec}
\pmb{Z}_{n}^{\tau}&=\begin{pmatrix}
Z_{n}^{\tau}(1,\sigma(2),\ldots , \sigma(n))
\end{pmatrix}_{
	\sigma\in S_{n{-}1}}
\end{align}
from \eqn{eqn:TauNpoint} leads to a closed differential equation upon differentiating $\pmb{Z}_{n}^{\tau}$ with respect to $\tau$ (and using IBP and Fay identities). This differential equation can be solved recursively to compute the $\ap$-expansion of $\pmb{Z}_{n}^{\tau}$ and, thus, the integrals from \eqn{eqn:TauNpoint}. This last fact about the integrals $	Z_{n}^\tau$ gives evidence that the vector of integrands described by the $ \tilde{\varphi}^\tau(1,\sigma(2),\dots, \sigma(n)) $ form a \textit{basis} of integrands at genus-one \cite{Mafra:2019ddf,Mafra:2019xms}. Chapter 6 of \cite{KaderliThesis1919660} contains a proof to this statement in the doubly-periodic version of these integrals.

\section{$Z_n^{\tau}$-integrals with multiple unintegrated punctures}
In this section, we introduce further classes of integrals defined on the punctured $A$-cycle of a torus\footnote{In the thesis \cite{KaderliThesis1919660} of one of the authors of this paper, various results of this section solely worked out by the author of the thesis have been published already, referring to this paper.}. On the one hand, as shown in detail in the thesis \cite{KaderliThesis1919660}, they generalize the open-string configuration-space integrals $Z_n^{\tau}$ from \eqn{eqn:TauNpoint} at genus one to multiple unintegrated punctures. On the other hand, they are the genus-one analogues of the genus-zero integrals investigated in \rcite{Britto:2021prf}.

\subsection{Construction of $Z^{\tau}_{n,p}$-integrals}
Let us consider $n$ distinct punctures on the unit interval (i.e.\ the $A$-cycle), where the first puncture $z_1=0$ is fixed at the origin (by the translation invariance of the torus) and the remaining ones are ordered according to\footnote{This order defines the branch choice considered. However, the integrals can be analytically continued away from this specific choice, see  \secref{sec:AnalyticConti}.}
\begin{align}\label{eqn:domain}
0&=z_1<z_2<z_3<\dots<z_n<1\,.
\end{align}
%

In \rcite{Broedel:2020tmd}, the $Z_n^\tau$-integrals of \eqn{eqn:TauNpoint} were modified by introducing an additional unintegrated puncture $z_0$ on the A-cycle, which  lead to a recursive method to calculate the $Z_n^\tau$-integrals based on the elliptic KZB associator \cite{Broedel:2019gba,Broedel:2020tmd}. We extend this modification and define for each 
\begin{align}
0&\leq p\leq n-1
\end{align}
a class of iterated integrals, where the $p$ punctures $z_2,z_3,\dots, z_{p+1}$ out of the $n$ punctures from \eqn{eqn:domain} are integrated over and the remaining $n{-}p$ punctures $z_1,z_{p+2},z_{p+3},\dots,z_n$  are kept unintegrated. The integrals $Z^{\tau}_{n,p}$ defined this way will be the central objects of this work, and are given by \cite{KaderliThesis1919660}:
\begin{align}\label{def:Znp}
	&Z^{\tau}_{n,p}((1,A^1),(p+2,A^{p+2}),(p+3,A^{p+3}),\dots,(n,A^n); z_{p+2},z_{p+3},\dots,z_n)\nonumber\\
	&=\int_{0<z_i<z_{i+1}<z_{p+2}}\prod_{i=2}^{p+1}\dd z_i\, \KN^{\tau}_{12\dots n}\,\tilde{\varphi}^\tau(1,A^1)\prod_{k=p+2}^n\tilde{\varphi}^\tau(k,A^k)\,,
\end{align}
where for $k\in \{1,p+2,p+3,\dots, n\}$ the sequences $A^k$ are disjoint, possibly empty subsequences of all the possible permutations of the $p$ labels $(2,3,\dots, p+1)$ of the integrated punctures. In other words, for each $(n{-}p)$-tuple of (sub)sequences $(A^1,A^{p+2},\dots,A^n)$ there exists exactly one permutation $\sigma\in S_p$ (and vice-versa) acting on the $p$-tuple $(2,3,\dots, p+1)$ such that the following identity of $p$-tuples holds
\begin{align}
	(A^1,A^{p+2},\dots,A^n)&=\sigma (2,3,\dots, p+1)\,.
\end{align}
This defines\footnote{This dimension is an  $n\rightarrow n+2$ offset from the genus-zero result from \rcite{Britto:2021prf}. A reason for this is that the Parke-Taylor factors \eqref{eqn:parkeTaylorIntegrand} in the genus-zero setting have a genus-one counterpart in chains of Kronecker-Eisenstein series \eqref{eqn:EKchain}, but at genus-zero we have in addition to the fixed puncture at $0$ two further unintegrated punctures that are usually $SL_2(\mathbb{R})$-fixed at $1$ and $\infty$, explaining the offset of two. One can also define $d_{n,p}$ natural integration contours along the $A-$cycle, following the analogy with the genus-zero case of \rcite{Britto:2021prf}.} 
\begin{align}\label{eqn:dimnp}
d_{n,p}&=\frac{(n-1)!}{(n-1-p)!}
\end{align}
distinct integrals, which can be written as one vector of integrals
\begin{align}\label{def:ZnpVector}
&\pmb{Z}^{\tau}_{n,p}(z_{p+2},z_{p+3},\dots,z_n)\nnl
&=\begin{pmatrix}
Z^\tau_{n,p}((1,A^1),(p+2,A^{p+2}),\dots,(n,A^n); z_{p+2},z_{p+3},\dots,z_n)
\end{pmatrix}_{
	(A^1,A^{p+2},\dots,A^n)=\sigma (2,3,\dots, p+1)}\,,
\end{align}
where $\sigma\in S_p$ is a permutation of the $p$ indices ${2,3,\dots, p+1}$. We remark that the integrals $\pmb{Z}^{\tau}_{n,p}$ defined above coincide (with the relabeling $z_n=z_0$ ) with the $\pmb{Z}^{\tau}_{0,n-1}$ integrals of \rcite{Broedel:2020tmd} for the case $(n,p)=(n,n-2)$. Additionally, the case $(n,p)=(n,n-1)$ yields the integrals $Z_n^\tau$ of \eqn{eqn:TauNpoint}:
\begin{align}
\pmb{Z}^{\tau}_{n,n-1}&=\pmb{Z}_n^\tau\,.
\end{align}

\subsection{Differential system}\label{subsec:diffSystem}
The vector $\pmb{Z}^{\tau}_{n,p}(z_{p+2},z_{p+3},\dots,z_n)$ from \eqn{def:ZnpVector} satisfies a closed differential system \cite{KaderliThesis1919660} which is reminiscent of an elliptic KZB system on the $(n{-}p)$-punctured torus \cite{Enriquez:EllAss}: 
\begin{subequations}
	\label{eqn:ellKZBeq}
	\begin{align}\label{eqn:ellKZBeqzi}
	\partial_i \boldsymbol{Z}^\tau_{n,p}&=\left(\boldsymbol{x}_{i}^{(0)}+\sum_{k\geq 1}\sum_{\begin{smallmatrix}
		r\in \{1,p+2,\dots,n\}\\r\neq i
		\end{smallmatrix}}\boldsymbol{x}_{ir}^{(k)}g^{(k)}_{ir}\right)\boldsymbol{Z}^\tau_{n,p}\,,\\
	\label{eqn:ellKZBeqtau}
	2\pi i\partial_{\tau} \boldsymbol{Z}^\tau_{n,p}&=\left(-\boldsymbol{\epsilon}^{(0)}+\sum_{k\geq 4}(1-k)\boldsymbol{\epsilon}^{(k)} G_k +\sum_{\begin{smallmatrix}
		r,q\in \{1,p+2,\dots,n\}\\
		q< r
		\end{smallmatrix}}\sum_{k\geq 2 }(k-1)\boldsymbol{x}_{qr}^{(k-1)}g^{(k)}_{qr}\right)\boldsymbol{Z}^\tau_{n,p}\,,
	\end{align}
\end{subequations}
where $\partial_i=\partial_{z_i}$ for $p+2\leq i\leq n$ . The \mbox{$(d_{n,p}\times d_{n,p})$}-dimensional square matrices $\boldsymbol{x}_{i}^{(0)}$, $\boldsymbol{x}_{ir}^{(k)}$ and $\boldsymbol{\epsilon}^{(k)}$ are explicitly known and calculated in \appref{app:ziDeriv} and \appref{app:tauDeriv}, respectively. The explicit formulae are given in \eqn{eqn:derZnpClosed} and \eqn{eqn:tauDerivClosed}, respectively. For $k\geq 1$ their entries are polynomials of degree $k{-}1$ and $k{-}2$ in the auxiliary variables $\eta_i$. The entries of $\boldsymbol{x}_{i}^{(0)}$ are linear combinations of $\eta_i^{-1}$ and first-order derivatives $\partial_{\eta_{i}}$. The matrix $\boldsymbol{\epsilon}^{(0)}$ is a linear combination of $\eta_i^{-2}$, second-order derivatives with respect to $\eta_i$ and factors of $\zeta_2$ in the diagonals. Moreover, all the entries of the matrices $\boldsymbol{x}_{i}^{(0)}$, $\boldsymbol{x}_{ir}^{(k)}$ and $\boldsymbol{\epsilon}^{(k)}$ are linear combinations of the Mandelstam variables $s_{ij}$. Accordingly, the matrices are proportional to $\alpha'$:
\begin{align}
\boldsymbol{x}_{i}^{(0)},\,\boldsymbol{x}_{ir}^{(k)},\,\boldsymbol{\epsilon}^{(k)}&\propto \alpha'\,.
\end{align}
\subsubsection{Commutation relations}\label{subsec:comRel}
In \appref{app:comRel}, we show how the Schwarz integrability conditions
\begin{align}\label{eqn:comij}
[\partial_j,\partial_i] \boldsymbol{Z}^\tau_{n,p}&=0
\end{align}
and
\begin{align}\label{eqn:comTaui}
[\partial_{\tau},\partial_i] \boldsymbol{Z}^\tau_{n,p}&=0
\end{align}
for a general system of differential equations of the form
\begin{subequations}\label{eqn:ellKZBeqGeneral}
\begin{align}
\partial_i \boldsymbol{Z}^\tau_{n,p}&=\left(\boldsymbol{x}_{i}^{(0)}+\sum_{k\geq 1}\sum_{\begin{smallmatrix}
	r\in \{1,p+2,\dots,n\}\\r\neq i
	\end{smallmatrix}}\boldsymbol{x}_{ir}^{(k)}g^{(k)}_{ir}\right)\boldsymbol{Z}^\tau_{n,p}\,,\\
2\pi i\partial_{\tau} \boldsymbol{Z}^\tau_{n,p}&=\left(-\boldsymbol{\epsilon}^{(0)}+\sum_{k\geq 4}(1-k)G_k \boldsymbol{\epsilon}^{(k)}+\sum_{\begin{smallmatrix}
	r,q\in \{1,p+2,\dots,n\}\\
	q< r
	\end{smallmatrix}}\sum_{k\geq 2 }(k-1)\boldsymbol{b}_{qr}^{(k)}g^{(k)}_{qr}\right)\boldsymbol{Z}^\tau_{n,p}\,.
\end{align}
\end{subequations}
are used to extract commutation relations among the matrices $\boldsymbol{x}^{(k)}_{ir},\boldsymbol{\epsilon}^{(k)}, \boldsymbol{b}_{qr}^{(k)}$. Note that this system is a priori more general than the system \eqref{eqn:ellKZBeq}, due to the appearence of $\boldsymbol{b}_{qr}^{(k)}$. However, as shown below, the integrability conditions imply that they are the same. The statements in this subsection hold for any function $\boldsymbol{Z}^\tau_{n,p}$ satisfying the system \eqref{eqn:ellKZBeqGeneral}. In particular, we do not use the explicit construction \eqref{def:ZnpVector} of $\boldsymbol{Z}^\tau_{n,p}$.

The restriction \eqref{eqn:comij} for an a priori unknown function $\boldsymbol{Z}^\tau_{n,p}$ leads to the following relations for distinct labels $|{i,j,q,r}|=4$ and $k,l\geq 1$:
\begin{align}\label{eqn:comRealij}
\boldsymbol{x}^{(k)}_{ij}&=(-1)^{k+1}\boldsymbol{x}^{(k)}_{ji}\,,\nonumber\\
[\boldsymbol{x}^{(0)}_{i},\boldsymbol{x}^{(0)}_{j}]&=0\,,\nonumber\\
[\boldsymbol{x}^{(k)}_{iq},\boldsymbol{x}^{(l)}_{jr}]&=0\,,\nonumber\\
[\boldsymbol{x}^{(0)}_{i}+\boldsymbol{x}^{(0)}_{j},\boldsymbol{x}^{(k)}_{ij}]&=\sum_{
	l=1}^{k-1}\sum_{\begin{smallmatrix}
	q\in \{1,p+2,\dots,n\}\\q\neq i,j
	\end{smallmatrix}}[\boldsymbol{x}^{(l)}_{iq},\boldsymbol{x}^{(k-l)}_{qj}]\,,\nonumber\\
 [\boldsymbol{x}^{(k)}_{ij},\boldsymbol{x}^{(l)}_{qi}+\boldsymbol{x}^{(l)}_{qj}]&=\sum_{m=1}^{k-1}\binom{k-1}{m}[\boldsymbol{x}^{(l+m)}_{qj},\boldsymbol{x}^{(k-m)}_{ij}] \, ,\nonumber\\
 [\boldsymbol{x}^{(k)}_{ij},\boldsymbol{x}^{(0)}_{q}]&=\sum_{m=1}^{k-1}\binom{k-1}{m}[\boldsymbol{x}^{(m)}_{qj},\boldsymbol{x}^{(k-m)}_{ij}]  \, .
\end{align}

The second integrability condition \eqref{eqn:comTaui} leads to the third and the last relation in \eqn{eqn:comRealij} and to the following new relations: for distinct labels $|{i,j,q}|$ and $k\geq 2$ 
\begin{align}
\boldsymbol{b}^{(k)}_{ij}&=\boldsymbol{x}^{(k-1)}_{ij}\,,
\end{align}
which ensures that the differential systems \eqref{eqn:ellKZBeq} and \eqref{eqn:ellKZBeqGeneral} are equivalent, and for $k\geq 4$, $l\geq 1$ 
\begin{align} \label{eqb_xijEpskSome}
[\boldsymbol{x}^{(0)}_{i},\boldsymbol{\epsilon}^{(0)}]&=0\,,\nonumber\\
\ph{=}[ \boldsymbol{x}^{(0)}_{i},\boldsymbol{\epsilon}^{(k)} ]&=\sum_{l= k/2}^{k-2}\sum_{\begin{smallmatrix}
	j\in \{1,p+2,\dots,n\}\\
	j\neq i
	\end{smallmatrix}}(-1)^l [\boldsymbol{x}_{ij}^{(l)},\boldsymbol{x}^{(k-l-1)}_{ij}]\,,\nnl
\ph{=}[\boldsymbol{x}^{(l)}_{ij},\boldsymbol{\epsilon}^{(k)}]&=\sum_{m=1}^l{ l-1 \choose m-1 }
[\boldsymbol{x}^{(l+k-m-1)}_{ij},\boldsymbol{x}_{ij}^{(m)}]
\end{align}
and for $k\geq 1$ 
\begin{align}\label{eqn_xijEX}
[\boldsymbol{x}^{(k)}_{ij},\boldsymbol{\epsilon}^{(0)}]&=(k-1)[\boldsymbol{x}_{ij}^{(k-1)},\boldsymbol{x}^{(0)}_{i}]+\sum_{\begin{smallmatrix}
	q\in \{1,p+2,\dots,n\}\\
	q\neq i
	\end{smallmatrix}}\sum_{l=1}^{k-2} \binom{k-1}{l-1}[\boldsymbol{x}^{(l)}_{iq},\boldsymbol{x}_{ij}^{(k-l-1)}]\,,\nonumber\\
[\boldsymbol{x}_{ij}^{(k)},\boldsymbol{x}^{(0)}_{q}]&=\sum_{l=1}^{k-1}[\boldsymbol{x}_{iq}^{(l)},\boldsymbol{x}^{(k-l)}_{qj}]\,.
\end{align}

The above relations are all the information we can extract from the integrability conditions \eqns{eqn:comij}{eqn:comTaui}: they are sufficient for the commutators \eqns{eqn:comij}{eqn:comTaui} to vanish. We remark that the last lines of $\eqref{eqn:comRealij}$ of $\eqref{eqn_xijEX}$ give two different, but compatible, equations for $[\boldsymbol{x}_{ij}^{(k)},\boldsymbol{x}^{(0)}_{q}]$.

Further relations can be deduced by combining the above relations. We can for example combine the fourth or fifth relation in \eqn{eqn:comRealij} and the last relation in \eqn{eqn_xijEX} to obtain for distinct labels $i,j,q$ and $k\geq 1$:
\begin{align}
[\boldsymbol{x}^{(k)}_{ij},\sum_{\begin{smallmatrix}
	q\in \{1,p+2,\dots,n\}
	\end{smallmatrix}}\boldsymbol{x}^{(0)}_{q}]&=0
\end{align}
and
\begin{align}
\sum_{l=1}^k[\boldsymbol{x}^{(l)}_{qj},\boldsymbol{x}^{(k+1-l)}_{iq}+\binom{k}{l} \boldsymbol{x}^{(k+1-l)}_{ij}]&=0
\end{align}
respectively. We remark that the matrices in a KZB system satisfy some extra relations that are not implied by the its integrability. See e.g.\ \eqns{TsunigaiDepth1}{TsunogaiDepth2}, and the text below these equations.

\subsubsection{Boundary values}\label{subsec:boundaryValues}
The asymptotic behaviour of $\boldsymbol{Z}^\tau_{n,p}$ for iteratively merging $z_{p+2}\to0$, then $z_{p+3}\to 0$ up to $z_n\to 0$, respecting the order defined by the domain \eqref{eqn:domain} is calculated 
in this subsection.

The degeneration of the Koba--Nielsen factor can be determined using the change of variables 
\begin{align}
z_i=z_{p{+}2}\, x_i
\end{align}
for $1\leq i\leq p+2$, leading to 
\begin{align}\label{eqn:limKNTau}
\lim_{n,p,k}\KN^{\tau}_{12\dots n}
&=e^{s_{12\dots p+k}\omega(1,0)}\KN_{12\dots p+2}\prod_{
	p+k<j\leq n}e^{-s_{(12\dots p+k),j}\CG_{j1}^{\tau}}\KN^{\tau}_{p+k+1\dots n}\,,
\end{align}
for $k=2,\dots, n-p$ and the regularised limit\footnote{The factors of  $(-2 \pi i z_{m})^{s_{ij...r}}$ cancel the nonanalytic behavior of $\lim_{z_{m}\to 0} e^{-s_{ij...r}\tilde{\Gamma}\big(\begin{smallmatrix}1\\0\end{smallmatrix};z_m\,,\tau\big)  }$. The choice of normalization follows the regularization of the divergent eMPL $\tilde{\Gamma}\big(\begin{smallmatrix}1\\0\end{smallmatrix};z\,,\tau\big) $ of \eqn{eqn:Gamma01reg}.} 
\begin{align}\label{eqn:limReg}
\lim_{n,p,k}&=\lim_{z_{p+k}\to 0}(- 2 \pi i z_{p+k})^{s_{(12\dots p+k-1),p+k}}\dots \lim_{z_{p+3\to 0}}(- 2 \pi i z_{p+3})^{s_{(12\dots p+2),p+3}}\lim_{z_{p+2\to 0}} (- 2 \pi i z_{p+2})^{s_{12\dots p+2}}\,,
\end{align}
and where we have used the following two definitions of sums of Mandelstam variables:
\begin{align}
	s_{A}&=\sum_{\begin{smallmatrix}
			i,j\in A\\
			i<j
	\end{smallmatrix}}s_{ij}\,,\quad
	s_{A,B}=\sum_{i\in A,j\in B}s_{ij}\,.
\end{align}
The punctures and associated momenta $s_{ij}$ appearing in the genus-zero Koba--Nielsen factor $\KN_{12\dots p+2}$ in \eqn{eqn:limKNTau} correspond to the genus-zero punctures
\begin{align}
0=x_1<x_2<\dots<x_{p+1}<x_{p+2}=1\,.
\end{align}
Note that this gives the same result as simply merging $z_{p+k}\to 0$ and squeezing the punctures $z_{p+2},\dots,z_{p+k-1}$ in between:
\begin{align}
\lim_{z_{p+k}\to 0}( -2 \pi i z_{p+k})^{s_{12\dots p+k}}\KN^{\tau}_{12\dots n}&=e^{s_{12\dots p+k}\omega(1,0)}\KN_{12\dots p+2}\prod_{
	p+k<j\leq n}e^{-s_{(12\dots p+k),j}\CG_{j1}^{\tau}}\KN^{\tau}_{p+k+1\dots n}\,,
\end{align}
in particular
\begin{align}
\lim_{n,p,n-p}\KN^{\tau}_{12\dots n}&=e^{s_{12\dots n}\omega(1,0,\tau)}\KN_{12\dots p+2}\,.
\end{align}

Using the same change of variables, the remaining differential form in the integral $Z^\tau_{n,p}$ from \eqn{def:Znp} degenerates as follows: for $z_{p+2}\to 0$
\begin{align}
&\tilde{\varphi}^\tau(1,A^1)\prod_{k=p+2}^n\tilde{\varphi}^\tau(k,A^k)\prod_{i=2}^{p+1}\dd z_i\nnl
&=\begin{cases}
\textrm{pt}(1,A^1 )\textrm{pt}(p+2,A^{p+2} )\prod_{i=2}^{p+1}\dd x_i&\text{if }A^k=\emptyset \text{ for }k=p+3,\dots, n\,,\\
0&\text{otherwise,}
\end{cases}
\end{align}
%
where $\textrm{pt}(a_1,\ldots,a_m)$ is an open-chain version of the Parke-Taylor factor of \eqn{eqn:parkeTaylorIntegrand}, defined by
\begin{align}
\textrm{pt}(a_1,a_2, \ldots, a_m)&= \prod_{i=2}^{m}\frac{1}{x_{a_{i-1}} - x_{a_i}}\,.
\end{align}
The remaining limits $z_{p+k}\to 0$ for $2 < k\leq n-p$ do not further affect these differential forms. Putting all together, we find 
\begin{align}\label{eqn:calculation}
&\lim_{n,p,k}Z^\tau_{n,p}((1,A^1),(p+2,A^{p+2}),(p+3),\dots,(n); z_{p+2},z_{p+3},\dots,z_n)\nnl
&=e^{s_{12\dots p+k}\omega(1,0)}\prod_{
	p+k<j\leq n}e^{-s_{(12\dots p+3),j}\CG_{j1}^{\tau}}\KN^{\tau}_{p+k+1\dots n} \nnl 
 &\quad \times  \int_{0<x_i<x_{i+1}<x_{p+2}}\prod_{i=2}^{p+1}\dd x_i\, \KN_{12\dots p+2}\,\textrm{pt}(1,A^1 )\textrm{pt}(p+2,A^{p+2} )\nnl
&=e^{s_{12\dots p+k}\omega(1,0)}\prod_{
	p+k<j\leq n}e^{-s_{(12\dots p+3),j}\CG_{j1}^{\tau}}\KN^{\tau}_{p+k+1\dots n}Z^\textrm{tree}_{p+3,p+2}((1,A^1 ),(p+2,A^{p+2} ))\,,
\end{align}
while the remaining integrals $Z^\tau_{n,p}$ from \eqn{def:Znp}, violating \mbox{$A^{p+3}=A^{p+4}=\dots =A^n=\emptyset$}, vanish. The integral on the last line is an $\SL_2$-fixed genus-zero, open-string integral given in \eqn{eqn:treeNpoint}, i.e.\
\begin{align}\label{eqn:ZTree}
Z^\textrm{tree}_{p+3,p+2}((1,A^1 ),(p+2,A^{p+2} ))
&=\int_{0<x_i<x_{i+1}<x_{p+2}}\prod_{i=2}^{p+1}\dd x_i\, \KN_{12\dots p+2}\,\textrm{pt}(1,A^1 )\textrm{pt}(p+2,A^{p+2} )\,,
\end{align}
where $(A^1,A^{p+2})$ is a partition of a permutation $\sigma\in S_{p}$ of $\{2,3,\dots,p+1\}$, i.e.\
\begin{align*}
(A^1,A^{p+2})&=\sigma(2,3,\dots,p+1)\,,
\end{align*}
and
\begin{align}
(x_1,x_{p+2},x_{p+3})&=(0,1,\infty)
\end{align}
the three fixed genus-zero punctures. Thus for $k=2,\dots, n-p$
\begin{align}\label{eqn:limitZTaunp}
\lim_{n,p,k}\boldsymbol{Z}^\tau_{n,p}
&=e^{s_{12\dots p+k}\omega(1,0)}\prod_{
	p+k<j\leq n}e^{-s_{(12\dots p+k),j}\CG_{j1}^{\tau}}\KN^{\tau}_{p+k+1\dots n}\begin{pmatrix}\boldsymbol{U}_{p+3,p+1}
\\
0
\end{pmatrix}\boldsymbol{Z}^\textrm{tree}_{p+3,p+2}\,,
\end{align}
where
\begin{align}
\boldsymbol{Z}^\textrm{tree}_{p+3,p+2}&=\begin{pmatrix}\int_{0<x_i<x_{i+1}<x_{p+2}}\prod_{i=2}^{p+1}\dd x_i\, \KN_{12\dots p+2}\,\textrm{pt}(1,A^1 )
\end{pmatrix}_{A^1=\sigma(2,3,\dots,p+1)}
\end{align}
is a $p!$-dimensional basis vector of the integrals $Z^\textrm{tree}_{p+3,p+2}$ from \eqn{eqn:ZTree} with $A^{p+2}=\emptyset$. Using integration by parts and partial fractioning, the remaining integrals from \eqn{eqn:ZTree} with \mbox{$A^{p+2}\neq \emptyset$} can be written in terms of the integrals in $\boldsymbol{Z}^\textrm{tree}_{p+3,p+2}$, which is implemented by the $((p{+}1)! \times p!)$-dimensional matrix $\boldsymbol{U}_{p+3,p+2}$.

\subsection{Example: $Z^{\tau}_{4,1}$-integrals}
For $p=n-1$, the integrals $Z_n^\tau$ of \rcite{Mafra:2019xms} and for $p=n-2$ the augmented integrals $Z_{0,n-1}^\tau$ of \rcite{Broedel:2020tmd} (with $z_n=z_0$) are obtained from the integrals $Z^\tau_{n,p}$ of \eqn{def:Znp}. The next case to consider is $p=n-3$, with two additional unintegrated punctures, $z_{n-1},$ $z_n$ besides $z_1=0$. Accordingly, the simplest non-trivial example is the class of $Z^{\tau}_{n,p}$-integrals with $(n,p)=(4,1)$. 

The construction in \eqn{def:Znp} defines the $d_{4,1}=3$ integrals
\begin{align}
Z_{4,1}^\tau((1,A_1),(3,A_3),(4,A_4);z_3,z_4)&=\int_0^{z_3}\dd z_2\, \KN_{1234}^{\tau}\, \tilde{\varphi}^\tau(1,A_1)\tilde{\varphi}^\tau(3,A_3)\tilde{\varphi}^\tau(4,A_4)\,,\nonumber\\
(A_1,A_2,A_3)&=(2)\,.
\end{align}
The corresponding vector from \eqn{def:ZnpVector} is given by
\begin{align}
\label{eqn:Z43Vector}
\boldsymbol{Z}_{4,1}^{\tau}(z_3,z_4)=\begin{pmatrix}
Z_{4,1}^\tau((1,2),(3),(4);z_3,z_4)\\
Z_{4,1}^\tau((1),(3,2),(4);z_3,z_4)\\
Z_{4,1}^\tau((1),(3),(4,2);z_3,z_4)
\end{pmatrix}
= \int_0^{z_3} \dd z_2\, \KN_{1234}^\tau 
\begin{pmatrix}
	F(z_{12},\eta_2,\tau )\\
    F(z_{32},\eta_2,\tau )\\
	F(z_{42},\eta_2,\tau )
\end{pmatrix}
\,.
\end{align}
\subsubsection{Differential system of $Z^{\tau}_{4,1}$}\label{subsec:exDEQ}
Now, let us write down the differential system satisfied by $ \boldsymbol{Z}_{4,1}^{\tau}(z_3,z_4)$. The closed formul\ae{} for the matrices in the differential \eqn{eqn:ellKZBeqzi} is given in \eqn{eqn:derZnpClosed} and yields for $(n,p)=(4,1)$ the $z_3$ derivative 
\begin{align}
\partial_3 \boldsymbol{Z}_{4,1}^{\tau}(z_3,z_4) &= \left(\boldsymbol{x}^{(0)}_{3}+\sum_{k\geq 1}\left(\boldsymbol{x}^{(k)}_{31} g^{(k)}_{31}+\boldsymbol{x}^{(k)}_{34} g^{(k)}_{34}\right)\right)\boldsymbol{Z}_{4,1}^{\tau}(z_3,z_4)\,,
\end{align}
the following matrices
\begin{align}
\boldsymbol{x}^{(0)}_{3}&=\begin{pmatrix}
-s_{23}\partial_{\eta_2}&-s_{23}\frac{1}{\eta_2}&0\\s_{12}\frac{1}{\eta_2}&(s_{12}+s_{24})\partial_{\eta_2}&s_{24}\frac{1}{\eta_2}\\0&-s_{23}\frac{1}{\eta_2}&-s_{23}\partial_{\eta_2}
\end{pmatrix}\,,\nonumber\\
\boldsymbol{x}^{(1)}_{31}&=\begin{pmatrix}
-s_{13}-s_{23}&s_{23}&0\\s_{12}&-s_{13}-s_{12}&0\\0&0&-s_{13}
\end{pmatrix}\,,\nonumber\\
\boldsymbol{x}^{(k)}_{31}&=\begin{pmatrix}
0&s_{23}(-\eta_2)^{k-1}&0\\s_{12}\eta_2^{k-1}&0&0\\0&0&0
\end{pmatrix}\,,\nnl
\boldsymbol{x}^{(1)}_{34}&=\begin{pmatrix}
-s_{34}&0&0\\0&-s_{34}-s_{24}&s_{24}\\0&s_{23}&-s_{34}-s_{23}
\end{pmatrix}\,,\nonumber\\
\boldsymbol{x}^{(k)}_{34}&=\begin{pmatrix}
0&0&0\\0&0&s_{24}(\eta_2)^{k-1}\\0&s_{23}(-\eta_2)^{k-1}&0
\end{pmatrix}\,,
\end{align}
where $k\geq 2$. Similarly, the partial differential equation with respect to $z_4$ takes the form
\begin{align}
\partial_4 \boldsymbol{Z}_{4,1}^{\tau}(z_3,z_4) &= \left(\boldsymbol{x}^{(0)}_{4}+\sum_{k\geq 1}\left(\boldsymbol{x}^{(k)}_{41} g^{(k)}_{41}+\boldsymbol{x}^{(k)}_{43} g^{(k)}_{43}\right)\right)\boldsymbol{Z}_{4,1}^{\tau}(z_3,z_4)\,,
\end{align}
where for $k\geq 2$
\begin{align}
\boldsymbol{x}^{(0)}_{4}&=\begin{pmatrix}
-s_{24}\partial_{\eta_2}&0&-s_{24}\frac{1}{\eta_2}\\0&-s_{24}\partial_{\eta_2}&-s_{24}\frac{1}{\eta_2}\\s_{12}\frac{1}{\eta_2}&s_{23}\frac{1}{\eta_2}&(s_{12}+s_{23})\partial_{\eta_2}
\end{pmatrix}\,,\nonumber\\
\boldsymbol{x}^{(1)}_{41}&=\begin{pmatrix}
-s_{14}-s_{24}&0&s_{24}\\0&-s_{14}&0\\s_{12}&0&-s_{14}-s_{12}
\end{pmatrix}\,,\nonumber\\
\boldsymbol{x}^{(k)}_{41}&=\begin{pmatrix}
0&0&s_{24}(-\eta_2)^{k-1}\\0&0&0\\s_{12}\eta_2^{k-1}&0&0
\end{pmatrix}\,,
\nnl
\boldsymbol{x}^{(1)}_{43}&=\begin{pmatrix}
-s_{34}&0&0\\0&-s_{34}-s_{24}&s_{24}\\0&s_{23}&-s_{34}+s_{23}
\end{pmatrix}\,,\nonumber\\
\boldsymbol{x}^{(k)}_{43}&=\begin{pmatrix}
0&0&0\\0&0&s_{24}(-\eta_2)^{k-1}\\0&s_{23}\eta_2^{k-1}&0
\end{pmatrix}\,.
\end{align}
These matrices indeed satisfy the commutation relations from \subsecref{subsec:comRel}.
 Moreover, they satisfy an additional relation
\begin{align}\label{eqn:bonusRelationRepresentation}
 		[\boldsymbol{x}^{(k)}_{ij},\boldsymbol{x}^{(k+2s)}_{ij}] \big|_{(n,p)=(4,1)}&=0\,,\quad |\{i,j\}|=2\, , \, k\geq 1 \, , s \in \mathbb{Z}^+\,,
\end{align}
which is not implied by the integrability of the KZB connection, and in fact does not hold for the matrices one obtains from $\boldsymbol{Z}_{n,p}^{\tau}$ with $p\geq 2$.  This kind of ``accidental  relations'' among matrices obtained in differential equations of integrals are common when the number of integrated punctures $p$ is small ($p=1$ here), see e.g. the matrices $e_0$ and $e_1$ in equation (23) of \rcite{Broedel:2013aza}, which do not give rise to higher-depth MZVs when used as arguments of the Drinfeld associator\footnote{We give  the definition of the Drinfeld associator in \eqn{eqn:drinfAssoc}.} $\Phi(e_0,e_1)$. An equivalent phenomena was observed in the genus-one setup of \rcite{Britto:2021prf}, for $(n,p)=(n,1)$.

The vector $\boldsymbol{Z}^\tau_{4,1}(z_3,z_4)$ also satisfies a differential equation with respect to $\tau$:

\begin{align}
2\pi i\partial_{\tau} \boldsymbol{Z}^\tau_{4,1}&=\left(-\boldsymbol{\epsilon}^{(0)}+\sum_{k\geq 4}(1-k)G_k \boldsymbol{\epsilon}^{(k)}+\sum_{k\geq 2 }(k-1)\left[ \boldsymbol{b}_{13}^{(k)}g^{(k)}_{13}+\boldsymbol{b}_{14}^{(k)}g^{(k)}_{14}+\boldsymbol{b}_{34}^{(k)}g^{(k)}_{34}\right]\right)\boldsymbol{Z}^\tau_{4,1}\,,
\end{align}
where  we find that $\boldsymbol{b}_{qr}^{(k)}=\boldsymbol{x}_{qr}^{(k-1)}$, and where for $k\geq 4$ the matrices $\boldsymbol{\epsilon}^{(0)}$ and $\boldsymbol{\epsilon}^{(k)}$ are given by:
\begin{align}
	\boldsymbol{\epsilon}^{(0)}&=\mathbb{I}_3(2 s_{1234}\zeta_2 -\frac{1}{2}(s_{12}+s_{23}+s_{24}) \partial^2_{\eta_2})+\begin{pmatrix}
		s_{12}\frac{1}{\eta_2^2}&s_{23}\frac{1}{\eta_2^2}&s_{24}\frac{1}{\eta_2^2}\\
		s_{12}\frac{1}{\eta_2^2}&s_{23}\frac{1}{\eta_2^2}&s_{24}\frac{1}{\eta_2^2}\\
		s_{12}\frac{1}{\eta_2^2}&s_{23}\frac{1}{\eta_2^2}&s_{24}\frac{1}{\eta_2^2}
	\end{pmatrix}\,, \nnl
	\boldsymbol{\epsilon}^{(k)}&=\begin{pmatrix}
		s_{12}\eta_2^{k-2}&0&0\\0&s_{23}\eta_2^{k-2}&0\\0&0&s_{24}\eta_2^{k-2}
	\end{pmatrix}\,.
\end{align}

The matrices $\boldsymbol{x}^{(k)}_{ij}$ for $(n,p)=(4,1)$ satisfy the additional relation \eqref{eqn:bonusRelationRepresentation} that is not implied by the commutation relations  of \eqn{eqn:comRealij}, but is rather due to the specific form of the matrices. To write an analogous set of commutation relations for the matrices $\boldsymbol{\epsilon}^{(k)}$, it is convenient to first define some matrices  $\tilde{\boldsymbol{\epsilon}}^{(k)}$  for $k\geq 2$ an even number, as follows:
\begin{align}
	\tilde{\boldsymbol{\epsilon}}^{(0)}&=\boldsymbol{\epsilon}^{(0)} \, ,& \nnl
	\tilde{\boldsymbol{\epsilon}}^{(k)}&=\boldsymbol{\epsilon}^{(k)} +\boldsymbol{x}_{13}^{(k-1)}+\boldsymbol{x}_{14}^{(k-1)}+\boldsymbol{x}_{34}^{(k-1)} \, , \label{eqn:EpsilonTildeN4p1}
\end{align} 
leading to the explicit matrices
\begin{align}
	\tilde{\boldsymbol{\epsilon}}^{(2)}&=-s_{1234} \mathbb{I}_3 +\begin{pmatrix}
		s_{12} &s_{23} &s_{24} \\
		s_{12} &s_{23} &s_{24} \\
		s_{12} &s_{23} &s_{24} 
	\end{pmatrix}\,, \nnl
	\tilde{\boldsymbol{\epsilon}}^{(k)}&= \eta_2^{k-2}\begin{pmatrix}
		s_{12} &s_{23} &s_{24} \\
		s_{12} &s_{23} &s_{24} \\
		s_{12} &s_{23} &s_{24} 
	\end{pmatrix}\,.
\end{align}
This definition yields the following extra commutation relations for $(n,p)=(4,1)$:
\begin{align}\label{eqn:bonusRelationRepresentationEpsilonsN4P1}
	[	\tilde{\boldsymbol{\epsilon}}^{(2m)},	\tilde{\boldsymbol{\epsilon}}^{(2n)} ]\big{|}_{(n,p)=(4,1)}&=0\,,\quad m,n \in \mathbb{Z}_{\geq 1} \, . 
\end{align}
This last relation is not implied by the commutation relations we have computed in \eqns{eqb_xijEpskSome}{eqn_xijEX}, \textit{nor} are they implied by the relations of Tsunogai's derivation algebra (or Pollack's relations) \cite{Tsunogai,Pollack} . Nonetheless, the definition of the $\tilde{\boldsymbol{\epsilon}}^{(k)}$ above will be related to Pollack's relations in \secref{subsec:Pollack}, for general values of $(n,p)$. 

\subsubsection{Boundary values of $Z^{\tau}_{4,1}$}
 Let us specialise  the analysis from \subsecref{subsec:boundaryValues} to  the example $\boldsymbol{Z}_{4,1}^{\tau}(z_3,z_4)$. Thus, we first consider the degeneration for $z_3\to 0$, then additionally $z_4\to 0$ of the integrals in \eqn{eqn:Z43Vector}
\begin{align}
\label{eq:eqnZn4p1Defn}
\boldsymbol{Z}_{4,1}^{\tau}(z_3,z_4)=\int_0^{z_3} \dd z_2\, \KN_{1234}^\tau 
\begin{pmatrix}
	F(z_{12},\eta_2,\tau )\\
	F(z_{32},\eta_2,\tau )\\
	F(z_{42},\eta_2,\tau )
\end{pmatrix}
\,.
\end{align}

For the limit $z_3\to 0$, we can use the change of variables $z_i=z_3 x_i$ for $i=1,2,3$, such that $x_1=0$, $x_3=1$, keeping $z_4$ fixed. Then, the Koba--Nielsen factor degenerates to 
\begin{align}
\KN_{1234}^{\tau}&=\prod_{1\leq j<i\leq 3}e^{-s_{ij}\CG^{\tau}_{ij}}\prod_{1\leq i\leq 3}e^{-s_{ij}\CG^{\tau}_{4i}}\nnl
&=(-2\pi i z_3)^{-s_{123}}e^{s_{123}\omega(1,0)}\prod_{1\leq i<j\leq 3}x_{ij}^{-s_{ij}}e^{-s_{(123),4}\CG^{\tau}_{41}}(1+\mathcal{O}(z_3))\nnl
&=(-2\pi i z_3)^{-s_{123}}e^{s_{123}\omega(1,0)}\KN_{123}\KN^{\tau}_{14}|_{\tilde s_{14}=s_{(123),4}}(1+\mathcal{O}(z_3))\,,
\end{align}
where $\KN_{123}$ is  the genus-zero, four-point Koba--Nielsen factor of \eqn{eqn:treeKN}, with an additional puncture $x_4=\infty$. The differential form without the Koba--Nielsen factor degenerates to 
\begin{align}
\tilde{\varphi}^\tau(1,A_1)\tilde{\varphi}^\tau(3,A_3)\tilde{\varphi}^\tau(4,A_4)\,\dd z_2&=\begin{cases}
\frac{\dd x_2}{-x_2}&\text{if }A_1=(2)\\
\frac{\dd x_2}{1-x_2}&\text{if }A_3=(2)\\
\frac{z_3 x_2}{z_4 -z_3 x_2} \rightarrow 0&\text{if }A_4=(2)\, ,
\end{cases}
\end{align}
where the last entry tends to $0$ as $x_3 \rightarrow 0$ at fixed $z_4$.

Putting all together, we find that\footnote{Here and  throughout this work, we  the Mandelstam variables $s_{ij}$ to be analytically continued from a region in which  $\Re(s_{ij})<0$.} 
\begin{align}\label{eqn:exampleBoundaryValue}
&\lim_{z_3\to 0}(-2 \pi i z_3)^{s_{123}}\boldsymbol{Z}_{4,3}^{\tau}(z_3,z_4)\nnl
&=e^{s_{123}\omega(1,0)}e^{-s_{(123),4}\CG^{\tau}_{41}}\begin{pmatrix}
\int_0^{1}\dd x_2\, \KN_{123}\frac{1}{-x_2}\\
\int_0^{1}\dd x_2\, \KN_{123}\frac{1}{1-x_2}\\
0
\end{pmatrix}\nnl
&=e^{s_{123}\omega(1,0)}e^{-s_{(123),4}\CG^{\tau}_{41}}\begin{pmatrix}
Z^\textrm{tree}_{4,3}((1,2),(3))\\
Z^\textrm{tree}_{4,3}((1),(3,2))\\
0
\end{pmatrix}\nnl
&=e^{s_{123}\omega(1,0)}e^{-s_{(123),4}\CG^{\tau}_{41}}\begin{pmatrix}
1\\
-\frac{s_{12}}{s_{23}}\\
0
\end{pmatrix}Z^\textrm{tree}_{4,3}((1,2),(3))\,,
\end{align}
where 
the integrals $Z^\textrm{tree}_{4,3}$ are related by the integration by parts relation
\begin{align}\label{eqn:ibpRelation}
s_{12}\,Z^\textrm{tree}_{4,3}((1,2),(3))+s_{23}\,
Z^\textrm{tree}_{4,3}((1),(3,2))&=0
\end{align}
and can be expressed in terms of the Veneziano amplitude
\begin{align}
Z^\textrm{tree}_{4,3}((1,2),(3))&=\frac{1}{s_{12}}\frac{\Gamma(1-s_{12})\Gamma(1-s_{23})}{\Gamma(1-s_{12}-s_{23})}\,.
\end{align}
The relation \eqref{eqn:ibpRelation} leads to the last step in \eqn{eqn:exampleBoundaryValue} and defines the matrix $\boldsymbol{U}_{4,2}$ from \eqn{eqn:limitZTaunp}:
\begin{align}
\boldsymbol{U}_{4,2}&=\begin{pmatrix}
1\\
-\frac{s_{12}}{s_{23}}
\end{pmatrix}\,,
\end{align}
such that
\begin{align}
\begin{pmatrix}
Z^\textrm{tree}_{4,3}((1,2),(3))\\Z^\textrm{tree}_{4,3}((1),(3,2))
\end{pmatrix}&=\boldsymbol{U}_{4,2}Z^\textrm{tree}_{4,3}((1,2),(3))\,.
\end{align}
The additional limit $z_4\to 0$ is now straightforward:
\begin{align}
&\lim_{z_4\to 0} (-2 \pi i z_4)^{s_{(123),4}}\lim_{z_3\to 0}(-2 \pi i z_3)^{s_{123}}\boldsymbol{Z}_{4,1}^{\tau}(z_3,z_4)\nnl
&=\lim_{z_4\to 0} (-2 \pi i z_4)^{s_{(123),4}}e^{s_{123}\omega(1,0)}e^{-s_{(123),4}\CG^{\tau}_{41}}\begin{pmatrix}
\boldsymbol{U}_{4,2}\\0
\end{pmatrix}Z^\textrm{tree}_{4,3}((1,2),(3))\nnl
&=e^{s_{1234}\omega(1,0)}\begin{pmatrix}
\boldsymbol{U}_{4,2}\\0
\end{pmatrix}Z^\textrm{tree}_{4,3}((1,2),(3))
\nnl
&=e^{s_{1234}\omega(1,0)}
\begin{pmatrix}
	1/s_{12}\\
	-1/s_{23}\\
	0
\end{pmatrix} \frac{\Gamma(1-s_{12})\Gamma(1-s_{23})}{\Gamma(1-s_{12}-s_{23})}\, .
\end{align}

\section{Expansion of $Z^{\tau}_{n,p}$-integrals}
\label{sec:alphaExpansion}
\subsection{Solving the differential system}
The elliptic KZB system \eqref{eqn:ellKZBeq} satisfied by $\boldsymbol{Z}^\tau_{n,p}=\boldsymbol{Z}^\tau_{n,p}(z_{p+2},\dots,z_n)$ can be solved in the domain \eqref{eqn:domain}
using the genus-zero methods in section 3 of \rcite{Britto:2021prf}, leading to a representation of the $\alpha'$-expansion of $\boldsymbol{Z}^\tau_{n,p}$ in terms of generating series of eMZVs: First, the corresponding differential
\begin{align}\label{eqn:diff_form}
\dd \boldsymbol{Z}^\tau_{n,p}&=\sum_{i=p+2}^n\boldsymbol{\Omega}^i_{n,p} \boldsymbol{Z}^\tau_{n,p}\dd z_i+\boldsymbol{\Omega}^\tau_{n,p}\boldsymbol{Z}^\tau_{n,p}\dd\tau
\end{align}
will be integrated along the path
\begin{align}\label{eqn:pathZero}
(0,\dots,0,\tau)\overset{\gamma_n}{\to} (0,\dots,0,z_n,\tau)\overset{\gamma_{n-1}}{\to}\dots  \overset{\gamma_{p+2}}{\to} (z_{p+2},\dots,z_n,\tau)\, ,
\end{align}
where the path $\gamma_i$ refers to the path $(0,\ldots,0,t,z_{i+1},\ldots,z_n)$ where $t\in (0,z_i)$. The matrices $\boldsymbol{\Omega}^i_{n,p}$ and $\boldsymbol{\Omega}^\tau_{n,p}  $  in \eqn{eqn:diff_form} are the operator-valued $d_{n,p}\times d_{n,p}$ matrices appearing on the right hand side of \eqns{eqn:ellKZBeqzi}{eqn:ellKZBeqtau} in the form $\partial_{i} \boldsymbol{Z}^\tau_{n,p} = \boldsymbol{\Omega}^i_{n,p}  \boldsymbol{Z}^\tau_{n,p}$  and $\partial_{\tau} \boldsymbol{Z}^\tau_{n,p} = \boldsymbol{\Omega}^\tau_{n,p}  \boldsymbol{Z}^\tau_{n,p}$ respectively. Second, the corresponding initial value at $z_{p+2}, \ldots , z_n \rightarrow 0$ will be determined. And third, homotopy invariance ensures that the solution is valid on the whole domain \eqref{eqn:domain}.  

In order to do so, let us consider for $i\in \{p+2,\dots,n\}$ the following generating series of eMPLs
\begin{align}\label{eqn:Gammai}
\boldsymbol{\Gamma}_i&=\boldsymbol{\Gamma}_i(z_i,\dots,z_n)\nnl
&=\sum_{r\geq 0}\sum_{k_1,\dots, k_r\geq 0}\sum_{j_{1},\dots,j_r\in \{1,i+1,\dots, n\}}\tilde{\Gamma} \left(\begin{smallmatrix}
k_1&\dots&k_r\\
z_{j_i}&\dots&z_{j_r}
\end{smallmatrix}; z_i,\tau\right)\boldsymbol{X}_{ij_1 }^{(k_1)}\dots \boldsymbol{X}_{ij_r}^{(k_r)}\,.
\end{align}
The matrices $\boldsymbol{X}_{ij}^{(k)}$ are defined for $j\in \{1,p+2,\dots,n\}\setminus\{i\}$  and $k \geq 0$ from the matrices $\boldsymbol{x}_{ij}^{(k)}$ appearing in \eqn{eqn:ellKZBeqzi} by
\begin{align}
\boldsymbol{X}_{ij}^{(k)}&=\begin{cases}
\boldsymbol{x}_{ij}^{(k)}& j\neq 1\,,\\
\boldsymbol{x}_{i1}^{(k)}+\sum_{r=p+2}^{i-1}\boldsymbol{x}_{ir}^{(k)}& j=1\,.
\end{cases}\
\end{align}
By construction, for each $i\in \{p+2,\dots,n\}$ the generating series $\boldsymbol{\Gamma}_i$ satisfies along the path $\gamma_i$ the partial differential equations 
\begin{align}\label{eqn:pdEqGamma}
\partial_i \boldsymbol{\Gamma}_i |_{\gamma_{i}}&=\boldsymbol{\Omega}^{i}_{n,p}|_{\gamma_{i}}\,\boldsymbol{\Gamma}_i |_{\gamma_{i}}\,,
\end{align}
where the matrix $\boldsymbol{\Omega}^{i}_{n,p}$ from the differential form \eqref{eqn:diff_form} is evaluated at $\gamma_i$, i.e.\ for $z_{p{+}2}=\dots z_{i{-}1}=0$. For $i<j$, we obtain by definition
\begin{align}
\label{eqn:pdEqGammaTrivial}
\partial_i \boldsymbol{\Gamma}_j&=0\,.
\end{align}
Moreover, the asymptotic behaviour of $\boldsymbol{\Gamma}_i$ on $\gamma_{i}$ follows from the regularisation of the eMPLs and is given by\footnote{We use the same normalization as in \eqn{eqn:limReg}.} 
\begin{align}\label{eqn:asympt}
\lim_{\epsilon\to 0}\boldsymbol{\Gamma}_i(\gamma_{i}(\epsilon))(-2\pi i \epsilon)^{-\boldsymbol{X}_{i1}^{(1)}}&=\mathds{1}\,.
\end{align}

With these definitions, we can define the $d_{n,p}$-component vector
\begin{align}
\boldsymbol{\Gamma}_{n,p}&=\boldsymbol{\Gamma}_{n,p}(z_{p{+}2},\dots,z_n)\nnl
&=\boldsymbol{\Gamma}_{p{+}2}\dots \boldsymbol{\Gamma}_n \,\boldsymbol{v}_{n,p}\,,
\end{align}
where $\boldsymbol{v}_{n,p}$ is a vector independent of $z_{p{+}2},\dots,z_n$. Remarkably, the vector asymptotically satisfies for each path $\gamma_i$ the partial differential equation \eqref{eqn:pdEqGamma}, i.e.\
\begin{align}\label{eqn:asymptPDE}
\partial_i \boldsymbol{\Gamma}_{n,p} |_{\gamma_{i}}
&\sim \boldsymbol{\Omega}^{i}_{n,p}|_{\gamma_{i}}\,\boldsymbol{\Gamma}_{n,p} |_{\gamma_{i}} \,,
\end{align}
which can be shown using the commutation relations \eqref{eqn:comRealij} as follows: they imply that for $i>j$ 
and $s\in \{i+1,\dots,n\}$
\begin{align}
[\boldsymbol{X}_{j1}^{(1)},\boldsymbol{X}_{is}^{(k)}]&=[\boldsymbol{x}_{j1}^{(1)}+\sum_{r=p+2}^{j-1}\boldsymbol{x}_{jr}^{(1)},\boldsymbol{x}_{is}^{(k)}]=0\,.
\end{align}
For the special case $i>j$ and $s=1$, the last relation of \eqn{eqn:comRealij} ensures that also the following commutator vanishes
\begin{align}\label{eqn:XX}
[\boldsymbol{X}_{j1}^{(1)},\boldsymbol{X}_{i1}^{(k)}]&=[\boldsymbol{x}_{j1}^{(1)}+\sum_{r=p+2}^{j-1}\boldsymbol{x}_{jr}^{(1)},\boldsymbol{x}_{i1}^{(k)}+\sum_{s=p+2}^{i-1}\boldsymbol{x}_{is}^{(k)}]\nnl
&=\sum_{r\in \{1,p{+}2,\dots j{-}1 \}}[\boldsymbol{x}_{jr}^{(1)},\boldsymbol{x}_{ir}^{(k)}+\boldsymbol{x}_{ij}^{(k)}]
\nnl
&=0\,,
\end{align}
such that
\begin{align}\label{eqn:comRelOmega}
[\boldsymbol{X}_{j1}^{(1)},\boldsymbol{\Omega}^i_{n,p}|_{\gamma_i}]&=\sum_{k\geq 0}[\boldsymbol{X}_{j1}^{(1)},\boldsymbol{X}_{i1}^{(k)}]g_{1i}^{(k)}=0\,.
\end{align}
\Eqn{eqn:comRelOmega} ensures that $\boldsymbol{\Gamma}_{n,p}$ asymptotically satisfies the following partial differential equation in the limit $z_{i{-}1}\to z_{i{-}2}\to \dots \to z_{p{+}2}\to 0$:
\begin{align}
\partial_i\boldsymbol{\Gamma}_{n,p}
&\sim (-2\pi i z_{p{+}2})^{\boldsymbol{X}_{p{+}2,1}^{(1)}}\dots  (-2\pi i z_{i{-}1})^{\boldsymbol{X}_{i{-}1,1}^{(1)}}(\partial_i \boldsymbol{\Gamma}_{i})\boldsymbol{\Gamma}_{i{+}1}\dots \boldsymbol{\Gamma}_n \,\boldsymbol{v}_{n,p}\nnl
&=(-2\pi i z_{p{+}2})^{\boldsymbol{X}_{p{+}2,1}^{(1)}}\dots  (-2\pi i z_{i{-}1})^{\boldsymbol{X}_{i{-}1,1}^{(1)}}(\boldsymbol{\Omega}^i_{n,p}|_{\gamma_i} \boldsymbol{\Gamma}_{i})\boldsymbol{\Gamma}_{i{+}1}\dots \boldsymbol{\Gamma}_n \,\boldsymbol{v}_{n,p}\nnl
&=\boldsymbol{\Omega}^i_{n,p}|_{\gamma_i}(-2\pi i z_{p{+}2})^{\boldsymbol{X}_{p{+}2,1}^{(1)}}\dots  (-2\pi i z_{i{-}1})^{\boldsymbol{X}_{i{-}1,1}^{(1)}} \boldsymbol{\Gamma}_{i}\boldsymbol{\Gamma}_{i{+}1}\dots \boldsymbol{\Gamma}_n \,\boldsymbol{v}_{n,p}\nnl
&=\boldsymbol{\Omega}^i_{n,p}|_{\gamma_i}\boldsymbol{\Gamma}_{n,p}\,,
\end{align}
which proves the asymptotic behaviour \eqref{eqn:asymptPDE}.
Of course, $\boldsymbol{\Gamma}_{n,p}$ is singular on $\gamma_i$ for $i<n$. However, \eqn{eqn:asymptPDE} ensures that for the initial-value vector
\begin{align}\label{eqn:initialv}
\boldsymbol{v}_{n,p}= \lim_{z_{n}\to 0}(-2\pi i z_{n})^{-\boldsymbol{X}_{n,1}^{(1)}}\dots \lim_{z_{p{+}2}\to 0}(-2\pi i z_{p{+}2})^{-\boldsymbol{X}_{p{+}2,1}^{(1)}}\boldsymbol{Z}^{\tau
}_{n,p}\,,
\end{align}
the following representation is indeed the integral of $\dd \boldsymbol{Z}^{\tau
}_{n,p}$ starting from the initial value $\boldsymbol{v}_{n,p}$ along $\gamma_n\dots\gamma_{p{+}2}$:
\begin{align}\label{eqn:repZTaunp}
\boldsymbol{Z}^{\tau
}_{n,p}&=\boldsymbol{\Gamma}_{n,p}=\boldsymbol{\Gamma}_{p{+}2}\dots \boldsymbol{\Gamma}_n\,\boldsymbol{v}_{n,p}\,.
\end{align}
By homotopy invariance, \eqn{eqn:repZTaunp} expresses the dependence of $\boldsymbol{Z}^{\tau
}_{n,p}$ on the punctures $z_{p{+}2},\dots,z_n$ on the whole domain \eqref{eqn:domain}. Since the matrices $\boldsymbol{x}^{(k)}_{ij}$ are proportional to $\alpha'$, the expansion of the generating series of eMPLs $\boldsymbol{\Gamma}_{i}$ in the word length $r$ \textit{is} its $\alpha'$-expansion. Thus, in practice, \eqn{eqn:repZTaunp} can be used to calculate the $\alpha'$-expansion of $\boldsymbol{Z}^{\tau
}_{n,p}$ up to any desired order. 

The initial vector $\boldsymbol{v}_{n,p}$ from \eqn{eqn:initialv} can be determined using the following eigenvalue equations

\begin{align}\label{eqn:eigenvalueEq}
\boldsymbol{X}_{p{+}2,1}^{(1)}\begin{pmatrix}
\boldsymbol{U}_{p+3,p+1}\\0
\end{pmatrix}&=-s_{12\dots p{+}2}\begin{pmatrix}
\boldsymbol{U}_{p+3,p+1}\\0
\end{pmatrix}\,,\nnl
\boldsymbol{X}_{p{+}k,1}^{(1)}\begin{pmatrix}
\boldsymbol{U}_{p+3,p+1}\\0
\end{pmatrix}&=-s_{12\dots p{+}k{-}1,p{+}k}\begin{pmatrix}
\boldsymbol{U}_{p+3,p+1}\\0
\end{pmatrix}\,,\quad 2<k\leq n-p\,,
\end{align}
derived in \appref{app:evalEq}, and where the matrices $\boldsymbol{U}_{p+3,p+1}$ are defined in \eqn{eqn:limitZTaunp} by integration-by-parts relations of genus-zero integrals. They imply that it is simply given by the limit from \eqn{eqn:limitZTaunp} for $k=n-p$:
\begin{align}\label{eqn:initialVector}
\boldsymbol{v}_{n,p}&=\lim_{n,p,n-p}\boldsymbol{Z}^{\tau
}_{n,p}=e^{s_{12\dots n}\omega(1,0)}\begin{pmatrix}\boldsymbol{U}_{p+3,p+1}
\\
0
\end{pmatrix}\boldsymbol{Z}^\textrm{tree}_{p+3,p+2}\,.
\end{align}

Putting all together, in the domain \eqref{eqn:domain}, the vector $\boldsymbol{Z}^{\tau
}_{n,p}$ can be represented in terms of genus-zero $Z_{n,p}$-integrals and generating series of eMPLs as follows\footnote{We have checked that this hold for the $(n,p)=(4,1)$ case up to eMPLs of depth 3 and weight 5.}: 
\begin{align}\label{eqn:ZnpTauRep}
\boldsymbol{Z}^{\tau
}_{n,p}&=e^{s_{12\dots n}\omega(1,0)}\,\boldsymbol{\Gamma}_{p{+}2}\dots \boldsymbol{\Gamma}_n\,\begin{pmatrix}\boldsymbol{U}_{p+3,p+1}
\\
0
\end{pmatrix}\boldsymbol{Z}^\textrm{tree}_{p+3,p+2}\,.
\end{align}
A considerably simpler formula for $\boldsymbol{Z}^{\tau
}_{n,p}$ will be given in the next subsection, in \eqn{eqn:ZnpTauRepShort}.

\subsection{An alternative solution strategy}
Instead of using the above arguments with the asymptotic behaviour of the integrals leading to the representation \eqref{eqn:repZTaunp}, a simpler representation can be deduced for the integration domain \eqref{eqn:domain}.

On the one hand, from the calculation \eqref{eqn:limitZTaunp} we see that already for merging only one puncture $z_{p+2}\to 0$, the dependence on the other punctures $z_{p+3},\dots, z_n$ becomes quite simple: it only involves the prefactor formed by the genus-one Koba--Nielsen factor $\KN^{\tau}_{p+3\dots n}$ and the exponentials involving $\CG_{j1}^{\tau}$:
\begin{align}
	\label{eqn:simpleDependenceDroppingOut}
&\lim_{z_{p+2\to 0}}(-2 \pi i z_{p+2})^{s_{12\dots p+2}}\boldsymbol{Z}^\tau_{n,p}\nnl
&=e^{s_{12\dots p+2}\omega(1,0)}\prod_{
	p+2<j\leq n}e^{-s_{(12\dots p+2),j}\CG_{j1}^{\tau}}\KN^{\tau}_{p+3\dots n}\begin{pmatrix}\boldsymbol{U}_{p+3,p+1}
\\
0
\end{pmatrix}\boldsymbol{Z}^\textrm{tree}_{p+3,p+2}\,.
\end{align}
The eigenvalue equation \eqref{eqn:eigenvalueEq} and the definition \eqref{eqn:initialVector} of the initial vector $ \boldsymbol{v}_{n,p}$ can be used for the following alternative representation of the above limit
\begin{align}\label{eqn:limitZTaunpAlt}
\lim_{z_{p+2\to 0}}(-2 \pi i z_{p+2})^{-\boldsymbol{X}_{p{+}2,1}^{(1)}}\boldsymbol{Z}^\tau_{n,p}&=e^{s_{p+3\dots n}\omega(1,0)}\prod_{
	p+2<j\leq n}e^{-s_{(12\dots p+2),j}(\CG_{j1}^{\tau}-\omega(1,0))}\KN^{\tau}_{p+3\dots n}\boldsymbol{v}_{n,p}\,.
\end{align}
On the other hand, the vector $\boldsymbol{Z}^\tau_{n,p}$ and the matrix $\boldsymbol{\Gamma}_{p{+}2}$ defined in \eqn{eqn:Gammai} satisfy on the domain \eqref{eqn:domain} the same partial differential equation \eqref{eqn:pdEqGamma} with respect to $z_{p+2}$:
\begin{align}
\partial_{p+2} \boldsymbol{\Gamma}_{p+2}&=\boldsymbol{\Omega}^{p+2}_{n,p}\,\boldsymbol{\Gamma}_{p+2}\,,\nnl
\partial_{p+2} \boldsymbol{Z}^\tau_{n,p}&=\boldsymbol{\Omega}^{p+2}_{n,p}\,\boldsymbol{Z}^\tau_{n,p}\,.
\end{align}
Since $\boldsymbol{\Gamma}_{p+2}$ is a path-ordered matrix exponential, it is invertible, such that we can consider the vector
\begin{align}\label{eqn:vnDef}
\hat{\boldsymbol{v}}_{n}&=(\boldsymbol{\Gamma}_{p+2})^{-1}\boldsymbol{Z}^\tau_{n,p}\,.
\end{align}
Differentiation with respect to $z_{p+2}$ and using the above partial differential equations, we find $\hat{\boldsymbol{v}}_{n}$ is independent of $z_{p+2}$,
\begin{align}
\partial_{p+2}\hat{\boldsymbol{v}}_{n}&=\partial_{p+2}\left((\boldsymbol{\Gamma}_{p+2})^{-1}\boldsymbol{Z}^\tau_{n,p}\right)\nnl
&=-(\boldsymbol{\Gamma}_{p+2})^{-1}\boldsymbol{\Omega}^{p+2}_{n,p}\,(\boldsymbol{\Gamma}_{p+2})^{-1}\boldsymbol{\Gamma}_{p+2}\boldsymbol{Z}^\tau_{n,p}+(\boldsymbol{\Gamma}_{p+2})^{-1}\boldsymbol{\Omega}^{p+2}_{n,p}\,\boldsymbol{Z}^\tau_{n,p}\nnl
&=0\,.
\end{align}
Therefore, we can evaluate $\hat{\boldsymbol{v}}_{n}$ in the limit $z_{p+2}\to 0$, where the asymptotic behaviour \eqref{eqn:asympt} of $\boldsymbol{\Gamma}_{p+2}$ yields the limit \eqref{eqn:limitZTaunpAlt}, i.e.
\begin{align}
\hat{\boldsymbol{v}}_{n}&=\lim_{z_{p+2}\to 0}\hat{\boldsymbol{v}}_{n}\nnl
&=\lim_{z_{p+2}\to 0}(\boldsymbol{\Gamma}_{p+2})^{-1}\boldsymbol{Z}^\tau_{n,p}\nnl
&=\lim_{z_{p+2\to 0}}(-2 \pi i z_{p+2})^{-\boldsymbol{X}_{p{+}2,1}^{(1)}}\boldsymbol{Z}^\tau_{n,p}\nnl
&=e^{s_{p+3\dots n}\omega(1,0)}\prod_{
	p+2<j\leq n}e^{-s_{(12\dots p+2),j}(\CG_{j1}^{\tau}-\omega(1,0))}\KN^{\tau}_{p+3\dots n}\boldsymbol{v}_{n,p}\,.
\end{align}
Using this representation of $\hat{\boldsymbol{v}}_{n}$ and multiplying \eqn{eqn:vnDef} from the left by $\boldsymbol{\Gamma}_{p+2}$ leads to the following matrix representation of $\boldsymbol{Z}^\tau_{n,p}$:
\begin{align}\label{eqn:ZnpTauRepShort}
\boldsymbol{Z}^\tau_{n,p}&=e^{s_{p+3\dots n}\omega(1,0)}\prod_{
	p+2<j\leq n}e^{-s_{(12\dots p+2),j}(\CG_{j1}^{\tau}-\omega(1,0))}\KN^{\tau}_{p+3\dots n}\boldsymbol{\Gamma}_{p+2}\boldsymbol{v}_{n,p}\nnl
&=e^{s_{12\dots p+2}\omega(1,0)}\prod_{
	p+2<j\leq n}e^{-s_{(12\dots p+2),j}\CG_{j1}^{\tau}}\KN^{\tau}_{p+3\dots n}\boldsymbol{\Gamma}_{p+2}\,\begin{pmatrix}\boldsymbol{U}_{p+3,p+1}
\\
0
\end{pmatrix}\boldsymbol{Z}^\textrm{tree}_{p+3,p+2}\,.
\end{align}
Comparing this representation with the previous result \eqref{eqn:ZnpTauRep}, we have achieved a considerable simplification for the dependence on the punctures $z_{p+3},\dots, z_n$. In particular, the following eigenvalue equation can be read off:
\begin{align}
\boldsymbol{\Gamma}_{p{+}3}\dots \boldsymbol{\Gamma}_n\,\begin{pmatrix}\boldsymbol{U}_{p+3,p+1}
\\
0
\end{pmatrix}&=e^{(s_{12\dots p+2}-s_{12\dots n})\omega(1,0)}\prod_{
	p+2<j\leq n}e^{-s_{(12\dots p+2),j}\CG_{j1}^{\tau}}\KN^{\tau}_{p+3\dots n}\begin{pmatrix}\boldsymbol{U}_{p+3,p+1}
\\
0
\end{pmatrix}\,.
\end{align}
\subsection{$Z^{\tau}_{4,1}$-integrals}
Let us check the formul\ae{} \eqref{eqn:ZnpTauRep} and \eqref{eqn:ZnpTauRepShort} for $(n,p)=(4,1)$, i.e.\
\begin{align}\label{eqn:Z41TauRep}
\boldsymbol{Z}^{\tau
}_{4,1}(z_3,z_4)&=e^{s_{1234}\omega(1,0)}\,\boldsymbol{\Gamma}_{3}\boldsymbol{\Gamma}_4\,\begin{pmatrix}
1/s_{12}\\
-1/s_{23}\\
0
\end{pmatrix} \frac{\Gamma(1-s_{12})\Gamma(1-s_{23})}{\Gamma(1-s_{12}-s_{23})}\nnl
&=e^{s_{123}\omega(1,0)-s_{(123,4)}\CG_{41}^{\tau}}\,\boldsymbol{\Gamma}_{3}\,\begin{pmatrix}
1/s_{12}\\
-1/s_{23}\\
0
\end{pmatrix} \frac{\Gamma(1-s_{12})\Gamma(1-s_{23})}{\Gamma(1-s_{12}-s_{23})}\,,
\end{align}
where the generating series
\begin{align}
\label{eq:Gamma4normalOrdering}
\boldsymbol{\Gamma}_4(z_4)&=\sum_{r\geq 0}\sum_{k_1,\dots, k_r\geq 0}\tilde{\Gamma}\left(\begin{smallmatrix}
k_1&\dots&k_r\\
0&\dots&0
\end{smallmatrix}; z_4,\tau\right)(\boldsymbol{x}_{41 }^{(k_1)}+\boldsymbol{x}_{43}^{(k_1)})\dots (\boldsymbol{x}_{41}^{(k_r)}+\boldsymbol{x}_{43}^{(k_r)})
\end{align}
and
\begin{align}
\label{eq:Gamma3normalOrdering}
\boldsymbol{\Gamma}_3(z_3,z_4)=\sum_{r\geq 0}\sum_{k_1,\dots, k_r\geq 0}\sum_{j_{1},\dots,j_r\in \{1,4\}}\tilde{\Gamma}\left(\begin{smallmatrix}
k_1&\dots&k_r\\
z_{j_i}&\dots&z_{j_r}
\end{smallmatrix}; z_3,\tau\right)\boldsymbol{x}_{3j_1 }^{(k_1)}\dots \boldsymbol{x}_{3j_r}^{(k_r)}
\end{align}
involve the matrices from \subsecref{subsec:exDEQ}. The first few terms in the $\ap$-expansion of $\boldsymbol{Z}^{\tau
}_{4,1}(z_3,z_4)$ by the use of \eqn{eqn:Z41TauRep} are given by:

\begin{align}\label{eqn:Z41TauRepAPexpansion}
	\boldsymbol{Z}^{\tau
	}_{4,1}(z_3,z_4)&=
	e^{s_{123}\omega(1,0)-s_{(123,4)}\CG_{41}^{\tau}}\,  
	\begin{pmatrix}
		W^\tau_1\\
		W^\tau_2\\
		W^\tau_3
	\end{pmatrix} \,,
\end{align}
where the non-trivial $\ap$-expansions of the components $W^\tau_1$, $W^\tau_2$ and $W^\tau_3$ are
\begin{align}
	 \label{eq:vector1APandEta}
   	W^\tau_1   =&  \frac{1}{s_{12}}   \nnl  
   	                &+ \left[ \frac{1}{\eta_2}\tilde{\Gamma}\left(\begin{smallmatrix}
   	                	0\\
   	                	0
   	                \end{smallmatrix}; z_3,\tau\right) - \frac{s_{12}+s_{23}+s_{13}}{s_{12}}\tilde{\Gamma}\left(\begin{smallmatrix}
   	                1\\
   	                0
                   \end{smallmatrix}; z_3,\tau\right) -\frac{s_{34}}{s_{12}}\tilde{\Gamma}\left(\begin{smallmatrix}
                   1\\
                   z_4
               \end{smallmatrix}; z_3,\tau\right)
                 +\mathcal{O}(\eta_2)   \right]  \nnl
                & - \left[ \frac{1}{\eta_2} \left( ( s_{12}+s_{23}+s_{13}) \tilde{\Gamma}\left(\begin{smallmatrix}
                	0 & 1\\
                	0 & 0
                \end{smallmatrix}; z_3,\tau\right)
                 +(s_{24}+s_{34}) \tilde{\Gamma}\left(\begin{smallmatrix}
                 	0 & 1\\
                 	0 & z_4
                 \end{smallmatrix}; z_3,\tau\right)
                 \right. 
                 \right.
                 \nnl
                &
                \left.
                \left.
                \quad + s_{13} \tilde{\Gamma}\left(\begin{smallmatrix}
                	1 & 0\\
                	0 & 0
                \end{smallmatrix}; z_3,\tau\right)
            +s_{34} \tilde{\Gamma}\left(\begin{smallmatrix}
            	1 & 0\\
            	z_4 & 0
            \end{smallmatrix}; z_3,\tau\right)
                \right)
                  +\mathcal{O}(\eta_2^0)\right]
                \nnl
                & +\mathcal{O}(\ap^2) \,, \\
 \label{eq:vector2APandEta}
   	W^\tau_2   =&  -\frac{1}{s_{23}}   \nnl  
&+ \left[ \frac{1}{\eta_2}\tilde{\Gamma}\left(\begin{smallmatrix}
	0\\
	0
\end{smallmatrix}; z_3,\tau\right) + \frac{s_{12}+s_{23}+s_{13}}{s_{23}}\tilde{\Gamma}\left(\begin{smallmatrix}
	1\\
	0
\end{smallmatrix}; z_3,\tau\right) +\frac{s_{24}+s_{34}}{s_{23}}\tilde{\Gamma}\left(\begin{smallmatrix}
	1\\
	z_4
\end{smallmatrix}; z_3,\tau\right)
+\mathcal{O}(\eta_2)   \right]  \nnl
& - \left[ \frac{1}{\eta_2} \left( ( s_{12}+s_{23}+s_{13}) \tilde{\Gamma}\left(\begin{smallmatrix}
	0 & 1\\
	0 & 0
\end{smallmatrix}; z_3,\tau\right)
+(s_{24}+s_{34}) \tilde{\Gamma}\left(\begin{smallmatrix}
	0 & 1\\
	0 & z_4
\end{smallmatrix}; z_3,\tau\right)
\right. 
\right.
\nnl
&
\left.
\left.
\quad + s_{13} \tilde{\Gamma}\left(\begin{smallmatrix}
	1 & 0\\
	0 & 0
\end{smallmatrix}; z_3,\tau\right)
+s_{34} \tilde{\Gamma}\left(\begin{smallmatrix}
	1 & 0\\
	z_4 & 0
\end{smallmatrix}; z_3,\tau\right)
\right)
+\mathcal{O}(\eta_2^0)\right]
\nnl
& +\mathcal{O}(\ap^2) \,, \\
\label{eq:vector3APandEta}
	W^\tau_3 =&  \left[ \frac{1}{\eta_2} \tilde{\Gamma}\left(\begin{smallmatrix}
		0\\
		0
	\end{smallmatrix}; z_3,\tau\right) - \tilde{\Gamma}\left(\begin{smallmatrix}
	1\\
	z_4
\end{smallmatrix}; z_3,\tau\right) + \eta_2   \tilde{\Gamma}\left(\begin{smallmatrix}
2\\
z_4
\end{smallmatrix}; z_3,\tau\right) +\mathcal{O}(\eta_2^2) \right]   \nnl
	&-\left[(s_{12}+s_{23}+s_{13})\tilde{\Gamma}\left(\begin{smallmatrix}
		0&1\\
		0&0
	\end{smallmatrix}; z_3,\tau\right) +(s_{24}+s_{34})\tilde{\Gamma}\left(\begin{smallmatrix}
	0&1\\
	0&z_4
\end{smallmatrix}; z_3,\tau\right)  \right.    \nnl
	&\left. \quad + s_{13} \tilde{\Gamma}\left(\begin{smallmatrix}
		1&0\\
		0&0
	\end{smallmatrix}; z_3,\tau\right) + s_{14} \tilde{\Gamma}\left(\begin{smallmatrix}
	1&0\\
	z_4&0
\end{smallmatrix}; z_3,\tau\right)
	 + \mathcal{O}(\eta_2^0) \right] \nnl
	 &+\mathcal{O}(\ap^2)  \, .
\end{align}

There are three  key features of the components $W^\tau_i$ above. Both $W^\tau_1$ and $W^\tau_2$ have a kinematic pole (in $s_{12}$ and $s_{23}$ respectively) at their $\mathcal{O}(\eta^0_2)$ coefficients, while every other coefficients have no kinematic pole. The $W^\tau_3$ component, however, has no kinematic pole. Some of these poles are expected from the form of the initial values -- however, the fact that several of these cancel is not obvious from this differential equation method. Lastly, another fact not apparent from the differential equation method shown  here is that none of the $W^\tau_i$ have poles $\eta^{-2}_2$, $\eta^{-3}_2, \, \ldots$ of order $\geq 2$ in $\eta_2$. This is not obvious from the generating series of $\boldsymbol{\Gamma}_3(z_3,z_4)$, but is clear from the original definition of the integrand of $\boldsymbol{Z}^\tau_{4,1}(z_3,z_4)$ in \eqn{eq:eqnZn4p1Defn}.

In the next subsection, we will review an alternative way to obtain the $\ap$-expansions of $\boldsymbol{Z}^\tau_{4,1}(z_3,z_4)$ via the direct $\ap$-expansion of the integral. This will serve both as a sanity check for the values of $W^\tau_i$ found above, but will also explain the   appearance or absence of the kinematic poles in these components.

\subsection{$Z^{\tau}_{4,1}$-integrals by direct integration}

Now, we want to cross-check \eqref{eqn:Z41TauRepAPexpansion} via direct integration, i.e. $\ap$-expanding the Koba--Nielsen factor $\KN^\tau_{12\ldots n}$, $\eta_2$-expanding the Kronecker-Einsenstein series and performing the integrations of every term in this expansion. However, we can see that the vector of integrals $Z^{\tau}_{4,1}$ contains entries ($W^\tau_1$ and $W^\tau_2$) with poles in the Mandelstam variables $s_{ij}$, which cannot be generated by $\ap$-expanding the genus-one Koba--Nielsen factor of the integrand.  These poles only appear a the order $\eta^{0}_2$, and are an indication that we have to deal with  the simple poles of the worldsheet functions $g^{(1)}(z_{j}-z_2)$ , for  $j=1,3$.  In the presence of these simple poles, one needs the pole subtraction method of Section 5 of \rcite{Broedel:2019vjc}. We will detail the results of using these methods in \appref{app:PoleSubraction}. 

For the direct integration, we need to $\ap$-expand the integrand. A first identity for the $\ap-$expansion of the Koba--Nielsen factor $\KN^\tau_{1234}$ with $z_i>z_j$ reads 
\begin{align} \label{eqn:keyKNidentityZij}
\tilde{\Gamma}\left(\begin{smallmatrix}
	1\\
	0
\end{smallmatrix}; z_i-z_j,\tau\right) =  \tilde{\Gamma}\left(\begin{smallmatrix}
1\\
z_i
\end{smallmatrix}; z_j,\tau\right) + \tilde{\Gamma}\left(\begin{smallmatrix}
1\\
0
\end{smallmatrix}; z_i,\tau\right) 
	\, ,
\end{align}
which we use for all $\mathcal{G}^\tau_{ji}$ with $i \not = 0$. With this identity, we note that  the genus-one Koba--Nielsen factor $\KN^\tau_{1234}$  factorizes as follows:

\begin{align} \label{eqn:expandingKNtau}
		\KN^\tau_{1234} =& e^{s_{1234}\omega(1,0;\tau)}\, e^{-s_{123,4} \tilde{\Gamma}\left(\begin{smallmatrix}
				1\\
				0
			\end{smallmatrix}; z_4,\tau\right) } \times \exp[-s_{34} \tilde{\Gamma}\left(\begin{smallmatrix}
		1\\
	z_4
\end{smallmatrix}; z_3,\tau\right) -(s_{13}+s_{23})\tilde{\Gamma}\left(\begin{smallmatrix}
1\\
0
\end{smallmatrix}; z_3,\tau\right) ] \nnl
&\times \exp[-s_{12}\tilde{\Gamma}\left(\begin{smallmatrix}
	1\\
	0
\end{smallmatrix}; z_2,\tau\right) -s_{23}\tilde{\Gamma}\left(\begin{smallmatrix}
1\\
z_3
\end{smallmatrix}; z_2,\tau\right) -s_{24} \tilde{\Gamma}\left(\begin{smallmatrix}
1\\
z_4
\end{smallmatrix}; z_2,\tau\right)] \, ,
\end{align}
from which we can write $\boldsymbol{Z}^{\tau}_{4,1}(z_3,z_4)$ as:

\begin{align} \label{eqn:Zn4p1ExpandingKN}
	&\boldsymbol{Z}^{\tau
	}_{4,1}(z_3,z_4) \nnl
	&=e^{s_{1234}\omega(1,0;\tau)}\, e^{-s_{123,4} \tilde{\Gamma}\left(\begin{smallmatrix}
			1\\
			0
		\end{smallmatrix}; z_4,\tau\right) }   \nnl
	&\times \exp[-s_{34} \tilde{\Gamma}\left(\begin{smallmatrix}
		1\\
		z_4
	\end{smallmatrix}; z_3,\tau\right) -(s_{13}+s_{23})\tilde{\Gamma}\left(\begin{smallmatrix}
		1\\
		0
	\end{smallmatrix}; z_3,\tau\right) ] \nnl
 &\times \int_0^{z_3}\dd z_2\,  \exp[-s_{12}\tilde{\Gamma}\left(\begin{smallmatrix}
 	1\\
 	0
 \end{smallmatrix}; z_2,\tau\right) -s_{23}\tilde{\Gamma}\left(\begin{smallmatrix}
 	1\\
 	z_3
 \end{smallmatrix}; z_2,\tau\right) -s_{24} \tilde{\Gamma}\left(\begin{smallmatrix}
 	1\\
 	z_4
 \end{smallmatrix}; z_2,\tau\right)] 
	 \begin{pmatrix}
		F(z_{12},\eta_2,\tau)\\
		F(z_{32},\eta_2,\tau)\\
		F(z_{42},\eta_2,\tau)
	\end{pmatrix} \, .
\end{align}
The purpose of writing the Koba--Nielsen factor $\KN^\tau_{1234}$ as in \eqn{eqn:expandingKNtau} becomes clear when we compare \eqn{eqn:Zn4p1ExpandingKN} with \eqn{eqn:Z41TauRepAPexpansion}: the last two lines of \eqn{eqn:Zn4p1ExpandingKN} correspond to the nontrivial vector $(W^\tau_1,W^\tau_2,W^\tau_3)^\intercal$ of \eqn{eqn:Z41TauRepAPexpansion}. Furthermore, the last line of \eqn{eqn:Zn4p1ExpandingKN} contains an integral in $z_2$ that we can now start to $\ap$-expand and $\eta_2$-expand its integrand.

For their importance in this subsection, we will define the last line of \eqn{eqn:Zn4p1ExpandingKN} as:

\begin{align} \label{eqn:mu1mu2mu3Defn}
\begin{pmatrix}
	V^\tau_1\\
	V^\tau_2\\
	V^\tau_3
\end{pmatrix} 
 =  \int_0^{z_3}\dd z_2\,  \exp[-s_{12}\tilde{\Gamma}\left(\begin{smallmatrix}
		1\\
		0
	\end{smallmatrix}; z_2,\tau\right) -s_{23}\tilde{\Gamma}\left(\begin{smallmatrix}
		1\\
		z_3
	\end{smallmatrix}; z_2,\tau\right) -s_{24} \tilde{\Gamma}\left(\begin{smallmatrix}
		1\\
		z_4
	\end{smallmatrix}; z_2,\tau\right)] 
	\begin{pmatrix}
		F(z_{12},\eta_2,\tau)\\
		F(z_{32},\eta_2,\tau)\\
		F(z_{42},\eta_2,\tau)
	\end{pmatrix} \, .
\end{align}

We now report the first few  $\eta_2$,$\ap$- coefficients of these $V^\tau_j$ integrals, up to $\mathcal{O}(\ap^2)$ and $\mathcal{O}(\eta_2)$ terms\footnote{It is natural to organize this computation by first expanding in $\eta_2$, and then taking different orders in $\ap$. This is contrary to the differential equation method, in which one naturally expands in $\ap$ first, and afterwards in $\eta_2$.}:

\begin{subequations}
\begin{align} \label{eq:vector1muAPandEta}
	V^\tau_1 =&  \frac{1}{\eta_2} \left[  \tilde{\Gamma}\left(\begin{smallmatrix}
		0\\
		0
	\end{smallmatrix}; z_3,\tau\right) -   s_{12 }\tilde{\Gamma}\left(\begin{smallmatrix}
		0&1\\
		0&0
	\end{smallmatrix}; z_3,\tau\right) - s_{24 }  \tilde{\Gamma}\left(\begin{smallmatrix}
		0&1\\
		0&z_4
	\end{smallmatrix}; z_3,\tau\right) + s_{23 }  \tilde{\Gamma}\left(\begin{smallmatrix}
		1&0\\
		0&0
	\end{smallmatrix}; z_3,\tau\right)   \right]   \nnl
	&+ \left[ \frac{1}{s_{12}} -  \tilde{\Gamma}\left(\begin{smallmatrix}
		1\\
		0
	\end{smallmatrix}; z_3,\tau\right) -2 s_{23}\tilde{\Gamma}\left(\begin{smallmatrix}
		0&2\\
		0&0
	\end{smallmatrix}; z_3,\tau\right) +s_{12}\tilde{\Gamma}\left(\begin{smallmatrix}
		1&1\\
		0&0
	\end{smallmatrix}; z_3,\tau\right)   +s_{24}\tilde{\Gamma}\left(\begin{smallmatrix}
		1&1\\
		0&z_4
	\end{smallmatrix}; z_3,\tau\right)    \right.     \nnl
	&  \left. \quad  - s_{23}\tilde{\Gamma}\left(\begin{smallmatrix}
		2&0\\
		0&0
	\end{smallmatrix}; z_3,\tau\right) - s_{23} \zeta_2  \right]   \nnl
	&+\mathcal{O}(\eta_2)     +\mathcal{O}(\ap^2)   \, , \\
\label{eq:vector2muAPandEta}
	V^\tau_2 =&  \frac{1}{\eta_2} \left[  \tilde{\Gamma}\left(\begin{smallmatrix}
		0\\
		0
	\end{smallmatrix}; z_3,\tau\right) -   s_{12 }\tilde{\Gamma}\left(\begin{smallmatrix}
		0&1\\
		0&0
	\end{smallmatrix}; z_3,\tau\right) - s_{24 }  \tilde{\Gamma}\left(\begin{smallmatrix}
		0&1\\
		0&z_4
	\end{smallmatrix}; z_3,\tau\right) + s_{23 }  \tilde{\Gamma}\left(\begin{smallmatrix}
		1&0\\
		0&0
	\end{smallmatrix}; z_3,\tau\right)   \right]   \nnl
	& +\left[ -\frac{1}{s_{23}} +\frac{s_{12}}{s_{23}}  \tilde{\Gamma}\left(\begin{smallmatrix}
		1\\
		0
	\end{smallmatrix}; z_3,\tau\right) +\frac{s_{24}}{s_{23}}  \tilde{\Gamma}\left(\begin{smallmatrix}
		1\\
		0
	\end{smallmatrix}; z_3,\tau\right) - \frac{s^{2}_{12}}{s_{23}}\tilde{\Gamma}\left(\begin{smallmatrix}
		1&1\\
		0&0
	\end{smallmatrix}; z_3,\tau\right) - \frac{s_{12} s_{24}}{s_{23}}\tilde{\Gamma}\left(\begin{smallmatrix}
		1&1\\
		0&z_4
	\end{smallmatrix}; z_3,\tau\right)     \right.     \nnl
	&  \left.  \quad - \frac{s_{12} s_{24}}{s_{23}}\tilde{\Gamma}\left(\begin{smallmatrix}
		1&1\\
		z_4&0
	\end{smallmatrix}; z_3,\tau\right)   - \frac{s^{2}_{24}}{s_{23}}\tilde{\Gamma}\left(\begin{smallmatrix}
		1&1\\
		z_4&z_4
	\end{smallmatrix}; z_3,\tau\right) +(2 s_{12} +s_{24}) \tilde{\Gamma}\left(\begin{smallmatrix}
		0&2\\
		0&0
	\end{smallmatrix}; z_3,\tau\right)  \right.   \nnl
	&  \left.  \quad +  s_{24} \tilde{\Gamma}\left(\begin{smallmatrix}
		0&2\\
		0&z_4
	\end{smallmatrix}; z_3,\tau\right)   +s_{24}  \tilde{\Gamma}\left(\begin{smallmatrix}
		1&1\\
		0&z_4
	\end{smallmatrix}; z_3,\tau\right) + s_{12} \tilde{\Gamma}\left(\begin{smallmatrix}
		2&0\\
		0&0
	\end{smallmatrix}; z_3,\tau\right) - s_{24} \tilde{\Gamma}\left(\begin{smallmatrix}
		1&1\\
		z_4&z_4
	\end{smallmatrix}; z_3,\tau\right) \right.   \nnl
	&  \left.  \quad +  s_{24} \tilde{\Gamma}\left(\begin{smallmatrix}
		2&0\\
		z_4&0
	\end{smallmatrix}; z_3,\tau\right)   +s_{12} \zeta_2 \right]   \nnl
	&+\mathcal{O}(\eta_2)     +\mathcal{O}(\ap^2)   \, , \\
 \label{eq:vector3muAPandEta}
	V^\tau_3 =&  \frac{1}{\eta_2} \left[  \tilde{\Gamma}\left(\begin{smallmatrix}
		0\\
		0
	\end{smallmatrix}; z_3,\tau\right) -   s_{12 }\tilde{\Gamma}\left(\begin{smallmatrix}
		0&1\\
		0&0
	\end{smallmatrix}; z_3,\tau\right) - s_{24 }  \tilde{\Gamma}\left(\begin{smallmatrix}
		0&1\\
		0&z_4
	\end{smallmatrix}; z_3,\tau\right) + s_{23 }  \tilde{\Gamma}\left(\begin{smallmatrix}
		1&0\\
		0&0
	\end{smallmatrix}; z_3,\tau\right)   \right]   \nnl
	&+ \left[ - \tilde{\Gamma}\left(\begin{smallmatrix}
		1\\
		z_4
	\end{smallmatrix}; z_3,\tau\right) - s_{23}\tilde{\Gamma}\left(\begin{smallmatrix}
		0&2\\
		0&0
	\end{smallmatrix}; z_3,\tau\right) - s_{23}\tilde{\Gamma}\left(\begin{smallmatrix}
		0&2\\
		0&z_4
	\end{smallmatrix}; z_3,\tau\right)   -s_{23}\tilde{\Gamma}\left(\begin{smallmatrix}
		1&1\\
		0&z_4
	\end{smallmatrix}; z_3,\tau\right)    \right.     \nnl
	&  \left. \quad  + s_{12}\tilde{\Gamma}\left(\begin{smallmatrix}
		1&1\\
		z_4&0
	\end{smallmatrix}; z_3,\tau\right) + (s_{23}+s_{24})\tilde{\Gamma}\left(\begin{smallmatrix}
		1&1\\
		z_4&z_4
	\end{smallmatrix}; z_3,\tau\right) -  s_{23}\tilde{\Gamma}\left(\begin{smallmatrix}
		2&0\\
		z_4&0
	\end{smallmatrix}; z_3,\tau\right)  \right]   \nnl
	&+\mathcal{O}(\eta_2)     +\mathcal{O}(\ap^2)   \, .
\end{align}
\end{subequations}

In computing these  $V^\tau_j$ integrals  we need to use $z$-removal identities that are detailed in appendix B.2 of \rcite{Broedel:2014vla}. Computing these integrals is  mostly straightforward, except for the $\mathcal{O}(\eta^0_2)$ coefficients of $V^\tau_1$ and $V^\tau_2$, which are spelled out in \eqns{eq:V1tauExplicitFormula}{eq:V2tauExplicitFormula}.

We note that the expansions of obtained here for \eqn{eqn:mu1mu2mu3Defn} is consistent with the one found in \eqn{eqn:Z41TauRepAPexpansion}. This serves as an important sanity check for our method of computing Selberg integrals by the differential equation of $\boldsymbol{Z}^{\tau
}_{n,p}$. We have performed this check up to and including some $\mathcal{O}(\eta^{2}_2)$ and $\ap^2$ expressions. More explicitly, we have found both computations for 
$\boldsymbol{Z}^{\tau
}_{4,1}$ to agree up to eMPLs of depth 3 and total weight 5.

\section{Analytic continuation and alternative initial conditions}
\label{sec:AnalyticConti}

In the previous section, we have studied the $\ap$-expansion of $\boldsymbol{Z}^\tau_{n,p}$-integrals  $\boldsymbol{\Gamma}_{n,p}$, a generating function of eMPLs defined on the domain

\begin{align}\label{eqn:domainGamma}
	0&=z_1<z_{p+2}<z_{p+3}<\dots<z_{n}<1\,.
\end{align}

\begin{figure}
\centering
\begin{tikzpicture}
		\braid[ number of strands=3, 
		border height=2pt,
		style strands={3}{line width=1pt},style strands={1}{line width=0.5pt},style strands={2}{line width=1.5pt}] (braid) at (1,0), it's a name
		a_1 a_2 a_1;
		\node[ at=(braid-1-s),  label=north : $z_1$ ] {} ;
		\node[ at=(braid-2-s),  label=north : $z_2$ ] {} ;
		\node[ at=(braid-3-s),  label=north : $z_3$ ] {} ;
		\node[ at=(braid-1-e),  label=south : $z_1$ ] {} ;
		\node[ at=(braid-2-e),  label=south : $z_2$ ] {} ;
		\node[ at=(braid-3-e),  label=south : $z_3$ ] {} ;
		\node at ([yshift=-3cm]braid) {\((a)\)};
\end{tikzpicture}
\quad \quad \quad \quad \quad \quad
\begin{tikzpicture}
	\braid[ number of strands=3, 
	border height=2pt,
	style strands={3}{line width=1pt},style strands={1}{line width=0.5pt},style strands={2}{line width=1.5pt}] (braid) at (1,0), it's a name
	a_1 a_2 a_1;
	\node[ at=(braid-1-s),  label=north : $z_1$ ] {} ;
	\node[ at=(braid-2-s),  label=north : $z_2$ ] {} ;
	\node[ at=(braid-3-s),  label=north : $z_3$ ] {} ;
	\node[ at=(braid-1-e),  label=south : $z_1$ ] {} ;
	\node[ at=(braid-2-e),  label=south : $z_2$ ] {} ;
	\node[ at=(braid-3-e),  label=south : $z_3$ ] {} ;
	\node at ([yshift=-3cm]braid) {\((b)\)};
	\draw [fill=white] (0.7,-3.0) rectangle (3.25,-0.2);
	\node at (2,-1.5) {pr};
\end{tikzpicture}
\caption{(a) A graphical representation of the braiding $\sigma_{1,2} \sigma_{2,3} \sigma_{1,2}$ of punctures $z_1$, $z_2$, and $z_3$. (b) The projection $\textrm{pr}$ forgets the details of the braiding, and just remembers the permutation performed on the punctures: the permutation $(13)$, which exchanges punctures $z_1$ with $z_3$.}\label{fig:projectionPR}
\end{figure}
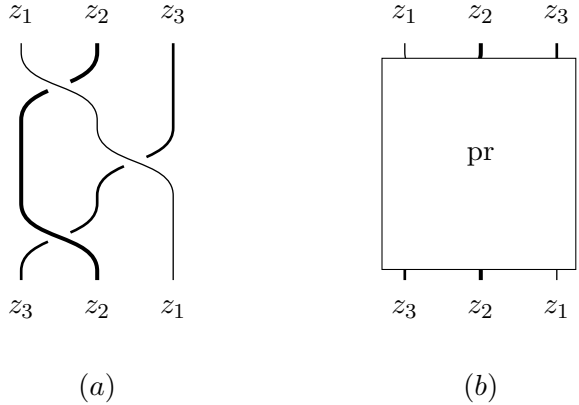

In this section, we make use of elements of the theory of the universal KZB equation to describe how to extend this generating function to a domain

\begin{align}\label{eqn:domainGammaSigmaPermutted}
	0&=z_1<z_{\beta(p+2)}<z_{\beta(p+3)}<\dots<z_{\beta(n)}<1\,,
\end{align}
where $\beta \in S_{n-p-1}$. We can relate these two domains by continuously braiding the unintegrated punctures around each other, while keeping them distinct along the way. This process defines a certain \textit{braiding} of the punctures.
We need to make a choice of how to braid the punctures $\{z_{p+2},z_{p+3},\ldots,z_n\}$ around each other when relating these two domains. That is, the permutation $\beta$  described above is actually obtained from a  projection of a braiding $g \in B_{n-p-1}$, where $B_N$ is the braid group on $N$ strands. We denote by $\textrm{pr}$ projection from the braid group $B_{n-p-1}$ into the permutation group $S_{n-p-1}$: 

\begin{align}
\textrm{pr}: &B_{n-p-1}\rightarrow S_{n-p-1}\,  \nnl
\textrm{pr}: &g \mapsto g^{\textrm{pr}} \, ,
\end{align}
where the permutation $g^{\textrm{pr}} $ only remembers the endpoints of the braiding.  \figref{fig:projectionPR} exemplifies what the projection $\textrm{pr}$ does. The theory of the KZB equation will allow us to describe such a braiding by a simple matrix multiplication acting on our generating function $\boldsymbol{\Gamma}_{n,p}$.

\subsection{Analytic continuation of  $\boldsymbol{\Gamma}_{n,p}$ and relating fibration bases of eMPLs}

We will describe the analytic continuation of $\boldsymbol{\Gamma}_{n,p}(z_{p+2},\ldots,z_n)$ given by an element $g\in B_N$, where $N=n-p$.\footnote{One can also consider the  braid group on the torus, $B_{1,N}$, with additional generators given by $A-$cycle and $B-$cycle monodromies. We will not consider these braiding elements in this work, but note that the theory of the KZB equation can also deal with these generators} The braid group $B_N$ on the strands $\{1,2,\ldots,N\}$ is the group generated by the elements $\sigma_{i}:=\sigma_{i,i+1}$ acting on nearest neighbors, where $1\leq i \leq N-1$, satisfying the relations \cite{kassel2008braid}

\begin{align}
\sigma_i \sigma_j &= \sigma_j \sigma_i \, , \; & |i-j| \geq 2 \, , \nnl
\sigma_i \sigma_{i+1} \sigma_i & = \sigma_{i+1} \sigma_i \sigma_{i+1} \, ,  \;& 1 \leq i \leq N-2\, .	
\end{align}

For convenience, we will label the generators of the braid group by the indices of the unintegrated punctures $\{p+2,p+3,\ldots,n\}$ \footnote{We note that $z_1$ is also unintegrated.}. Now, letting $\sigma_{i,i+1}$ denote the operation of braiding the puncture $z_{i+1}$ counterclockwise  half a turn around $z_i$, we can define the function $\sigma_{i,i+1} \boldsymbol{\Gamma}_{n,p}$,  on the region 
\begin{align}\label{eqn:domainGeneratoricommaiplus1}
	0&=z_1<z_{p+2}<z_{p+3}<\dots<z_{i+1}<z_i<\ldots<z_{n}<1\,.
\end{align}

This function is obtained from the  analytic continuation of $\boldsymbol{\Gamma}_{n,p}$ by performing the braiding operation $\sigma_{i,i+1}$ described above: we interchange $z_i$ and $z_{i+1}$ by moving the puncture  $z_{i+1}$ counterclockwise around $z_i$. We write the  function we obtain through this process of analytic continuation by \cite{KZB}:

\begin{align}\label{eqn:analyticConticommaiplus1}
	\sigma_{i,i+1} \boldsymbol{\Gamma}_{n,p} &:=\boldsymbol{\Gamma}_{p{+}2}\dots \boldsymbol{\Gamma}_n\,\mathbb{X}(\sigma_{i,i+1})   \boldsymbol{v}_{n,p} \nnl
	&\phantom{:}= \left( \boldsymbol{\Gamma}_{p{+}2}\dots \boldsymbol{\Gamma}_n\, \right)\big{|}_{(i,i+1)}  \boldsymbol{v}_{n,p} \, ,
\end{align}
where $\mathbb{X}(\sigma_{i,i+1})$ is a matrix that will be defined later, and $(i,i+1)$ is a permutation that  acts on the indices of the punctures $z_i$ and $z_{i+1}$  in the definition of $\boldsymbol{\Gamma}_{n,p}$.  This permutation acts in the arguments of the eMPLs, in the path-ordered integration of the KZB connection, shown in \eqn{eqn:pathZero}, and additionally on the lower indices of the matrices  $\boldsymbol{x}_{ri}^{(k)}$ and $\boldsymbol{x}_{i}^{(0)}$. However, this $(i,i+1)$ permutation does not act on the indices of the Mandelstam variables $s_{ij}$ appearing in the matrix entries of our representation, nor does it on the indices of the $\eta_i$ variables. 

The matrix $\mathbb{X}(\sigma_{i,i+1})$, in a sense, performs the analytic continuation above, and is given by 

\begin{align}\label{eqn:matrixXicommaiplus1}
\mathbb{X}(\sigma_{i,i+1}) &= \Phi \left(
 - \boldsymbol{x}_{1,i}^{(1)} - \sum_{j=p+2}^{i-1}\boldsymbol{x}_{j,i+1}^{(1)}  
 ,
 -\boldsymbol{x}_{i,i+1}^{(1)}  
  \right)
\exp\left(i \pi  \boldsymbol{x}_{i,i+1}^{(1)}   \right)
 \nnl
	& \; \; \times  \Phi \left(
	 - \boldsymbol{x}_{i,i+1}^{(1)} 
	 ,
	 - \boldsymbol{x}_{1,i+1}^{(1)} - \sum_{j=p+2}^{i-1}\boldsymbol{x}_{j,i+1}^{(1)}     
	 \right) 
 \, ,
\end{align}
where $\Phi (A,B)$ is the Drinfeld associator, a regularized holonomy of the KZ equation, and is given by the series in noncommuting variables $E_0$,$E_1$ \cite{Le}:
\begin{align}\label{eqn:drinfAssoc}
	\Phi (E_0,E_1)&=\sum^{\infty}_{r=0} \sum_{a_1,a_2,\ldots,a_r\in \{0,1\}}  G(a_r,\ldots,a_2,a_1;1)E_{a_1}E_{a_2}\ldots E_{a_r}  \; \;  \nnl 
	& =1 +  \zeta_2 [E_0,E_1] - \zeta_3[E_0+E_1, [E_0,E_1] ] + \ldots    \, ,
\end{align}
where we obtain the shuffle-regularized MZVs from the first line, as in \eqn{eqn:zetaFromMPLs}. We note that its multiplicative inverse is simply given by $[\Phi(E_0,E_1)]^{-1}=\Phi(E_1,E_0)$. The genus-0 case (i.e. the analytic continuation of two punctures for the KZ equation) has a form precisely analogous to  \eqn{eqn:analyticConticommaiplus1} here. The genus-0 case is written in equation 5.18 of \rcite{Britto:2021prf}, but with \textit{braid matrices} $e_{ij}$ instead of the matrices $\boldsymbol{x}^{(1)}_{ij}$ here.

Knowing how analytic continuation works for a generator of the braid group $\sigma_{i,i+1}$, we can furnish from this an action of any group element $g \in B_{n-p}$. The key is to find out how to describe the group action of a product of group elements $g= g_1 g_2$. In particular, it is sufficient to find a matrix $\mathbb{X}(g)$ that satisfies:
\begin{align}\label{eqn:analyticContGbraid}
 \boldsymbol{\Gamma}_{p{+}2}\dots \boldsymbol{\Gamma}_n\, \mathbb{X}(g)   = g^\textrm{pr}\left(  \boldsymbol{\Gamma}_{p{+}2}\dots \boldsymbol{\Gamma}_n\,  \right) \, ,
\end{align}
where $g^\textrm{pr}=\textrm{pr}(g) \in S_{n-p}$. By using the equation above and inserting $g=g_1 g_2$ we find the condition
\begin{align}\label{eqn:matrixXContGbraid}
 \mathbb{X}(g_1 g_2) = \mathbb{X}( g_1)    g_1^\textrm{pr} \left(  \mathbb{X}( g_2)  \right) ,
\end{align}
where we are using conventions of composition of braidings and permutations consistent with $\sigma_{34}^\textrm{pr}\sigma_{45}^\textrm{pr} = (34)(45)=(345)$. We can use \eqn{eqn:matrixXContGbraid} to obtain $ \mathbb{X}(g) $ for any group element $g\in B_{n-p}$.

\subsection{Analytic continuation of $\boldsymbol{\Gamma}_{4,1} $ and an alternative initial value} \label{subsec:AnalyticContn4p1}

The simplest example of an analytic continuation as described in the previous section is the case of $(n,p)=(4,1)$, with a braiding $\sigma_{34}$. From the \eqn{eqn:analyticConticommaiplus1} , we can see that the initial value $\boldsymbol{v}_{4,1}$ plays an spectating role in this equation, so we can write the analytic continuation of generating functions of eMPLs as follows\footnote{The explicit matrices $\boldsymbol{x}_{3}^{(0)}$ and $\boldsymbol{x}_{4}^{(0)}$  have $\eta_2-$derivatives, which turns the generating functions $\boldsymbol{\Gamma}_{i}$ into operators. However, there is no $\eta_2-$dependence in $\boldsymbol{v}_{4,1}$, so we can understand the following equation as an operator acting on a $3\times 3 $ identity matrix. The end result of this operation is a matrix.}:

\begin{align}\label{eqn:analyticContn4p1braid34}
	\boldsymbol{\Gamma}_{3}(z_3,z_4) \boldsymbol{\Gamma}_4(z_4)\, \mathbb{X}(\sigma_{34})   &= \left(  \boldsymbol{\Gamma}_{3}(z_3,z_4) \boldsymbol{\Gamma}_4(z_4)\,  \right)\big{|}_{3\leftrightarrow 4} \, =:  \boldsymbol{\Gamma}_{4}(z_4,z_3) \boldsymbol{\Gamma}_3(z_3) \,,
\end{align}

where the generating functions on the right-hand side are defined by the permutation $(34)$ of the middle term, and $\mathbb{X}(\sigma_{34})$ is given by
\begin{align}\label{eqn:matrixXicommaiplus1case34}
	\mathbb{X}(\sigma_{34}) &= \Phi \left(
	- \boldsymbol{x}_{13}^{(1)} 
	,
	-\boldsymbol{x}_{34}^{(1)}  
	\right)
	\exp\left(i \pi  \boldsymbol{x}_{34}^{(1)}   \right)  \Phi \left(
	- \boldsymbol{x}_{34}^{(1)} 
	,
	- \boldsymbol{x}_{14}^{(1)}   
	\right) 
	\, .
\end{align}
In particular, we have that 

\begin{align}\label{eqn:Gammaz3ofz3}
	\boldsymbol{\Gamma}_{3}(z_3) &=\sum_{r\geq 0}\sum_{k_1,\dots, k_r\geq 0}\Gamma\left(\begin{smallmatrix}
		k_1&\dots&k_r\\
		z_{1}&\dots&z_{1}
	\end{smallmatrix}; z_3,\tau\right)\tilde{\boldsymbol{X}}_{3,1 }^{(k_1)}\dots \tilde{\boldsymbol{X}}_{3,1}^{(k_r)}\,.
\end{align}
The matrices $\tilde{\boldsymbol{X}}_{3,1}^{(k)}$ are defined for  $k \geq 0$ from the matrices $\boldsymbol{x}_{ij}^{(k)}$ appearing in the elliptic KZB system \eqref{eqn:ellKZBeqzi,eqn:ellKZBeqtau} by
\begin{align}
	\tilde{\boldsymbol{X}}_{3,1}^{(k)}&=\boldsymbol{x}_{31}^{(k)}+\boldsymbol{x}_{34}^{(k)} \, .
\end{align}
Meanwhile, we have for the other generating function on the RHS of \eqn{eqn:analyticContn4p1braid34}:

\begin{align}\label{eqn:Gammaz4ofz4z3}
	\boldsymbol{\Gamma}_{4}(z_4,z_3) &=\sum_{r\geq 0}\sum_{k_1,\dots, k_r\geq 0}\sum_{j_{1},\dots,j_r\in \{1,3\}}\tilde{\Gamma}\left(\begin{smallmatrix}
		k_1&\dots&k_r\\
		z_{j_1}&\dots&z_{j_r}
	\end{smallmatrix}; z_4,\tau\right)\tilde{\boldsymbol{X}}_{4,j_1 }^{(k_1)}\dots \tilde{\boldsymbol{X}}_{4,j_r}^{(k_r)}\,.
\end{align}
The matrices $\tilde{\boldsymbol{X}}_{4,j}^{(k)}$ are defined for $j\in \{1,3\}$  and $k \geq 0$ from the matrices $\boldsymbol{x}_{ij}^{(k)}$ appearing in the elliptic KZB system \eqref{eqn:ellKZBeqzi,eqn:ellKZBeqtau} by
\begin{align}
	\tilde{\boldsymbol{X}}_{4,j}^{(k)}&=\boldsymbol{x}_{4,j}^{(k)} \, .
\end{align}

We have made explicit which  $\boldsymbol{x}_{ij}^{(k)}$ appear in \eqn{eqn:analyticContn4p1braid34}  to make explicit the action of the permutation $3\leftrightarrow 4$, which is consistent with performing the path-ordered integration of \secref{sec:alphaExpansion}. There is key difference between the definitions here, for $\boldsymbol{\Gamma}_3(z_3)$ and $\boldsymbol{\Gamma}_4(z_4,z_3)$, compared with the definitions of \eqns{eq:Gamma4normalOrdering}{eq:Gamma3normalOrdering} for 
$\boldsymbol{\Gamma}_4(z_4)$ and $\boldsymbol{\Gamma}_3(z_3,z_4)$. The series $\boldsymbol{\Gamma}_4(z_4,z_3)$ here contains eMPLs where the rightmost puncture is $z_4$, and the punctures to the left of the semicolon can be $z_3$ or $z_1$, e.g. $\Gamma\left(\begin{smallmatrix}
	1\\
	z_{3}
\end{smallmatrix}; z_4,\tau\right)$. Meanwhile, the series $\boldsymbol{\Gamma}_3(z_3,z_4)$ contains eMPLs where the rightmost puncture is $z_3$ and the punctures to the left of the semicolon can be $z_4$ or $z_1$, e.g. $\Gamma\left(\begin{smallmatrix}
1\\
z_{4}
\end{smallmatrix}; z_3,\tau\right)$.

Now, we  will look back at  \eqn{eqn:analyticContn4p1braid34} and note that it is relating eMPLs in different fibration bases. For example, the LHS contains $\tilde{\Gamma}\left(\begin{smallmatrix}
	1\\
	z_{3}
\end{smallmatrix}; z_4,\tau\right)$, while the RHS contains $\tilde{\Gamma}\left(\begin{smallmatrix}
1\\
z_{4}
\end{smallmatrix}; z_3,\tau\right)$. Because of this, we can obtain identities relating different fibration bases by taking coefficients in the $\ap-$ and $\eta_2-$expansion of both sides of \eqn{eqn:analyticContn4p1braid34}.

\subsubsection{Two-puncture fibration-basis-change identities from matrix representations} \label{subsec:TwoPuntureFibrationFromRep}

In this section, we use the explicit matrix representations that we found in \secref{subsec:exDEQ} into the definition of \eqref{eqn:analyticContn4p1braid34}. As explained in the previous subsection, we  insert an identity matrix as the initial condition on which the operator-valued LHS and RHS of \eqref{eqn:analyticContn4p1braid34} act. 

From the steps above, we obtain a  matrix equation for  \eqref{eqn:analyticContn4p1braid34} that can be expanded in both $\ap$ and $\eta_2$. From this equation, we  can now obtain identities by looking at a certain entries of these matrices, and looking at a certain Mandelstam monomial and $\eta_2$ power. An example of this is the following equation, that comes from looking at the $\eta^{0}_2 s_{34}$ coefficient of the $(1,1)$-entry:

\begin{align}\label{eqn:firstEasyFibrationn4p1}
\tilde{\Gamma}\left(\begin{smallmatrix}
	1\\
	0
\end{smallmatrix}; z_3,\tau\right)
+
\tilde{\Gamma}\left(\begin{smallmatrix}
	1\\
	z_3
\end{smallmatrix}; z_4,\tau\right)
=
i \pi 
+\tilde{\Gamma}\left(\begin{smallmatrix}
	1\\
	0
\end{smallmatrix}; z_4,\tau\right)
+\tilde{\Gamma}\left(\begin{smallmatrix}
	1\\
	z_4
\end{smallmatrix}; z_3,\tau\right)
.
\end{align}

This equation has been checked numerically in Mathematica, and it holds when $\arg (z_3) > \arg(z_4)$ for  $z_3$ and $z_4$ in the fundamental parallelogram. The identity  is very similar to the identity for MPLs, which can be found via PolyLogTools \cite{PolyLogTools}:
\begin{align}
G(0;z_3)+G(z_3;z_4)=i \pi + G(0;z_4)+G(z_4,z_3) \, ,
\end{align}
valid in the same domain. Note, however, that both of these last two fibration-basis-change formul\ae{} can be performed by hand. However, one can find higher length-and-weight fibration-basis-change formul\ae{}. For example, one can obtain a fibration-basis-change formula for the eMPL $\tilde{\Gamma}\left(\begin{smallmatrix}
	1&1&1\\
	0&z_3&z_3
\end{smallmatrix}; z_4,\tau\right)$  -- which contains  a  $\zeta_3$ -- by looking at the system of two equations formed by  the $\eta^{0}_2 s_{23} s_{24}$- and the $\eta^{0} s^2_{23} s_{24}$- coefficients of the $(1,3)$-entry of \eqn{eqn:analyticContn4p1braid34}. These identities are showcased in \eqns{eq:appFibrationLength2}{eq:appFibrationLength3}. We note that the RHS of these equations feature eMPLs of the form $\tilde{\Gamma}\left(\begin{smallmatrix}
k_1& \ldots& k_r\\
0 & \ldots &0
\end{smallmatrix}; z_4,\tau\right)$ and $\tilde{\Gamma}\left(\begin{smallmatrix}
k_1 &\ldots &k_r\\
a_1 &\ldots &a_r
\end{smallmatrix}; z_3,\tau\right)$ for $a_i \in \{0,z_4\}$, making these nontrivial identities.

We note that our numerical implementation has been cross-checked  with the one of GINAC whenever they coincide \cite{Weinzierl_numerical_elliptic}. We need our own implementation because identities like the one above always include at least one eMPL that requires an analytic continuation not supported by the current implementation of GINAC. We note that one can also use the elliptic symbol of \rcite{Broedel:2018iwv} as checks for these equations.

Following this method, one is  able to extract equations that relate eMPLs written in different fibration bases.  However,  the number of distinct eMPLs grows exponentially with powers of $\ap$ and $\eta_2$ -- which respectively correspond to the length and the weight of the eMPLs -- while the number of independent Mandelstam monomials grows polynomially with $\ap$. Thus, at some point the system of equations one is able to obtain  cannot necessarily be solved for. For example, the eMPLs $\tilde{\Gamma}\left(\begin{smallmatrix}
	2 & 3\\
	z_3 & 0
\end{smallmatrix}; z_4,\tau\right)$ and 
$\tilde{\Gamma}\left(\begin{smallmatrix}
	3 & 2\\
	z_3 & 0
\end{smallmatrix}; z_4,\tau\right)$ only appear in a single monomial, $s_{12} s_{24}$, in the $(2,1)-$entry of the left-hand-side of  \eqn{eqn:analyticContn4p1braid34}. Because of this, we can obtain an equation involving  these two eMPLs that can not be solved analytically for either of them.  Still, one can also insert in  \eqn{eqn:analyticContn4p1braid34}  larger matrix representations  with $(n,p)=(n,n-3)$. These larger matrix representations are obtained in scenarios that also leave the last two punctures unintegrated, but involve matrices of size ($\frac{(n-1)!}{2} \times \frac{(n-1)!}{2}$), more $\eta_i$ variables and more Mandelstam monomials. We invite the interested reader to use the ancillary files for the matrices for $(n,p)=(5,1),(5,2)$ to test this method. An alternative method to solve for the eMPLs of these analytic continuation identities is outlined in \appref{app:fibrationBasisChange}.

We have looked at several of the equations we can obtain from inserting the explicit matrix representations for $(n,p)=(4,1)$  in \eqn{eqn:analyticContn4p1braid34}. We have performed numerical checks on these equations up to weight $3$ and length $3$ in the eMPLs. All the equations so obtained have been found to be consistent with the same branch choice, namely $\arg (z_3) > \arg(z_4)$ for  $z_3$ an $z_4$ in the fundamental parallelogram. This is consistent with the definition of the braiding operation that performs the analytic continuation.

\subsubsection{Alternative initial value for $\boldsymbol{\Gamma}_{4,1} $}

In the previous subsection, we described the analytic continuation of $\boldsymbol{\Gamma}_{4,1} $ in an active way. However, there is an alternate interpretation of this analytic continuation, coming from the last two terms of  \eqn{eqn:analyticConticommaiplus1}. Because $\mathbb{X}(\sigma_{i,i+1})$ is invertible, the following equation holds:
\begin{align}\label{eqn:analyticConticommaiplus1PassiveVoice}
\boldsymbol{\Gamma}_{4,1}  = 	\boldsymbol{\Gamma}_{3}(z_3,z_4)  \boldsymbol{\Gamma}_4(z_4)\,   \boldsymbol{v}_{4,1} 
	= \boldsymbol{\Gamma}_{4}(z_4,z_3) \boldsymbol{\Gamma}_3(z_3)\, \big[ \mathbb{X}(\sigma_{3,4})\big]^{-1}  \boldsymbol{v}_{4,1} \, \, .
\end{align}

In the equation above, we remind the reader that the generating functions $\boldsymbol{\Gamma}_{4}(z_4,z_3) $ and $\boldsymbol{\Gamma}_{3}(z_3) $ are defined naturally from performing path-ordered integration in the order
\begin{align}
(0,0) \rightarrow (z_3,0) \rightarrow (z_3,z_4)\, \, ,
\end{align}
which suggests that with this order of integration in mind, we should use $\big[ \mathbb{X}(\sigma_{3,4})\big]^{-1}  \boldsymbol{v}_{4,1}$ as an initial value. That is:
\begin{align}\label{eqn:initialvAlternateN4P1}
	\big[ \mathbb{X}(\sigma_{3,4})\big]^{-1}  \boldsymbol{v}_{4,1}= \lim_{z_{3}\to 0}(-2\pi i z_{3})^{-\boldsymbol{X}_{3,1}^{(1)}} \lim_{z_{4}\to 0}(-2\pi i z_{4})^{-\boldsymbol{X}_{4,1}^{(1)}}\boldsymbol{Z}^{\tau
	}_{4,1}\, .
\end{align}

In this passive way of understanding the analytic continuation of \eqn{eqn:analyticConticommaiplus1}, we will be able to find the initial values from which to perform path-ordered integration along alternative integration paths. If we use this alternative initial value in \eqn{eqn:Z41TauRep}, now the $\ap$-expansion will include factors of $i \pi$ and additional MZVs.

\subsection{Outlook for analytic continuation for more punctures} \label{c}

In this short subsection, we  restate that \eqns{eqn:analyticContGbraid}{eqn:matrixXContGbraid} allow one to write more complicated analytic continuations (labeled by how one braids the punctures $z_i$ around each other) for any number of unintegrated punctures, and any braiding element $g$ --  this last fact is restating that the  $\sigma_{i,i+1}$ are the generators of $B_{1,n-p}$. We will show an example of such braidings in this section. We will write \eqn{eqn:analyticContGbraid} for the case $(n,p)=(5,1)$, i.e. with three unintegrated punctures: ($z_3$, $z_4$, $z_5$), and $g= \sigma_{34}\sigma_{45}$:
\begin{align}\label{eqn:analyticContGbraidForN5P1sigma34sigma45}
	\boldsymbol{\Gamma}_{3}(z_3,z_4,z_5) \boldsymbol{\Gamma}_{4}(z_4,z_5)\boldsymbol{\Gamma}_{5}(z_5) \mathbb{X}(\sigma_{34}) \left( \mathbb{X}(\sigma_{45}) \big|_{3\leftrightarrow 4} \right)  = \boldsymbol{\Gamma}_{4}(z_3,z_5,z_4) \boldsymbol{\Gamma}_{5}(z_3,z_5)\boldsymbol{\Gamma}_{3}(z_3) \, ,
\end{align}
where we have made  use of $\mathbb{X}(\sigma_{34}\sigma_{45})=\mathbb{X}(\sigma_{34}) \left( \mathbb{X}(\sigma_{45}) \big|_{3\leftrightarrow 4} \right)$ on the LHS and of $\textrm{pr}( \sigma_{34}\sigma_{45})=(345)$ on the RHS. $\mathbb{X}(\sigma_{34})$ is defined as in \eqn{eqn:matrixXicommaiplus1case34}, and
\begin{align}
 \mathbb{X}(\sigma_{45}) \big|_{3\leftrightarrow 4} &= \Phi \left(
 - \boldsymbol{x}_{14}^{(1)} -\boldsymbol{x}_{34}^{(1)} 
 ,
 -\boldsymbol{x}_{45}^{(1)}  
 \right)
 \exp\left(i \pi  \boldsymbol{x}_{45}^{(1)}   \right)  \Phi \left(
 - \boldsymbol{x}_{45}^{(1)} 
 ,
 -  \boldsymbol{x}_{15}^{(1)} -\boldsymbol{x}_{35}^{(1)} 
 \right)  \big|_{3\leftrightarrow 4}   \nnl
 &=  \Phi \left(
 - \boldsymbol{x}_{13}^{(1)} -\boldsymbol{x}_{34}^{(1)} 
 ,
 -\boldsymbol{x}_{35}^{(1)}  
 \right)
 \exp\left(i \pi  \boldsymbol{x}_{35}^{(1)}   \right)  \Phi \left(
 - \boldsymbol{x}_{35}^{(1)} 
 ,
 -  \boldsymbol{x}_{15}^{(1)} -\boldsymbol{x}_{45}^{(1)} 
 \right)  \, .
\end{align}
The LHS and RHS of \eqn{eqn:analyticContGbraidForN5P1sigma34sigma45} contain eMPLs in different fibration basis, and one can use e.g. the methods of \appref{app:fibrationBasisChange} to obtain change-of-fibration-basis identities for eMPLs in three variables. By using numerical methods, and checking the coefficients of \eqn{eqn:analyticContGbraidForN5P1sigma34sigma45} by the methods of \appref{app:fibrationBasisChange}, we have found that  this equation is valid for $\arg (z_3) > \arg(z_4)> \arg(z_5)$ for  $z_3$, $z_4$, and $z_5$ in the fundamental parallelogram. See \eqn{eqn:threeVariableFibration} for an example of a change-of-fibration-basis identity we can find, associated to this braiding. We remark that this is a similar domain of validity for $g=\sigma_{34}\sigma_{45}$ as in the genus-0 case of appendix B.3 of \rcite{Britto:2021prf}.

\section{Connections with the mathematical literature} \label{sec:ConnectionMathLit}

We remark that the elements $\boldsymbol{x}_{i,j}^{(n)}$ and $\boldsymbol{x}^{(0)}_i$ that we have described in our differential system have been checked to satisfy several of the relations of elements of the genus-one Drinfeld-Kohno algebra $\bar{\mathfrak{t}}_{1,N}$, for $N=n-p$ unintegrated punctures, which mathematicians use for the study of the  KZB equation. In this section, we will follow the notation of \rcite{KZB}, and refer to some related work on derivation algebras by \rcites{Tsunogai,Enriquez:EllAss, Pollack}. More precisely, the KZB equation in the so-called universal case is a system of differential equations for a function of $(n-p)$ punctures $\{z_i\}$ , $F^{\textrm{KZB}}(z_1,z_{p+2},\ldots , z_n,\tau)$, written in the following form\footnote{We are using a numbering system for indices matching our convention for $\pmb{Z}^{\tau}_{n,p}(z_{p+2},z_{p+3},\dots,z_n)$.}:

\begin{subequations}
\label{eqn:universalKZBeqnForF}
\begin{align}\label{eqn:universalKZBeqnForFzi}
	\partial_i F^{\textrm{KZB}} &=\left(-y_i+\sum_{k\geq 1}\sum_{\begin{smallmatrix}
			r\in \{1,p+2,\dots,n\}\\r\neq i
	\end{smallmatrix}} \ad^{k}_{x_i}(y_r) g^{(k)}_{ir}\right)F^{\textrm{KZB}}\,,\\
\label{eqn:universalKZBeqnForFtau}
	2\pi i\partial_{\tau} F^{\textrm{KZB}} &=\left(- \Delta_{0}+\sum_{k\geq 4}(1-k) \delta_{k-2} G_k +\sum_{\begin{smallmatrix}
			r,q\in \{1,p+2,\dots,n\}\\
			q< r
	\end{smallmatrix}}\sum_{k\geq 2 }(k-1) \ad^{k-1}_{x_q}(y_r)  g^{(k)}_{qr}\right) F^{\textrm{KZB}}  \,,
\end{align}
\end{subequations}

where we are using the fact that the Eisenstein series $G_k$ vanish for odd values of $k$, so the only terms that contribute in the second sum are $\delta_{2k}$, for $k \geq 2$. In  \eqn{eqn:universalKZBeqnForF} above, the $\{x_i\}$ and $\{y_i\}$ are generators of the algebra $\bar{\mathfrak{t}}_{1,n-p}$, with indices $\{1,p+2,p+3,\ldots, n\}$. A word of caution: these generators $x_i$ should not be confused with the matrices $\boldsymbol{x}^{(0)}_i$ and $\boldsymbol{x}^{(k)}_{ij}$, which we write in boldface letters to avoid confusion. $\Delta_0$ and the  $\delta_{2k}$ are elements of a lie Algebra $\mathfrak{d}$, which acts as a derivation algebra on $\bar{\mathfrak{t}}_{1,n-p}$ (i.e. there is a Lie algebra morphism $\mathfrak{d} \rightarrow \textrm{Der}(\bar{\mathfrak{t}}_{1,n-p})$. We refer to the image of this morphism with the same notation as elements in $\mathfrak{d}$ by abuse of notation). Moreover, we are denoting nested commutators or repeated adjoint actions by $\ad^n_a(b) = [a, \ad^{n-1}_a(b)]$, and  $ad^1_a(b)=\ad_a(b)=[a,b]$.

The genus-one Drinfeld-Kohno algebra $\bar{\mathfrak{t}}_{1,n-p}$ is generated by the elements $\{x_i\}$ and $\{y_i\}$, with indices $i \in \{1,p+2,p+3,\ldots, n\}$, and subject to the following relations\footnote{The last of these relations is what separates  $\mathfrak{t}_{1,n-p}$ from $\bar{\mathfrak{t}}_{1,n-p}$. The generators of the latter are written with horizontal bars on top of them in the mathematics literature, e.g. $\bar{x}_i$. We will denote these without the bar in this work.}:
\begin{subequations}
		\label{subalign:t1commanRelations}
\begin{align}\label{eqn:relationsOfGenusOneDrinfeldKohno}
	[x_i,x_j]=0 = [y_i,y_j] \,,  && \forall i,j \, ,  \\
	t_{ij} = [x_i,y_j] \,, &&  i \neq j \, ,  \label{eqn:tij}\\
	t_{ij} = t_{ji} \, && \\
	[t_{ij},t_{kl}] = 0 \, , &&  |\{i,j,k,l\}|=4 \, , \\
	[x_i,y_i] = - \sum_{ j\neq i} t_{ij}
	[x_i , t_{jk}] = 0 = [y_i, t_{jk}] \, , && |\{i,j,k\}|=3 \, , \\
	[x_i + x_j, t_{ij }] = 0 = [y_i + y_j, t_{ij}] \, ,&& |\{i,j,k\}|=3 \, , \\
	\sum_i x_i = 0 = \sum_i y_i \, \label{eqn:barAlgebraT1n}.
\end{align}
\end{subequations}
Note that  \eqn{eqn:tij} is nothing else as to name a certain commutator $t_{ij}$ to simplify the rest of the relations the generators of the algebra satisfy. Moreover, we note that \eqn{eqn:barAlgebraT1n} basically states that $F^{\textrm{KZB}}$ is translation-invariant, which reflects the case for our integrals $\boldsymbol{Z}^{\tau}_{n,p}$: $ \sum_i y_i =0$  translates to $ \sum_i \partial_{z_i} \boldsymbol{Z}^{\tau}_{n,p}=0$. Meanwhile, $ \sum_i x_i =0$ is related to an infinitesimal translation of all the punctures of $\boldsymbol{Z}^{\tau}_{n,p}$ along the $\tau$ direction, which should also vanish by translation-invariance. The derivations $\Delta_0$ and $\delta_{2k}$ have to satisfy the single relation among them in $\mathfrak{d}$:
\begin{align} \label{eqn:SL2orTsunogai}
	\ad^{2k+1}_{\Delta_0}(\delta_{2k})=0 \, ,
\end{align}
and as derivations on $\bar{\mathfrak{t}}_{1,n-p}$, the have to satisfy the following relations\footnote{For concreteness, we use here commutators for what in the \rcite{KZB} would be written in functional relation, e.g. $\Delta_0(x_i)=y_i $.}:
\begin{subequations}
	\label{subalign:DeltaRelations}
\begin{align}
	&[\Delta_0,x_i]=y_i  \, ,	\\
	&[\Delta_0,y_i] = [\Delta_0,t_{ij}] = 0  \, ,\\
	&[\delta_{2m},x_i]=0  \, ,\\
	&[\delta_{2m},t_{ij}] = [t_{ij}, \ad^{2m}_{x_i}(t_{ij}) ] \, , \\
	&[\delta_{2m},y_{i}] =  \frac{1}{2}  \sum_{ j \neq i} \sum_{p+q=2m-1} [\ad^p_{x_i}(t_{ij}), (-1)^q \ad^q_{x_i}(t_{ij})] \, . 
\end{align}
\end{subequations}
All in all, this means that the algebra elements appearing in \eqn{eqn:universalKZBeqnForFzi} are all elements of $\bar{\mathfrak{t}}_{1,n-p} \rtimes \mathfrak{d}$.\footnote{The semidirect product $\rtimes$ reflects the fact that the commutators among elements in $\mathfrak{d}$ and $\bar{\mathfrak{t}}_{1,n-p}$ written in  the equations \eqref{subalign:DeltaRelations} can all be expressed in terms of commutators of elements of $\bar{\mathfrak{t}}_{1,n-p}$ only.}  In this section, we will explain how the matrices $\boldsymbol{x}_{ij}^{(k)}$, $\boldsymbol{x}^{(0)}_i$ and $\epsilon^{(k)}$ we find in this work form a representation of a subalgebra of $  \bar{\mathfrak{t}}_{1,n-p} \rtimes \mathfrak{d}$, and discuss some nontrivial properties of $\bar{\mathfrak{t}}_{1,n-p} \rtimes \mathfrak{d}$ that are reflected on these matrices that appear in \eqn{eqn:ellKZBeq}.

\subsection{Building a dictionary between these languages} \label{subsection:Dictionary}

By comparing \eqn{eqn:universalKZBeqnForF}  with  \eqn{eqn:ellKZBeq}, we can readily postulate a dictionary that relates the matrices $\boldsymbol{x}_{i,j}^{(n)}$, $\boldsymbol{x}^{(0)}_i$ and $\boldsymbol{\epsilon}^{(k)}$ to elements in $\bar{\mathfrak{t}}_{1,n-p} \rtimes \mathfrak{d}$, i.e. with the commutation relations in \eqref{subalign:t1commanRelations} and \eqref{subalign:DeltaRelations}. The dictionary goes as follows:

\begin{subequations}
\begin{align}
	&\boldsymbol{x}^{(0)}_i =  -y_i  \, , \\
	&\boldsymbol{x}_{ij}^{(k)} = \ad^{k}_{x_i}(y_j) \, ,  \label{eqn:dictionaryAdx} \\
	&\boldsymbol{\epsilon}^{(0)} = \Delta_0   \, ,  \\
	&\boldsymbol{\epsilon}^{(2k)} = \delta_{2k-2} \, .
\end{align}
\end{subequations}
We first note the lack of a matrix representative for the generator $x_i \in \bar{\mathfrak{t}}_{1,n-p} \rtimes \mathfrak{d}$. Because of this, even if we were to forget about the action of $\mathfrak{d}$, the matrices $\boldsymbol{x}_{ij}^{(k)}$ and $\boldsymbol{x}^{(0)}_i$ furnish a representation of only a subalgebra $\mathcal{L} \subset \bar{\mathfrak{t}}_{1,n-p}$. We would like to emphasize that we tried to find a matrix representative for this $x_i \in \bar{\mathfrak{t}}_{1,n-p} $ but failed to find one in the same using $d_{(n,p)} \times d_{(n,p)}$ matrices. However, one can formally add Lie algebra generators that act like the $x_i \in \bar{\mathfrak{t}}_{1,n-p}$ and commute with each other, and thus obtain the whole of $\bar{\mathfrak{t}}_{1,n-p} $. \footnote{We thank Benjamin Enriquez for pointing this out}

One key property of $ \bar{\mathfrak{t}}_{1,n-p} \rtimes \mathfrak{d}$ is that its elements satisfy commutation relations sufficient for the integrability of the differential  system in \eqref{eqn:universalKZBeqnForFzi,eqn:universalKZBeqnForFtau} (i.e. the commutation of mixed partial derivatives). In the original writing of this equation of \rcite{KZB}, these properties are used succinctly via holomorphic Kronecker-Eisenstein series  that have been expanded in the writing of this equation here. In our analysis of the integrability of \eqns{eqn:ellKZBeqzi}{eqn:ellKZBeqtau}, we also arrived at some sufficient conditions that have to be imposed on commutators of $\boldsymbol{x}_{ij}^{(k)}$, $\boldsymbol{x}^{(0)}_i$ and $\epsilon^{(k)}$. We should expect these two results to match, but this matching is nontrivial, basically, because the commutation relations we found are written in terms of $\boldsymbol{x}_{ij}^{(k)}$, e.g. \eqn{eqn:comRealij} instead of the simpler generators $x_i$ and $y_i$. A satisfying fact to us was that most of these commutations relations among the $\boldsymbol{x}_{ij}^{(k)}$ (or rather, among the $\ad^{k}_{x_i}(y_j)$) are computed and used in \rcite{KZB}, the most complicated of these when proving the compatibility of $\mathfrak{d}$ as an algebra of derivations on $ \bar{\mathfrak{t}}_{1,n-p}$ and its own Lie algebraic relations.

To best of our knowledge, among the $\bar{\mathfrak{t}}_{1,n-p}$  relations, only the commutation relations for $ [\boldsymbol{x}_{i,q}^{(k)},\boldsymbol{x}_{j,r}^{(l)}]$ and $ [\boldsymbol{x}_{i,j}^{(k)},\boldsymbol{x}_{q,i}^{(l)}]$ are not included explicitly in \rcite{KZB}. However, given the dictionary of \eqn{eqn:dictionaryAdx}, it is straightforward to prove that the commutation relations of  these that appear in \eqn{eqn:comRealij} are implied by the relations among the generators of $\bar{\mathfrak{t}}_{1,n-p}$ .  Thus, we remark that to the best of our knowledge, the relations among the generators of $\bar{\mathfrak{t}}_{1,n-p} \rtimes \mathfrak{d}$, and the properties of $\mathfrak{d}$ as a derivation  imply the relations we have found in \secref{subsec:comRel}. We have additionally checked that the explicit matrices that we build out of the process described in \appref{app:DerivationOfDifferentialSystem} have been checked to satisfy the relations tabulated in  \secref{subsec:comRel}, for several different values of $(n,p)$ as sanity checks during the computation of the matrices themselves.

\subsection{Grading of $\bar{\mathfrak{t}}_{1,n-p}$}

We remark that $\bar{\mathfrak{t}}_{1,n-p} \rtimes \mathfrak{d}$  is a  graded algebra, with  grading of its elements given by
\begin{align} \label{eqn:degreesOfGrading}
\deg(x_i)&=(1,0) \, ,  &\deg(y_i) &= (0,1)  \, ,\nnl
\deg(\Delta_0) &= (-1,1)\, ,&\deg(\delta_{2m}) &= (2m+1,1)  \, .
\end{align}
The matrices $\boldsymbol{x}_{ij}^{(k)}$, $\boldsymbol{x}^{(0)}_i$ and $\boldsymbol{\epsilon}^{(k)}$ reflect this grading in the following way: According to our dictionary and the grading above, the second entry of the grading of these explicit matrices should be $1$. Noticing that these matrices are all proportional to Mandelstam variables $s_{ij}$ (which themselves are proportional to $\ap$), we posit that the $\ap-$order of the matrix representations that we have found follows precisely the grading of $y_i$. The grading of $x_i$ is slightly more subtle, but a glance at the matrices, for example, in the $(n,p)=(4,1)$ case should tell use that that this grading follows the homogeneity degree\footnote{We define this to follow the degree of the $\eta_i$ in the matrices, and to also be equal to $-k$ to $\partial^k_{\eta_i}$}  of the $\eta_i$ -- with an exception for  $\boldsymbol{\epsilon}^{(0)}$, which includes a term of order $\eta_i^0$ proportional to the identity matrix, which drops out from every commutator.  In particular, given that the matrices  $\boldsymbol{x}^{(0)}_i$ have a $(\eta_i,\ap)$ grading of $(-1,1)$, we conclude that the $\eta_i$ homogeneity power follows the abstract grading of $x_i$, with an offset of $1$. The homogeneity degrees in  $\eta_j$ and linearity in $\ap$ of the matrices $\boldsymbol{\epsilon}^{(k)}$ also follow the expected grading dictated by the grading given by this dictionary just described. In fact, this is the case generically for the matrices we find for our differential system in \eqref{eqn:ellKZBeqzi,eqn:ellKZBeqtau}, as can be read off from \eqns{eqn:derZnpClosed}{eqn:tauDerivClosed}. Thus, we posit that the  homogeneity degree of $\eta_i$ and the linearity in $\ap$  of the matrices $\boldsymbol{x}_{i,j}^{(n)}$, $\boldsymbol{x}^{(0)}_i$ and $\boldsymbol{\epsilon}^{(k)}$ is no coincidence, but rather is a consequence of these belonging to a representation of a graded algebra.\footnote{With this dictionary in mind, the two gradings in \eqn{eqn:degreesOfGrading} can be seen, up to a shift, as eigenvalues of the operators $[\sum_j \eta_j \partial_{\eta_j},\, \_\, ]$ and $[\alpha \partial_\alpha, \,\_\,]$, when dealing with the explicit matrix representations.}

The grading of $\bar{\mathfrak{t}}_{1,n-p}$ described above is relevant when considering equations such as \eqref{eqn:analyticContn4p1braid34}. If we give an interpretation of this equation in terms of generators of the algebra $\bar{\mathfrak{t}}_{1,n-p}$, then we notice that the eMPLs of weight $w$ and length $r$ in the generating function, say, $\boldsymbol{\Gamma}_{3}(z_3,z_4)$ appears as coefficient of an element of $\bar{\mathfrak{t}}_{1,n-p}$ with grading $(w,r)$. Moreover, every MZV $\zeta_{n_1,n_2, \ldots, n_r}$ in the Drinfeld associators appears as a coefficient of an element  of $\bar{\mathfrak{t}}_{1,n-p}$ with grading $(n_1+\ldots+n_r,n_1 + \ldots +n_r)$ -- this last fact means that we should assign a weight and length (as in eMPLs) of $n_1+\ldots+n_r$ to the MZV $\zeta_{n_1,n_2, \ldots, n_r}$  . With this in mind, we note that equations obtained from looking at components of \eqn{eqn:analyticContn4p1braid34} will contain eMPLs and MZV of uniform total weight and length as just described. An example of such an equation can be seen in the change-of-fibration-basis identities of \eqns{eqn:firstEasyFibrationn4p1}{eqn:threeVariableFibration}. As long as the  \eqn{eqn:analyticContn4p1braid34} indeed holds for generic values in $\bar{\mathfrak{t}}_{1,n-p}$, we should expect every such change-of-fibration-basis equation to respect the length-and-weight grading of eMPLs just described. 

\subsection{Pollack's relations}\label{subsec:Pollack}

We shall now comment on some extra commutation relations among the $\boldsymbol{\epsilon}^{(k)}$ not implied by the requirement of integrability. These relations have been studied by Tsunogai\cite{Tsunogai} and spelled out in great detail by Pollack\cite{Pollack}. These are commutation relations among a derivation algebra of elements $\boldsymbol{\epsilon}_{\textrm{Tsu}}^{(k)}$ that satisfy:
\begin{align}
	\ad^{k-1}_{\boldsymbol{\epsilon}_{\textrm{Tsu}}^{(0)}}(\boldsymbol{\epsilon}_{\textrm{Tsu}}^{(k)})=0 \, , \, \, \, k = 0,2,4,\ldots \, , \label{TsunigaiDepth1} 
\end{align}
which is a relation satisfied for example by the matrices $\boldsymbol{\epsilon}^{(k)}$ if we replace $\boldsymbol{\epsilon}_{\textrm{Tsu}}^{(k)} \rightarrow \boldsymbol{\epsilon}^{(k)}$,  according to \eqn{eqn:SL2orTsunogai}, after following the dictionary of  \subsecref{subsection:Dictionary}.  Moreover, the first lines of \eqns{eqb_xijEpskSome}{eqn_xijEX} implies that 
\begin{align}
	\ad^{k}_{\boldsymbol{\epsilon}^{(0)}}(\boldsymbol{x}_{ij}^{(k)}) =0  \label{TsunigaiISHxk}  \, .
\end{align}

However, there are further  nontrivial relations that these derivations of Tsunogai satisfy. For example, the following depth-two and depth-three identities are expected to be satisfied by such derivations of Tsunogai \cite{Broedel:2015hia}:

\begin{align}
 [\boldsymbol{\epsilon}_{\textrm{Tsu}}^{(10)} , \boldsymbol{\epsilon}_{\textrm{Tsu}}^{(4)}] - 3 [\boldsymbol{\epsilon}_{\textrm{Tsu}}^{(8)},\boldsymbol{\epsilon}_{\textrm{Tsu}}^{(6)}]&=0 \, , \nnl
 2[\boldsymbol{\epsilon}_{\textrm{Tsu}}^{(14)} , \boldsymbol{\epsilon}_{\textrm{Tsu}}^{(4)}] - 7 [\boldsymbol{\epsilon}_{\textrm{Tsu}}^{(12)} , \boldsymbol{\epsilon}_{\textrm{Tsu}}^{(6)}] + 11 [\boldsymbol{\epsilon}_{\textrm{Tsu}}^{(10)} , \boldsymbol{\epsilon}_{\textrm{Tsu}}^{(8)}] & =0 \, , \nnl
 -80[\boldsymbol{\epsilon}_{\textrm{Tsu}}^{(0)},[\boldsymbol{\epsilon}_{\textrm{Tsu}}^{(4)},\boldsymbol{\epsilon}_{\textrm{Tsu}}^{(12)}]]+250 [\boldsymbol{\epsilon}_{\textrm{Tsu}}^{(0)},[\boldsymbol{\epsilon}_{\textrm{Tsu}}^{(6)},\boldsymbol{\epsilon}_{\textrm{Tsu}}^{(10)}]] +96
 [\boldsymbol{\epsilon}_{\textrm{Tsu}}^{(4)},[\boldsymbol{\epsilon}_{\textrm{Tsu}}^{(0)},\boldsymbol{\epsilon}_{\textrm{Tsu}}^{(12)}]]& \nnl
 +462
 [\boldsymbol{\epsilon}_{\textrm{Tsu}}^{(4)},[\boldsymbol{\epsilon}_{\textrm{Tsu}}^{(4)},\boldsymbol{\epsilon}_{\textrm{Tsu}}^{(8)}]]-375 [\boldsymbol{\epsilon}_{\textrm{Tsu}}^{(6)},[\boldsymbol{\epsilon}_{\textrm{Tsu}}^{(0)},\boldsymbol{\epsilon}_{\textrm{Tsu}}^{(10)}]]-1725 [\boldsymbol{\epsilon}_{\textrm{Tsu}}^{(6)},[\boldsymbol{\epsilon}_{\textrm{Tsu}}^{(4)},\boldsymbol{\epsilon}_{\textrm{Tsu}}^{(6)}]]& \nnl 
 +280 [\boldsymbol{\epsilon}_{\textrm{Tsu}}^{(8)},[\boldsymbol{\epsilon}_{\textrm{Tsu}}^{(0)},\boldsymbol{\epsilon}_{\textrm{Tsu}}^{(8)}]]&=0 \label{TsunogaiDepth2}
\end{align} 

We have found that, while the matrices $\boldsymbol{\epsilon}^{(k)}$ that we found in this work do not satisfy these relations above, a certain combination of matrices do. If we define $\tilde{\boldsymbol{\epsilon}}^{(k)}$ by 

\begin{align}
	\tilde{\boldsymbol{\epsilon}}^{(0)}&=\boldsymbol{\epsilon}^{(0)} \, ,& \nnl
     \tilde{\boldsymbol{\epsilon}}^{(k)}&=\boldsymbol{\epsilon}^{(k)} + \sum_{i<j} \boldsymbol{x}_{i,j}^{(k-1)} \, , & k\geq 2\, , \textrm{even} \, , \label{eqn:EpsilonTilde}
\end{align} 
then the matrices $\tilde{\boldsymbol{\epsilon}}^{(k)}$ do satisfy the commutation relations of the Tsunogai special derivation algebra that we have exemplified above in \eqn{TsunogaiDepth2}. We have checked that these depth-two relations holds for different values of $(n,p)$, including $(4,1)$,  $(5,1)$, and $(5,2)$. Other higher-depth irreducible relations\footnote{\textit{Depth} here refers to the number of commutators not involving $\tilde{\boldsymbol{\epsilon}}_{\textrm{Tsu}}^{(0)}$ in the relation. \textit{Irreducible} relations are those that are not immediately implied by writing linear combinations of commutators of lower-depth relations.} that generalize the ones in \eqref{TsunogaiDepth2}, and described by  \rcite{Pollack} have been checked, including some depth-3 relations that have been written in \rcite{Pollack,Broedel:2015hia}. The formula of \eqn{eqn:EpsilonTilde} generalizes the result of \rcite{Broedel:2020tmd}\footnote{Reference \cite{Broedel:2020tmd} was the first to show that $\boldsymbol{\epsilon}^{(k)}$ by themselves failed to satisfy Pollack's relations at depth-3 and above, and found the combination of matrices that indeed satisfy Pollacks' relations.}, valid for  $(n,p)=(n,n-2)$, to the case of generic values of integrated and unintegrated punctures. 

The precise combinations that define  $\tilde{\boldsymbol{\epsilon}}^{(k)}$ have a geometrical meaning: If we take the limit $z_{ij}\rightarrow 0$ of \eqn{eqn:universalKZBeqnForFtau}, we obtain a differential equation in $\tau$  but for $\bar{\mathfrak{t}}_{1,1}$ instead of $\bar{\mathfrak{t}}_{1,n-p}$. Vectors of string integrals that satisfy such a differential equation have been studied in detail in \rcites{Mafra:2019xms,Mafra:2019ddf}. The differential equations studied in these papers gave rise to matrices $\tilde{\boldsymbol{\epsilon}}^{(k)}$ that are observed to satisfy the commutation relations of Pollack to high orders. We note that it was important in \rcites{Mafra:2019xms,Mafra:2019ddf} and in the present work to study cases of $(n,p)$ beyond $p=1$ to make sure that the accidental relations that appear in the $p=1$ case (see \eqn{eqn:bonusRelationRepresentationEpsilonsN4P1} ) don't spoil these tests. 

\section{Conclusion}

In this paper, we present vectors of genus-one open-string-like integrals with any number of unintegrated punctures, $\boldsymbol{Z}^\tau_{n,p}$ of length $d_{n,p}=\frac{(n-1)!}{(n-1-p)!}$  that satisfies a KZB system in multiple variables. We managed to do this by extending previous methods to handle integration-by-parts (IBP) and Fay identities -- a genus-1 version of partial-fractioning -- for chains of Kronecker-Eisenstein series used in \rcites{Mafra:2019xms,Mafra:2019ddf,Broedel:2020tmd}. The end result is that $\boldsymbol{Z}^\tau_{n,p}$ can be seen as a path-ordered integration of the KZB equation from an initial value of tree-level string integrals. More explicitly, the solution for $\boldsymbol{Z}^\tau_{n,p}$ is captured by some initial value given by a known series in MZVs, and a generating function of eMPLs. 

By showing that such a generalization of the vectors of integrals of \rcites{Mafra:2019xms,Mafra:2019ddf,Broedel:2020tmd} exists, we have provided further evidence towards the use of (generalized) chains of Kronecker-Eisenstein series as a basis of integrands that is complete under IBP and Fay identity. Because these Kronecker-Eisenstein series themselves are generating functions of the meromorphic but not doubly-periodic kernels  $g^{(k)}(z_{ij})$, the components of these  $\boldsymbol{Z}^\tau_{n,p}$ contain integrands closely related to the ones appearing in one-loop open-string amplitudes. Moreover, the IBP algorithms described in  \appref{app:DerivationOfDifferentialSystem} work perfectly well with the non-holomorphic version of the integrands of \rcites{Mafra:2019xms,Mafra:2019ddf,Broedel:2020tmd}. 
 
 The main results of this paper is the computation of integrals along the $A$-cycle (for unintegrated punctures $z_i \in \mathbb{R}$). Thus, most of the results here can be readily translated to ordinary doubly-periodic open-string integrals if we place the unintegrated punctures ${z_{p+2},\ldots,z_n}$ on the real axis. However, this is not the case for  \secref{sec:AnalyticConti} in which we use some results from the theory of the universal KZB equation -- itself defined with the meromorphic kernels $g^{(k)}(z_{ij})$ -- to determine  the analytic continuation of $\boldsymbol{Z}^\tau_{n,p}$  and their components, the eMPLs $\tilde{\Gamma}\left(\begin{smallmatrix}
 	k_1&\dots&k_r\\
 	z_{1}&\dots&z_{r}
 \end{smallmatrix}; z,\tau\right)$ of \rcite{Broedel:2017kkb}. This section also  suggests that the analytic continuation of eMPLs is itself dictated by properties of the universal KZB equation that can be read from  \rcite{KZB}. If so, this could be a valuable addition for the  numerical implementations of these eMPLs, e.g. of  \rcite{Weinzierl_numerical_elliptic}.
 
 We comment now on some of the questions that we have touched upon, but not gave a complete answer. We have described in  \appref{app:LastAppendixGenerlaIntegrationDomain} the steps needed to generalize the integrals $\boldsymbol{Z}^\tau_{n,p}$  to a different integration cycle for the \textit{integrated} punctures. We note that we do not currently know how to extend the content of this section towards a \textit{proof} showing that the dimension of the vector space of integration cycles  is\footnote{Because of the natural pairing of  integration cycles  and integrands given by integration, it is fair to  expect these to have the same dimension as vector spaces}  $d_{n,p}=\frac{(n-1)!}{(n-1-p)!}$. Given that we are working in a meromorphic setup, we believe that one can establish monodromy relations along the lines of \rcites{ManoWatanabe2012,Tourkine:2016bak,Hohenegger:2017kqy,OchirovTourkineVanhove2017,Casali:2019ihm,Stieberger:2021daa}, but we do not try to compute these in the present work\footnote{We expect that the material given in \appref{app:LastAppendixGenerlaIntegrationDomain} will help in checking such monodromy relations.}. Given all monodromy relations, one could find the number of independent integration cycles modulo these, which would give a starting candidate for a basis.  We expect that with such bases of integration cycles and integrands in hand, one could perform  the coaction checks for these genus-one integrals along the lines of \rcite{Britto:2021prf} \footnote{We note that such coaction checks have been done for $(n,p)=(n,n-1)$ in the section 7.2 of \rcite{Mafra:2019xms} using the coaction of \rcite{Broedel:2018iwv}}. 
 
We comment that a proof of the dimensionality\footnote{We could be using an overcomplete number of integrals or missing some  integrals in defining  the components of $\boldsymbol{Z}^\tau_{n,p}$} of the $\boldsymbol{Z}^\tau_{n,p}$  could be obtained by giving the entries of this vector  an interpretation as intersection numbers between (representatives of) twisted cohomology and homology with local coefficients, along the lines of  \rcites{Mizera:2017cqs,Mizera:2019gea,Britto:2021prf}, which build upon the theory of twisted (co)homology of Aomoto and Kita \cite{AomotoKita}. To the best of our knowledge, the closest construction in the literature to the one in this work are the ones by \rcites{FelderVarchenkoKZB95,ManoWatanabe2012}. In particular, the authors of  \rcite{ManoWatanabe2012} do manage to give a proof of the dimensionality of the integrands and integration contours for  the integrals $\boldsymbol{Z}^\tau_{n,p}$  for $(n,p)=(n,1)$ via methods of  twisted (co)homology. 
Moreover, it would be interesting to relate the KZB equations in this work
 to those of the  integral representations  in \rcite{FelderVarchenkoKZB95} and to investigate
 possible connections with conformal blocks on an elliptic curve.
 Finally, a  better understanding of these genus-one  integration contours -- in particular, their so-called ``twisted intersection numbers'' -- could lead to a genus-one generalization of the string theory KLT relations \cite{Kawai:1985xq}.

\section*{Acknowledgments}

The authors thank Johannes Broedel and Oliver Schlotterer for their encouragement and careful reading of this work in its final stages. The authors also thank Nils Matthes and Sebastian Mizera for their comments on this draft. CR thanks Benjamin Enriquez for a helpful and motivating discussion during Elliptics 2021. AK was supported by the IMPRS for Mathematical and Physical Aspects of Gravitation, Cosmology and Quantum Field Theory. CR is supported  by the European Research Council under ERC-STG-804286 UNISCAMP.

\appendix

\section{Meromorphic v.s. Doubly periodic integrals}
\label{app:DictionaryGenusOne}
In this appendix, we compare the integrals of this work with the ones of \rcites{Mafra:2019ddf,Mafra:2019xms,Broedel:2020tmd}. 

\subsection{Dictionary between the integrands}

The genus integrals described here and in \rcites{Mafra:2019ddf,Mafra:2019xms,Broedel:2020tmd} are schematically of the following form:

\begin{align}
\label{eqn:prototypicalIntegral}
Z_{n}^\tau= \int_\gamma \dd z_2 \dd z_3 \ldots \dd z_{p+1} \hat{\KN}^\tau_{12\ldots n}  \, \hat{\Phi}(\sigma,{z_{ij},\eta_{j}})   \, ,
\end{align}

where $\gamma$ is an integration contour, $\hat{\Phi}(\sigma,{z_{ij},\eta_{j}})$  is defined as in \eqn{eqn:EKchain}, but with either meromorphic $F(z,\eta, \tau)$ (in this work) or doubly-periodic $\Omega(z,\eta,\tau)$ (in \rcites{Mafra:2019ddf,Mafra:2019xms,Broedel:2020tmd}) Kronecker-Eisenstein series. The Koba--Nielsen factor  $\hat{\KN}^\tau_{12\ldots n} $  is defined as
\begin{align}
	\hat{\KN}^\tau_{12\ldots n}&=\exp\Big(-\sum\limits_{1\leq j<i\leq n} s_{ij}\hat{\CG}^\tau_{ij}\Big)\, ,
	\label{eqn:KNloopHat}
\end{align}
where the open-string Green's function $\hat{\CG}^\tau_{ij}$ is given by:

\begin{align}
	\hat{\CG}^\tau_{ij}&=\begin{cases}
		\tilde{\Gamma}\big(\begin{smallmatrix}1\\0\end{smallmatrix};z_{ij}\, , \tau\big)-\omega(1,0,\tau)  \, ,&  \textrm{this \, work} \, , \\
		{\Gamma}\big(\begin{smallmatrix}1\\0\end{smallmatrix};|z_{ij}|\, ,\tau\big)-\omega(1,0,\tau)  \, ,&  \textrm{\rcites{Mafra:2019ddf,Mafra:2019xms,Broedel:2020tmd}}  \, .
	\end{cases}
\end{align}
These differences means that with an integration contour $\gamma$ that corresponds to a planar integral (i.e. integrating along the real axis only), the integrals of our work and of \rcites{Mafra:2019ddf,Mafra:2019xms,Broedel:2020tmd} coincide, at least up to a harmless phase difference (due to the absolute value in $\hat{\CG}^\tau_{ij}$). We can further expand the dictionary thus far once we notice that the $\tau$ derivative of $\hat{\KN}^\tau_{12\ldots n}$ is given by

\begin{align}
2 \pi i \partial_{\tau}\hat{\KN}^\tau_{12\ldots n}&=\begin{cases}
		-\sum_{1\leq i \leq j \leq n} s_{ij} (g^{(2)}_{ij}+2 \zeta_2)\hat{\KN}^\tau_{12\ldots n}   \, ,&  \textrm{this \, work} \, , \\
			-\sum_{1\leq i \leq j \leq n} s_{ij} (f^{(2)}_{ij}+2 \zeta_2)\hat{\KN}^\tau_{12\ldots n} \, ,&  \textrm{\rcites{Mafra:2019ddf,Mafra:2019xms,Broedel:2020tmd}}  \, ,
	\end{cases}
\end{align}
and that 
\begin{align}
	\partial_{z_k} \hat{\KN}^\tau_{12\ldots n}&=\begin{cases}
		-\sum_{j \neq k} s_{kj} (g^{(1)}_{kj})\hat{\KN}^\tau_{12\ldots n}   \, ,&  \textrm{this \, work} \, , \\
			-\sum_{j \neq k} s_{kj} (f^{(1)}_{kj})\hat{\KN}^\tau_{12\ldots n} \, ,&  \textrm{\rcites{Mafra:2019ddf,Mafra:2019xms,Broedel:2020tmd}}  \, .
	\end{cases}
\end{align}
However, an important key identity, the mixed-heat  equation \eqref{eqn:KroneckerEisensteinMixedHeat} is not straightforwardly valid for the holomorphic $z-$derivative of $\Omega(z,\eta,\tau)$. Because of this, the differential equations in \rcite{Broedel:2020tmd} are taken when the extra, unintegrated puncture $z_0$ is purely real. A thorough discussion of the ways in which the mixed-heat equation can be written for the non-meromorphic $\Omega(z,\eta,\tau)$ can be found around eq. (2.9) of \rcite{Gerken:2019cxz}. 

Lastly, because the Fay relations and derivatives of the integrals in \eqn{eqn:prototypicalIntegral} can be made to work the same, there is one important corollary: The combinatorics used in this work to find the differential KZB system (in  \appref{app:DerivationOfDifferentialSystem}) will translate precisely to valid combinatorics  in the doubly-periodic setting of \rcites{Mafra:2019ddf,Mafra:2019xms,Broedel:2020tmd}: In the differential equation, simply replace all $g^{(k)}_{ij}$ by a $f^{(k)}_{ij}$, and take all the unintegrated punctures to be on the real axis. Because of this, we note that if one can \textit{prove} in this meromorphic setting that the basis of integrands is indeed $\frac{(n-1)!}{(n-p-1)!}$, then this would also prove this statement in the setting of \rcites{Mafra:2019ddf,Mafra:2019xms,Broedel:2020tmd}.

\section{Deriving the differential system} \label{app:DerivationOfDifferentialSystem}
In this appendix, explicit derivation of various results from \subsecref{subsec:diffSystem} are given.
\subsection{The $z_i$-derivatives for general (n,p)}\label{app:ziDeriv}
The derivation of the $z_i$-derivative in the differential system \eqref{eqn:ellKZBeqzi,eqn:ellKZBeqtau} is similar to the calculation in section 4 of \cite{Broedel:2020tmd}.

The Koba--Nielsen factor ensures that the partial derivative of the integration boundaries vanishes. Using the antisymmetry of $F_{ij}(\eta)=F(z_i-z_j,\eta,\tau)$ :
\begin{align}
\partial_i F_{ij}(\eta)&=-\partial_j F_{ij}(\eta)
\end{align}
and integration by parts,
the derivative of the integrand in \eqn{def:Znp} can be written as a sum of partial derivatives solely acting on the Koba--Nielsen factor
\begin{align}
&\partial_i Z^\tau_{n,p}((1,A^1),(p+2,A^{p+2}),(p+3,A^{p+3}),\dots,(n,A^n); z_{p+2},z_{p+3},\dots,z_n)\nonumber\\
&=\int_{0<z_i<z_{i+1}<z_{p+2}}\prod_{i=2}^{p+1}\dd z_i\, \partial_i\left(\KN^{\tau}_{12\dots n}\,\tilde{\varphi}^\tau(1,A^1)\prod_{k=p+2}^n\tilde{\varphi}^\tau(k,A^k)\right)\nonumber\\
&=\int_{0<z_i<z_{i+1}<z_{p+2}}\prod_{i=2}^{p+1}\dd z_i\, \left((\partial_i+\sum_{k\in A^i}\partial_k)\KN^{\tau}_{12\dots n}\right)\,\tilde{\varphi}^\tau(1,A^1)\prod_{k=p+2}^n\tilde{\varphi}^\tau(k,A^k)\nonumber\\
&=\int_{0<z_i<z_{i+1}<z_{p+2}}\prod_{i=2}^{p+1}\dd z_i\,\KN^{\tau}_{12\dots n} \left(-\sum_{\begin{smallmatrix}
	k\in \{i,A^i\}\\
	j\not \in \{i,A^i\}
	\end{smallmatrix}}s_{kj}g^{(1)}_{kj}\right)\,\tilde{\varphi}^\tau(1,A^1)\prod_{k=p+2}^n\tilde{\varphi}^\tau(k,A^k)\nonumber\\
&=-\sum_{\begin{smallmatrix}
	k\in \{i,A^i\}\\
	j\not \in \{i,A^i\}
	\end{smallmatrix}}s_{kj}\int_{0<z_i<z_{i+1}<z_{p+2}}\prod_{i=2}^{p+1}\dd z_i\,\KN^{\tau}_{12\dots n} \,F_{kj}(\xi)\tilde{\varphi}^\tau(1,A^1)\prod_{k=p+2}^n\tilde{\varphi}^\tau(k,A^k)|_{\xi^0}\nonumber\\
&=-\sum_{\begin{smallmatrix}
	r\in \{1,p+2,\dots,n\}\\r\neq i
	\end{smallmatrix}}\Bigg(s_{i r}g^{(1)}_{ir}Z^\tau_{n,p}((1,A^1),(p+2,A^{p+2}),(p+3,A^{p+3}),\dots,(n,A^n); z_{p+2},z_{p+3},\dots,z_n)\nonumber\\
&\phantom{=}+
\sum_{j=1}^{|A^r|}s_{i a^r_k}\int_{0<z_i<z_{i+1}<z_{p+2}}\prod_{i=2}^{p+1}\dd z_i\,\KN^{\tau}_{12\dots n} \,F_{i a^r_k}(\xi)\tilde{\varphi}^\tau(1,A^1)\prod_{k=p+2}^n\tilde{\varphi}^\tau(k,A^k)|_{\xi^0}
\nonumber\\
&\phantom{=}+
\sum_{k=1}^{|A^i|}s_{a^i_j r}\int_{0<z_i<z_{i+1}<z_{p+2}}\prod_{i=2}^{p+1}\dd z_i\,\KN^{\tau}_{12\dots n} \,F_{a^i_j r}(\xi)\tilde{\varphi}^\tau(1,A^1)\prod_{k=p+2}^n\tilde{\varphi}^\tau(k,A^k)|_{\xi^0}
\nonumber\\
&\phantom{=}+\sum_{k=1}^{|A^i|}\sum_{j=1}^{|A^r|}s_{a^i_j a^r_k}\int_{0<z_i<z_{i+1}<z_{p+2}}\prod_{i=2}^{p+1}\dd z_i\,\KN^{\tau}_{12\dots n} \,F_{a^i_j a^r_k}(\xi)\tilde{\varphi}^\tau(1,A^1)\prod_{k=p+2}^n\tilde{\varphi}^\tau(k,A^k)|_{\xi^0}\Bigg)\,,
\end{align}
where we denote the elements of a sequence $A^i$ by
\begin{align}
A^i&=(a^i_1,a^i_2,\dots,a^i_{|A^i|})\,.
\end{align}
The essential factors in the above sum over $r$ are 
\begin{equation}
s_{i a^r_k} F_{i a^r_k}(\xi)\tilde{\varphi}^\tau(r,A^r)\,,
\end{equation}
\begin{equation}
s_{a^i_j r}F_{a^i_j r}(\xi)\tilde{\varphi}^\tau(i,A^i)\,,
\end{equation}
and 
\begin{equation}
s_{a^i_j a^r_k}F_{a^i_k a^r_j}(\xi)\tilde{\varphi}^\tau(i,A^i)\tilde{\varphi}^\tau(r,A^r)\,.
\end{equation}
These are exactly the sums appearing in the derivatives of \rcite{Broedel:2020tmd} (with the notational identification $i=1, r=0$ and $A^i=B, A^r=A$, cf.\ equations (B.16), (B.18) and (B.14) of \rcite{Broedel:2020tmd}), where they have been rewritten in terms of linear combinations of products
\begin{align}
&\tilde{\varphi}^\tau(1,A^1)\prod_{k=p+2}^n\tilde{\varphi}^\tau(k,A^k)\,,
\end{align}
which appear in the definition \eqref{def:Znp} of the integrals $Z^\tau_{n,p}(A^1,A^{p+2},A^{p+3},\dots,A^n; z_{p+2},z_{p+3},\dots,z_n)$. Thus, we can simply copy the result eq. (4.21) of \rcite{Broedel:2020tmd} with the corresponding notational adaption to obtain the closed formula. Defining the decomposition of a sequence $C$ into subsequences $C_{ij}$ as
\begin{align}\label{eqn:subsequences}
C=(c_1,\dots,c_{i-1}, \underbrace{c_i,c_{i+1}\dots,c_{j-1}}_{C_{i,j}=C_{ij}},c_j,c_{j+1}\dots,c_m)\,,
\end{align}
with
\begin{align}
C_{ji}=\emptyset\,\text{ for }j\geq i\,, \ \ \ \ \ \
C_{1,m+1} =C\,,\ \ \ \ \ \ 
\tilde{C}_{ij}=(c_{j-1},c_{j-2},\dots, c_i)\,,
\end{align}
where a tilde denotes the reversal of a sequence and
for sequences $P=(p_1,p_2,\dots,p_l)$, $Q=(q_1,q_2,\dots,q_m)$ and the following sum of Mandelstam variables
\begin{equation}
s_{P,Q} = \sum_{i=1}^l \sum_{j=1}^m s_{p_i q_j}\,,
\label{shortmand}
\end{equation}
the closed formula can be written as
\small
\begin{align}\label{eqn:derZnpClosed}
&\partial_i Z^\tau_{n,p}((1,A^1),(p+2,A^{p+2}),(p+3,A^{p+3}),\dots,(n,A^n); z_{p+2},z_{p+3},\dots,z_n)\nonumber\\
&= \sum_{\begin{smallmatrix}
	r\in \{1,p+2,\dots,n\}\\r\neq i
	\end{smallmatrix}}\Bigg \{\left(s_{(i,A^i),(r,A^r)}g^{(1)}_{ri}+\sum_{j=1}^{|A^i|}s_{a^i_j,(r,A^r)}\partial_{\eta_{a^i_j}}-\sum_{k=1}^{|A^r|} s_{(i,A^i),a^r_k}\partial_{\eta_{a^r_k}}\right)Z^\tau_{n,p}(\dots)\nonumber\\
&\phantom{=}-\sum_{k=1}^{|A^r|}\sum_{j=1}^{|A^i|} s_{a^i_j,a^r_k}\sum_{m=1 }^{k}\sum_{l=1 }^{j}(-1)^{k+j-m-l}F_{ri}(-\eta_{A^i_{l,|A^i|+1}})\nonumber\\
&\phantom{=\sum_{m=1 }^{k}\sum_{l=1 }^{j}}Z^\tau_{n,p}\left(\dots,\left(r,A^r_{1m}\shuffle(a^r_k,(\tilde{A^r}_{m,k}\shuffle A^r_{k+1,|A^r|+1})\shuffle (a^i_j,\tilde{A^i}_{l,j}\shuffle A^i_{j+1,|A^i|+1}))\right),\dots,\left(i,A^i_{1l}\right),\dots\right)\nonumber\\
&\phantom{=}-\sum_{k=1}^{|A^r|}\sum_{j=1}^{|A^i|} s_{a^i_j,a^r_k}\sum_{m=1 }^{k}\sum_{l=1 }^{j}(-1)^{k+j-m-l}F_{ri}(\eta_{A^r_{m,|A^r|+1}})\nonumber\\
&\phantom{=\sum_{m=1 }^{k}\sum_{l=1 }^{j}}Z^\tau_{n,p}\left(\dots,\left(r,A^r_{1m}\right),\dots,\left(i,A^i_{1l}\shuffle(a^i_j,(\tilde{A^i}_{l,j}\shuffle A^i_{j+1,|A^i|+1})\shuffle (a^r_k,\tilde{A^r}_{m,k}\shuffle A^r_{k+1,|A^r|+1}))\right),\dots\right)\nonumber\\
&\phantom{=}-\sum_{j=1}^{|A^i|} s_{a^i_j,r}\sum_{l=1 }^{j}(-1)^{j-l}F_{ri}(-\eta_{A^i_{l,|A^i|+1}})Z^\tau_{n,p}\left(\dots,\left(r,A^r\shuffle (a^i_j,\tilde{A^i}_{l,j}\shuffle A^i_{j+1,|A^i|+1})\right),\dots,\left(i,A^i_{1l}\right)\dots\right)\nonumber\\
&\phantom{=}-\sum_{k=1}^{|A^r|} s_{i,a^r_k}\sum_{m=1 }^{k}(-1)^{k-m}F_{ri}(\eta_{A^r_{m,|A^r|+1}})Z^\tau_{n,p}\left(\dots,\left(r,A^r_{1m}\right),\dots, \left(i,A^i\shuffle (a^r_k,\tilde{A^r}_{m,k}\shuffle A^r_{k+1,|A^r|+1})\right)\dots\right)\Bigg \}\,,
\end{align}
\normalsize
where any empty slots in the argument of $Z^\tau_{n,p}$ denoted by $\dots$ are understood to be exactly the same as for the initial integral {\small $Z^\tau_{n,p}((1,A^1),(p+2,A^{p+2}),\dots,(n,A^n); z_{p+2},\dots,z_n)$} which is being differentiated.

Upon expanding the Eisenstein--Kronecker series $F_{ri}$ in \eqn{eqn:derZnpClosed}, one obtains the coefficients of the matrices $\boldsymbol{x}^{(0)}_i$ and $\boldsymbol{x}^{(k)}_{ir}$ in the partial differential equation
\begin{align}
\partial_i \boldsymbol{Z}^\tau_{n,p}&=\left(\boldsymbol{x}_i^{(0)}+\sum_{k \geq 1 }\sum_{\begin{smallmatrix}
	r\in \{1,p+2,\dots,n\}\\r\neq i
	\end{smallmatrix}}\boldsymbol{x}^{(k)}_{ir}g^{(k)}_{ir}\right)\boldsymbol{Z}^\tau_{n,p}\,.
\end{align}

\subsection{The $\tau$-derivative for general (n,p)}\label{app:tauDeriv}
The integrals $Z^\tau_{n,p}$ also depend on the modular parameter $\tau$. The partial derivative of the vector $\boldsymbol{Z}^\tau_{n,p}$ with respect to $\tau$ can also be expressed in a closed form, given in the differential system \eqref{eqn:ellKZBeqzi,eqn:ellKZBeqtau}. It is derived in this subsection and follows the calculation in section 4 of \cite{Broedel:2020tmd}.

First, the derivative of the Koba--Nielsen factor
\begin{align}
2 \pi i \pd_\tau \KN^\tau_{12\ldots n}&= -\sum\limits_{1\leq i<j\leq n}s_{ij}(g_{ij}^{(2)}+2\zeta_2)\KN^\tau_{12\ldots n}
\end{align}
and the heat equation
\begin{align}
2\pi i \partial_{\tau}F_{ij}(\eta)&=\partial_i \partial_{\eta}F_{ij}(\eta)=-\partial_j \partial_{\eta}F_{ij}(\eta)
\end{align}
are required. Denoting $A^k=(a^k_1,\dots,a^k_{|A^k|})$ and $k=a_0^k$, we find
{\small \begin{align}
	&2\pi i \partial_{\tau} Z^\tau_{n,p}((1,A^1),(p+2,A^{p+2}),(p+3,A^{p+3}),\dots,(n,A^n); z_{p+2},z_{p+3},\dots,z_n)\nonumber\\
	&=\int_{0<z_i<z_{i+1}<z_{p+2}}\prod_{i=2}^{p+1}\dd z_i\, 2\pi i \partial_{\tau}\left(\KN^{\tau}_{12\dots n}\,\tilde{\varphi}^\tau(1,A^1)\prod_{k=p+2}^n\tilde{\varphi}^\tau(k,A^k)\right)\nonumber\\
	\nonumber\\
	&=\int_{0<z_i<z_{i+1}<z_{p+2}}\prod_{i=2}^{p+1}\dd z_i\,  \bigg(\left(2\pi i\partial_{\tau}\KN^{\tau}_{12\dots n}\right)\tilde{\varphi}^\tau(1,A^1)\prod_{k=p+2}^n\tilde{\varphi}^\tau(k,A^k)\nonumber\\
	&\ph{=}+\KN^{\tau}_{12\dots n}\sum_{r\in \{1,p+2,\dots,n\}}\sum_{i=1}^{|A^r|}\prod_{\begin{smallmatrix}
		j=1\\
		j\neq i
		\end{smallmatrix}}^{|A^r|}F_{a^r_{j-1}a^r_{j}}(\eta_{a^r_{j}\dots a^r_{|A^r|}})(2\pi i \partial_\tau F_{a^r_{i-1}a^r_{i}}(\eta_{a^r_{i}\dots a^r_{|A^r|}})) \!\!\! \prod_{\begin{smallmatrix}
		k\in \{1,p+2,\dots,n\}\\
		k\neq r
		\end{smallmatrix}} \!\!\!\!\!\!  \tilde{\varphi}^\tau(k,A^k)\bigg)\nnl
	&=\int_{0<z_i<z_{i+1}<z_{p+2}}\prod_{i=2}^{p+1}\dd z_i\,  \bigg(\left(2\pi i\partial_{\tau}\KN^{\tau}_{12\dots n}\right)\tilde{\varphi}^\tau(1,A^1)\prod_{k=p+2}^n\tilde{\varphi}^\tau(k,A^k)\nonumber\\
	&\ph{=}-\KN^{\tau}_{12\dots n} \!\!\! \sum_{r\in \{1,p+2,\dots,n\}}\sum_{i=1}^{|A^r|} \prod_{\begin{smallmatrix}
		j=1\\
		j\neq i
		\end{smallmatrix}}^{|A^r|}F_{a^r_{j-1}a^r_{j}}(\eta_{a^r_{j}\dots a^r_{|A^r|}})(\partial_{a^r_{i}}\partial_{\eta_{a^r_{i}\dots a^r_{|A^r|}}} F_{a^r_{i-1}a^r_{i}}(\eta_{a^r_{i}\dots a^r_{|A^r|}})) \!\!\!\!\!   \prod_{\begin{smallmatrix}
		k\in \{1,p+2,\dots,n\}\\
		k\neq r
		\end{smallmatrix}} \!\!\!\!\!\!  \tilde{\varphi}^\tau(k,A^k)\bigg)\nnl
	&=\int_{0<z_i<z_{i+1}<z_{p+2}}\prod_{i=2}^{p+1}\dd z_i\,  \bigg(\left(2\pi i\partial_{\tau}\KN^{\tau}_{12\dots n}\right)\tilde{\varphi}^\tau(1,A^1)\prod_{k=p+2}^n\tilde{\varphi}^\tau(k,A^k)\nonumber\\
	&\ph{=}+\sum_{r\in \{1,p+2,\dots,n\}}\sum_{i=1}^{|A^r|}\left(\sum_{k=i}^{|A^r|}\partial_{a^r_{i}}\KN^{\tau}_{12\dots n}\right)(\partial_{\eta_{a^r_{i}}}-\theta_{i\geq 2}\partial_{\eta_{a^r_{i-1}}})\prod_{
		k\in \{1,p+2,\dots,n\}
	}\tilde{\varphi}^\tau(k,A^k)\bigg)\nnl
	&=-\sum\limits_{1\leq i<j\leq n}s_{ij}\int_{0<z_i<z_{i+1}<z_{p+2}}\prod_{i=2}^{p+1}\dd z_i\, \KN^\tau_{12\ldots n}\,(g_{ij}^{(2)}+2\zeta_2)\tilde{\varphi}^\tau(1,A^1)\prod_{k=p+2}^n\tilde{\varphi}^\tau(k,A^k)\nonumber\\
	&\ph{=}-\sum_{r\in \{1,p+2,\dots,n\}}\int_{0<z_i<z_{i+1}<z_{p+2}}\prod_{i=2}^{p+1}\dd z_i\,\KN^{\tau}_{12\dots n}\nnl
	&\ph{=}\sum_{i=1}^{|A^r|}\left(\sum_{k=i}^{|A^r|}\sum_{j=0}^{i-1}s_{a^r_ka^r_j}g^{(1)}_{a^r_ka^r_j}+\sum_{k=i}^{|A^r|}\sum_{\begin{smallmatrix}
		q\in \{1,p+2,\dots,n\}\\
		q\neq r
		\end{smallmatrix}}\sum_{j=0}^{|A^q|}s_{a^r_ka^q_j}g^{(1)}_{a^r_ka^q_j}\right)(\partial_{\eta_{a^r_{i}}}-\theta_{i\geq 2}\partial_{\eta_{a^r_{i-1}}}) \!\!\! \prod_{
		k\in \{1,p+2,\dots,n\}
	}\!\!\!\!\!\!  \tilde{\varphi}^\tau(k,A^k)\nnl
	&=-\sum\limits_{1\leq i<j\leq n}s_{ij}\int_{0<z_i<z_{i+1}<z_{p+2}}\prod_{i=2}^{p+1}\dd z_i\, \KN^\tau_{12\ldots n}\,(g_{ij}^{(2)}+2\zeta_2)\tilde{\varphi}^\tau(1,A^1)\prod_{k=p+2}^n\tilde{\varphi}^\tau(k,A^k)\nonumber\\
	&\ph{=}-\sum_{r\in \{1,p+2,\dots,n\}}\sum_{k=1}^{|A^r|}\sum_{j=0}^{k-1}s_{a^r_ka^r_j}\int_{0<z_i<z_{i+1}<z_{p+2}}\prod_{i=2}^{p+1}\dd z_i\,\KN^{\tau}_{12\dots n}\,g^{(1)}_{a^r_ka^r_j}(\partial_{\eta_{a^r_{k}}}-\theta_{j\geq 1}\partial_{\eta_{a^r_{j}}})\!\!\!\! \prod_{
		k\in \{1,p+2,\dots,n\}
	} \!\!\!\!\!\!\!\! \tilde{\varphi}^\tau(k,A^k)\nnl
	&\ph{=}-\sum_{r\in \{1,p+2,\dots,n\}}\sum_{\begin{smallmatrix}
		q\in \{1,p+2,\dots,n\}\\
		q\neq r
		\end{smallmatrix}}\sum_{k=1}^{|A^r|}\sum_{j=0}^{|A^q|}s_{a^r_ka^q_j}\int_{0<z_i<z_{i+1}<z_{p+2}}\prod_{i=2}^{p+1}\dd z_i\,\KN^{\tau}_{12\dots n}\,g^{(1)}_{a^r_ka^q_j}\partial_{\eta_{a^r_{k}}}\!\!\!\! \prod_{
		k\in \{1,p+2,\dots,n\}
	}\!\!\!\!\!\!\!\!\!  \tilde{\varphi}^\tau(k,A^k)\nnl
	&=-\sum\limits_{1\leq i<j\leq n}s_{ij}\int_{0<z_i<z_{i+1}<z_{p+2}}\prod_{i=2}^{p+1}\dd z_i\, \KN^\tau_{12\ldots n}\,(g_{ij}^{(2)}+2\zeta_2)\tilde{\varphi}^\tau(1,A^1)\prod_{k=p+2}^n\tilde{\varphi}^\tau(k,A^k)\nonumber\\
	&\ph{=}-\sum_{r\in \{1,p+2,\dots,n\}}\sum_{k=1}^{|A^r|}\sum_{j=0}^{k-1}s_{a^r_ka^r_j}\int_{0<z_i<z_{i+1}<z_{p+2}}\prod_{i=2}^{p+1}\dd z_i\,\KN^{\tau}_{12\dots n}\,g^{(1)}_{a^r_ka^r_j}(\partial_{\eta_{a^r_{k}}}-\theta_{j\geq 1}\partial_{\eta_{a^r_{j}}})\!\!\! \prod_{
		k\in \{1,p+2,\dots,n\}
	}\!\!\!\!\!\!\!\!\! \tilde{\varphi}^\tau(k,A^k)\nnl
	&\ph{=}- \!\!\!\!\!\!\!\!\!  \sum_{\begin{smallmatrix}
		r,q\in \{1,p+2,\dots,n\}\\
		q< r
		\end{smallmatrix}}\sum_{k=0}^{|A^r|}\sum_{j=0}^{|A^q|}s_{a^r_ka^q_j}\int_{0<z_i<z_{i+1}<z_{p+2}}\prod_{i=2}^{p+1}\dd z_i\,\KN^{\tau}_{12\dots n}\,g^{(1)}_{a^r_ka^q_j}(\theta_{k\geq 1}\partial_{\eta_{a^r_{k}}}-\theta_{j\geq 1}\partial_{\eta_{a^q_{j}}})\!\!\! \prod_{
		k\in \{1,p+2,\dots,n\}
	}\!\!\!\!\!\!\!\!\! \tilde{\varphi}^\tau(k,A^k)\nnl
	&=-s_{12\dots n}2\zeta_2\, Z^\tau_{n,p}((1,A^1),(p+2,A^{p+2}),(p+3,A^{p+3}),\dots,(n,A^n); z_{p+2},z_{p+3},\dots,z_n)\nonumber\\
	&\ph{=}-\sum_{r\in \{1,p+2,\dots,n\}}\sum_{k=1}^{|A^r|}\sum_{j=0}^{k-1}s_{a^r_ka^r_j}\int_{0<z_i<z_{i+1}<z_{p+2}}\prod_{i=2}^{p+1}\dd z_i\,\KN^{\tau}_{12\dots n}\nonumber\\
	&\ph{=}\left(g^{(1)}_{a^r_ka^r_j}(\partial_{\eta_{a^r_{k}}}-\theta_{j\geq 1}\partial_{\eta_{a^r_{j}}})+g^{(2)}_{a^r_ka^r_j}\right)\prod_{
		k\in \{1,p+2,\dots,n\}
	}\tilde{\varphi}^\tau(k,A^k)\nnl
	&\ph{=}-\sum_{\begin{smallmatrix}
		r,q\in \{1,p+2,\dots,n\}\\
		q< r
		\end{smallmatrix}}\sum_{k=0}^{|A^r|}\sum_{j=0}^{|A^q|}s_{a^r_ka^q_j}\int_{0<z_i<z_{i+1}<z_{p+2}}\prod_{i=2}^{p+1}\dd z_i\,\KN^{\tau}_{12\dots n}\nnl
	&\ph{=}\left(g^{(1)}_{a^r_ka^q_j}(\theta_{k\geq 1}\partial_{\eta_{a^r_{k}}}-\theta_{j\geq 1}\partial_{\eta_{a^q_{j}}})+g^{(2)}_{a^r_ka^q_j}\right)\prod_{
		k\in \{1,p+2,\dots,n\}
	}\tilde{\varphi}^\tau(k,A^k)
	\,.
	\end{align}}
 In \rcite{Broedel:2020tmd}, the following identities are proven
\begin{align}
&\sum_{k=1}^{|A^r|}\sum_{j=0}^{k-1}s_{a^r_ka^r_j}\left(g^{(1)}_{a^r_ka^r_j}(\partial_{\eta_{a^r_{k}}}-\theta_{j\geq 1}\partial_{\eta_{a^r_{j}}})+g^{(2)}_{a^r_ka^r_j}\right)\prod_{
	k\in \{1,p+2,\dots,n\}
}\tilde{\varphi}^\tau(k,A^k)\nnl
&=\sum_{k=1}^{|A^r|}\sum_{j=0}^{k-1}s_{a^r_ka^r_j}\left(\partial_{\eta_{a^r_{k}}}-\theta_{j\geq 1}\partial_{\eta_{a^r_{j}}}+\partial_{\xi}\right)F_{a^r_ka^r_j}(\xi)\prod_{
	k\in \{1,p+2,\dots,n\}
}\tilde{\varphi}^\tau(k,A^k)|_{\xi^0}\nnl
&=-\frac{1}{2}\sum_{k=1}^{|A^r|}\sum_{j=0}^{k-1}s_{a^r_ka^r_j} \left(\partial_{\eta_{a^r_k}}-  \theta_{j\geq 1} 
\partial_{\eta_{a^r_j}}\right)^2\prod_{
	k\in \{1,p+2,\dots,n\}
}\tilde{\varphi}^\tau(k,A^k)\nonumber\\
&\phantom{=}+\sum_{k=1}^{|A^r|}\sum_{j=0}^{k-1}s_{a^r_ka^r_j}\sum_{l=j+1}^k\wp(\eta_{A^r_{l,|A^r|+1}})(-1)^{k-l}\prod_{\begin{smallmatrix}
	k\in \{1,p+2,\dots,n\}\\
	k\neq r
	\end{smallmatrix}
}\tilde{\varphi}^\tau(k,A^k)\nnl
&\ph{=\sum_{k=1}^{|A^r|}}\phiChain(r,A^r_{1,j},a^r_j, A^r_{j+1,l}\shuffle (a_k,\tilde A^r_{l,k}\shuffle A^r_{k+1,|A^r|+1}))
\end{align}
and
\begin{align}
&\sum_{k=0}^{|A^r|}\sum_{j=0}^{|A^q|}s_{a^r_ka^q_j}\left(g^{(1)}_{a^r_ka^q_j}(\theta_{k\geq 1}\partial_{\eta_{a^r_{k}}}-\theta_{j\geq 1}\partial_{\eta_{a^q_{j}}})+g^{(2)}_{a^r_ka^q_j}\right)\prod_{
	k\in \{1,p+2,\dots,n\}
}\tilde{\varphi}^\tau(k,A^k)\nnl
&=\sum_{k=0}^{|A^r|}\sum_{j=0}^{|A^q|}s_{a^r_ka^q_j}\left(\theta_{k\geq 1}\partial_{\eta_{a^r_{k}}}-\theta_{j\geq 1}\partial_{\eta_{a^q_{j}}}+\partial_{\xi}\right)F_{a^r_ka^q_j}(\xi)\prod_{
	k\in \{1,p+2,\dots,n\}
}\tilde{\varphi}^\tau(k,A^k)|_{\xi^0}\nnl
&=\left(s_{(r,A^r),(q,A^q)}g^{(2)}_{a^r_ka^q_j}-\frac{1}{2}\sum_{k=0}^{|A^r|}\sum_{j=0}^{|A^q|}s_{a^r_ka^q_j}\left(\theta_{k\geq 1}\partial_{\eta_{a^r_{k}}}-\theta_{j\geq 1}\partial_{\eta_{a^q_{j}}}\right)^2\right)\prod_{
	k\in \{1,p+2,\dots,n\}
}\tilde{\varphi}^\tau(k,A^k)\nonumber\\
&\phantom{=}-\sum_{k=0}^{|A^r|}\sum_{j=0}^{|A^q|}s_{a^r_ka^q_j} \sum_{i=1 }^{k}\sum_{l=1 }^{j}(-1)^{k+j-i-l}F_{qr}^+(\eta_{A^q_{l,|A^q|+1}})\prod_{\begin{smallmatrix}
	k\in \{1,p+2,\dots,n\}\\
	k\neq r,q
	\end{smallmatrix}
}\tilde{\varphi}^\tau(k,A^k)\nonumber\\
&\phantom{=\sum_{i=1 }^{k}\sum_{l=1 }^{j}} \phiChain(q,A^q_{1l}) \phiChain(r,A^r_{1i}\shuffle(a^r_k,(\tilde{A}^r_{i,k}\shuffle A^r_{k+1,|A^r|+1})\shuffle (a^q_j,\tilde{A}^q_{l,j}\shuffle A^q_{j+1,|A^q|+1})))\nonumber\\
&\phantom{=}-\sum_{k=0}^{|A^r|}\sum_{j=0}^{|A^q|}s_{a^r_ka^q_j} \sum_{i=1 }^{k}\sum_{l=1 }^{j}(-1)^{k+j-i-l}F^-_{qr}(-\eta_{A^r_{i,|A^r|+1}})\prod_{\begin{smallmatrix}
	k\in \{1,p+2,\dots,n\}\\
	k\neq r,q
	\end{smallmatrix}
}\tilde{\varphi}^\tau(k,A^k)\nonumber\\
&\phantom{=\sum_{i=1 }^{k}\sum_{l=1 }^{j}}  \phiChain(r,A^r_{1i}) \phiChain(q,A^q_{1l}\shuffle(a^q_j,(\tilde{A}^q_{l,j}\shuffle A^q_{j+1,|A^q|+1})\shuffle (a^r_k,\tilde{A}^r_{i,k}\shuffle A^r_{k+1,|A^r|+1})))\,,
\end{align}
where 
\begin{align}
F_{ij}^{\pm}(\pm\xi)&=\pm \partial_{\xi}F_{ij}(\pm\xi)\nnl
&=\pm \partial_{\xi}\sum_{k\geq 0}g^{(k)}_{ij}(\pm\xi)^{k-1}
\nnl
&=\sum_{k\geq 0}(k-1) g^{(k)}_{ij}(\pm\xi)^{k-2}\,.
\end{align}
This leads to the closed formula
{\small \begin{align}\label{eqn:tauDerivClosed}
	& 2\pi i \partial_{\tau} Z^\tau_{n,p}((1,A^1),(p+2,A^{p+2}),(p+3,A^{p+3}),\dots,(n,A^n); z_{p+2},z_{p+3},\dots,z_n)\nonumber\\
	&= \bigg(\frac{1}{2}\sum_{j=2}^{p+1}(s_{(1,p+2,\dots,n),j}) \partial^2_{\eta_{j}} +\frac{1}{2}\sum_{2\leq i<j\leq p+1}s_{ij} (\partial_{\eta_{i}}{-}\partial_{\eta_{j}})^2- 2\zeta_2s_{12\dots n}- \!\!\!\!\!\!\!\!\!\!\!\!  \sum_{\begin{smallmatrix}
		r,q\in \{1,p+2,\dots,n\}\\
		q< r
		\end{smallmatrix}}s_{(k,A^k),(q,A^q)}g^{(2)}_{kq}\bigg)Z^\tau_{n,p}(\dots)
	\nonumber\\
	&\phantom{=}- \!\!\! \! \! \! \!\!\!\!\!\! \!\!  \sum_{r\in \{1,p+2,\dots,n\}}\sum_{k=1}^{|A^r|}\sum_{j=0}^{k-1} s_{a^r_k,a^r_j} \!\!\! \sum_{l=j+1}^k\wp(\eta_{A^r_{l,|A^r|+1}})(-1)^{k-l}Z^{\tau}_{n,p}\left(\dots,\left(r,A^r_{1,j},a^r_j, A^r_{j,l}\shuffle (a^r_k,\tilde A^r_{l,k}\shuffle A^r_{k+1,|A^r|+1})\right),\dots\right)\nonumber\\
	&\phantom{=}+\sum_{\begin{smallmatrix}
		r,q\in \{1,p+2,\dots,n\}\\
		q< r
		\end{smallmatrix}}\sum_{k=1}^{|A^r|}\sum_{j=1}^{|A^q|} s_{a^r_k,a^q_j}\sum_{i=1 }^{k}\sum_{l=1 }^{j}(-1)^{k+j-i-l}F^+_{qr}(\eta_{A^q_{l,|A^q|+1}})  \nonumber \\
	&\phantom{=\sum_{i=1 }^{k}\sum_{l=1 }^{j}}Z^\tau_{n,p}\left(\dots,\left(q,A^q_{1l}\right),\dots,\left(r,A^r_{1i}\shuffle(a^r_k,(\tilde{A}^r_{i,k}\shuffle A^r_{k+1,|A^r|+1})\shuffle (a^q_j,\tilde{A}^q_{l,j}\shuffle A^q_{j+1,|A^q|+1}))\right),\dots\right)\nonumber\\
	&\phantom{=}+\sum_{\begin{smallmatrix}
		r,q\in \{1,p+2,\dots,n\}\\
		q< r
		\end{smallmatrix}}\sum_{k=1}^{|A^r|}\sum_{j=1}^{|A^q|} s_{a^r_k,a^q_j}\sum_{i=1 }^{k}\sum_{l=1 }^{j}(-1)^{k+j-i-l}F^-_{qr}(-\eta_{A^r_{i,|A^r|+1}})\nonumber\\
	&\phantom{=\sum_{i=1 }^{k}\sum_{l=1 }^{j}}Z^\tau_{n,p}\left(\dots,\left(q,A^q_{1l}\shuffle(a^q_j,(\tilde{A}^q_{l,j}\shuffle A^q_{j+1,|A^q|+1})\shuffle (a^r_k,\tilde{A}^r_{i,k}\shuffle A^r_{k+1,|A^r|+1}))\right),\dots,\left(r,A^r_{1i}\right),\dots\right)\nonumber\\
	&\phantom{=}+\sum_{\begin{smallmatrix}
		r,q\in \{1,p+2,\dots,n\}\\
		q< r
		\end{smallmatrix}}\sum_{j=1}^{|A^q|} s_{r,a^q_j}\sum_{l=1 }^{j}(-1)^{j-l}F^+_{qr}(\eta_{A^q_{l,|A^q|+1}})\nonumber\\
	&\phantom{=\sum_{i=1 }^{k}\sum_{l=1 }^{j}}Z^\tau_{n,p}\left(\dots,\left(q,A^q_{1l}\right),\dots,\left(r,A^r\shuffle (a^q_j,\tilde{A}^q_{l,j}\shuffle A^q_{j+1,|A^q|+1})\right),\dots\right)\nonumber\\
	&\phantom{=}+\sum_{\begin{smallmatrix}
		r,q\in \{1,p+2,\dots,n\}\\
		q< r
		\end{smallmatrix}}\sum_{k=1}^{|A^r|} s_{a^r_k,q}\sum_{i=1 }^{k}(-1)^{k-i}F^-_{qr}(-\eta_{A^r_i,|A^r|+1})\nonumber\\
	&\phantom{=\sum_{i=1 }^{k}\sum_{l=1 }^{j}}Z^\tau_{n,p}\left(\dots,, \left(q,A^q\shuffle (a^r_k,\tilde{A}^r_{i,k}\shuffle A^r_{k+1,|A^r|+1})\right),\dots,\left(r,A^r_{1i}\right),\dots\right)\,,
	\end{align}}
where all arguments of the integrals $Z^\tau_{n,p}$ on the right-hand side which are only implicitly written by the dots, are the same as the ones on the left-hand side. Expanding the Weierstrass $\wp$-function in terms of the Eisenstein series $G_k=G_k(\tau)$ from \eqn{eqn:GkDef}\footnote{The Eisenstein series $G_2(\tau)$ is not absolutely convergent, and requires a summation prescription. This is given by $G_2(\tau)=\sum_{n\in \mathbb{Z}\backslash \{0\}}\frac{1}{n^2}+\sum_{m\in \mathbb{Z}\backslash \{0\} }\sum_{n\in\mathbb{Z}} \frac{1}{(n+m \tau )^2}$ \cite{Mafra:2018pll} .}
\begin{align}
\label{eqn:EisensteinSeriesDefn}
\wp(\eta)&=\wp(\eta,\tau)=\frac{1}{\eta^2}+\sum_{k\geq 4}(k-1)G_k \eta^{k-2}
\end{align}
and comparing with the closed formula for the $z_i$-derivative \eqref{eqn:derZnpClosed}, the $\tau$-derivative of the vector $\boldsymbol{Z}^\tau_{n,p}$ is of the form 
\begin{align}
2\pi i\partial_{\tau} \boldsymbol{Z}^\tau_{n,p}&=\left(-\boldsymbol{\epsilon}^{(0)}+\sum_{k\geq 4}(1-k)G_k \boldsymbol{\epsilon}^{(k)}+\sum_{\begin{smallmatrix}
	r,q\in \{1,p+2,\dots,n\}\\
	q< r
	\end{smallmatrix}}\sum_{k\geq 2 }(k-1)\boldsymbol{x}_{qr}^{k-1}g^{(k)}_{qr}\right)\boldsymbol{Z}^\tau_{n,p}\,,
\end{align}
where the matrices $\boldsymbol{\epsilon}^{(k)}$ are homogeneous of degree one in the Mandelstam variables and homogeneous of degree $k-2$ in the $\eta$-variables.

\subsection{Commutation relations}\label{app:comRel}
In this subsection, the (commutation) relations presented in \subsecref{subsec:comRel} satisfied by the matrices in the differential system \eqref{eqn:ellKZBeqGeneral}, i.e.
\begin{align}\label{eqn:ellKZBeqGeneralApp}
\partial_i \boldsymbol{Z}^\tau_{n,p}&=\left(\boldsymbol{x}_{i}^{(0)}+\sum_{k\geq 1}\sum_{\begin{smallmatrix}
	r\in \{1,p+2,\dots,n\}\\r\neq i
	\end{smallmatrix}}\boldsymbol{x}_{ir}^{(k)}g^{(k)}_{ir}\right)\boldsymbol{Z}^\tau_{n,p}\,,\\
2\pi i\partial_{\tau} \boldsymbol{Z}^\tau_{n,p}&=\left(-\boldsymbol{\epsilon}^{(0)}+\sum_{k\geq 4}(1-k)G_k \boldsymbol{\epsilon}^{(k)}+\sum_{\begin{smallmatrix}
	r,q\in \{1,p+2,\dots,n\}\\
	q< r
	\end{smallmatrix}}\sum_{k\geq 2 }(k-1)\boldsymbol{b}_{qr}^{(k)}g^{(k)}_{qr}\right)\boldsymbol{Z}^\tau_{n,p}
\end{align}
for an arbitrary solution $\boldsymbol{Z}^\tau_{n,p}$, are derived.

\subsubsection{Commutators from $[ \partial_{z_i},\partial_{z_j}]=0$}
The Schwarz integrability condition leads to commutation relations among the matrices $\boldsymbol{x}^{(k)}_{ri}$: differentiating
\begin{align}
\partial_i \boldsymbol{Z}^\tau_{n,p}&=\left(\boldsymbol{x}^{(0)}_{i}+\sum_{k\geq 1 }\sum_{\begin{smallmatrix}
	r\in \{1,p+2,\dots,n\}\\r\neq i
	\end{smallmatrix}}\boldsymbol{x}^{(k)}_{ir}g^{(k)}_{ir}\right)\boldsymbol{Z}^\tau_{n,p}
\end{align}
with respect to $x_j$, where $j\in \{p+2,\dots, n\}\setminus \{i\}$, leads to
\begin{align}
\partial_j\partial_i \boldsymbol{Z}^\tau_{n,p}&=\boldsymbol{x}^{(0)}_{i}\partial_j\boldsymbol{Z}^\tau_{n,p}+\sum_{k\geq 1}\sum_{\begin{smallmatrix}
	r\in \{1,p+2,\dots,n\}\\r\neq i
	\end{smallmatrix}}\boldsymbol{x}^{(k)}_{ir}\partial_j\left(g^{(k)}_{ir}\boldsymbol{Z}^\tau_{n,p}\right)\nnl
&=\boldsymbol{x}^{(0)}_{i}\left(\boldsymbol{x}^{(0)}_{j}+\sum_{k\geq 1 }\sum_{\begin{smallmatrix}
	r\in \{1,p+2,\dots,n\}\\r\neq j
	\end{smallmatrix}}\boldsymbol{x}^{(k)}_{jr}g^{(k)}_{jr}\right)\boldsymbol{Z}^\tau_{n,p}\nnl
&\ph{=}+\sum_{k\geq 1 }\sum_{\begin{smallmatrix}
	r\in \{1,p+2,\dots,n\}\\r\neq i
	\end{smallmatrix}}\boldsymbol{x}^{(k)}_{ir}\left(\delta_{jr}(\partial_jg^{(k)}_{ij})\boldsymbol{Z}^\tau_{n,p}+g^{(k)}_{ir}\partial_j\boldsymbol{Z}^\tau_{n,p}\right)\nnl
&=\boldsymbol{x}^{(0)}_{i}\boldsymbol{x}^{(0)}_{j}\boldsymbol{Z}^\tau_{n,p}+\sum_{k\geq 1 }\sum_{\begin{smallmatrix}
	r\in \{1,p+2,\dots,n\}\\r\neq j
	\end{smallmatrix}}\boldsymbol{x}^{(0)}_{i}\boldsymbol{x}^{(k)}_{jr}g^{(k)}_{jr}\boldsymbol{Z}^\tau_{n,p}\nnl
&\ph{=}+\sum_{k\geq 1 }\boldsymbol{x}^{(k)}_{ij}\partial_jg^{(k)}_{ij}\boldsymbol{Z}^\tau_{n,p}+\sum_{k\geq 1 }\sum_{\begin{smallmatrix}
	r\in \{1,p+2,\dots,n\}\\r\neq i
	\end{smallmatrix}}\boldsymbol{x}^{(k)}_{ir}\boldsymbol{x}^{(0)}_{j}g^{(k)}_{ir}\boldsymbol{Z}^\tau_{n,p}\nnl
&\ph{=}+\sum_{k,l\geq 1 }\sum_{\begin{smallmatrix}
	r,s\in \{1,p+2,\dots,n\}\\r\neq i\\s\neq j
	\end{smallmatrix}}\boldsymbol{x}^{(k)}_{ir}\boldsymbol{x}^{(l)}_{js}g^{(k)}_{ir} g^{(l)}_{js}\boldsymbol{Z}^\tau_{n,p}
\,.
\end{align}
Decomposing the double sum over $r,s$ as follows
\begin{align}
\sum_{\begin{smallmatrix}
	r,s\in \{1,p+2,\dots,n\}\\r\neq i\\s\neq j
	\end{smallmatrix}}
&=\sum_{\begin{smallmatrix}
	r\in \{1,p+2,\dots,n\}\\r\neq i,j
	\end{smallmatrix}}\sum_{\begin{smallmatrix}
	s\in \{1,p+2,\dots,n\}\\s\neq i,j,r
	\end{smallmatrix}}+\sum_{\begin{smallmatrix}
	r\in \{1,p+2,\dots,n\}\\r\neq i,j\\
	s=r
	\end{smallmatrix}}\nnl
&\ph{=}+\sum_{\begin{smallmatrix}
	r\in \{1,p+2,\dots,n\}\\r\neq i,j\\
	s=i
	\end{smallmatrix}}+\sum_{\begin{smallmatrix}
	s\in \{1,p+2,\dots,n\}\\s\neq i,j\\
	r=j
	\end{smallmatrix}}+\sum_{\begin{smallmatrix}
	s=i\\
	r=j
	\end{smallmatrix}}\nnl
\end{align}
yields 
\begin{align}
\partial_j\partial_i \boldsymbol{Z}^\tau_{n,p}
&=\boldsymbol{x}^{(0)}_{i}\boldsymbol{x}^{(0)}_{j}\boldsymbol{Z}^\tau_{n,p}\nnl
&\ph{=}+\sum_{k\geq 1 }\sum_{\begin{smallmatrix}
	r\in \{1,p+2,\dots,n\}\\r\neq j
	\end{smallmatrix}}\boldsymbol{x}^{(0)}_{i}\boldsymbol{x}^{(k)}_{jr}g^{(k)}_{jr}\boldsymbol{Z}^\tau_{n,p}+\sum_{k\geq 1 }\sum_{\begin{smallmatrix}
	r\in \{1,p+2,\dots,n\}\\r\neq i
	\end{smallmatrix}}\boldsymbol{x}^{(k)}_{ir}\boldsymbol{x}^{(0)}_{j}g^{(k)}_{ir}\boldsymbol{Z}^\tau_{n,p}\nnl
&\ph{=}+\sum_{k\geq 1 }\boldsymbol{x}^{(k)}_{ij}\partial_jg^{(k)}_{ij}\boldsymbol{Z}^\tau_{n,p}+\sum_{k,l\geq 1 }\sum_{\begin{smallmatrix}
	r\in \{1,p+2,\dots,n\}\\r\neq i,j
	\end{smallmatrix}}\sum_{\begin{smallmatrix}
	s\in \{1,p+2,\dots,n\}\\s\neq i,j,r
	\end{smallmatrix}}\boldsymbol{x}^{(k)}_{ir}\boldsymbol{x}^{(l)}_{js}g^{(k)}_{ir} g^{(l)}_{js}\boldsymbol{Z}^\tau_{n,p}\nnl
&\ph{=}+\sum_{k,l\geq 1 }\sum_{\begin{smallmatrix}
	r\in \{1,p+2,\dots,n\}\\r\neq i,j
	\end{smallmatrix}}\left(\boldsymbol{x}^{(k)}_{ir}\boldsymbol{x}^{(l)}_{jr}g^{(k)}_{ir} g^{(l)}_{jr}+\boldsymbol{x}^{(k)}_{ir}\boldsymbol{x}^{(l)}_{ji}g^{(k)}_{ir} g^{(l)}_{ji}+\boldsymbol{x}^{(k)}_{ij}\boldsymbol{x}^{(l)}_{jr}g^{(k)}_{ij} g^{(l)}_{jr}\right)\boldsymbol{Z}^\tau_{n,p}
\nnl
&\ph{=}+\sum_{k,l\geq 1 }\boldsymbol{x}^{(k)}_{ij}\boldsymbol{x}^{(l)}_{ji}g^{(k)}_{ij} g^{(l)}_{ji}\boldsymbol{Z}^\tau_{n,p}
\,.
\end{align}
Therefore
\begin{align}
0&=[\partial_j,\partial_i]\boldsymbol{Z}^\tau_{n,p}\nnl
&=[\boldsymbol{x}^{(0)}_{i},\boldsymbol{x}^{(0)}_{j}]\boldsymbol{Z}^\tau_{n,p}+\sum_{k\geq 1 }g^{(k)}_{ji}\left([\boldsymbol{x}^{(0)}_{i},\boldsymbol{x}^{(k)}_{ji}]-(-1)^k
[\boldsymbol{x}^{(0)}_{j},\boldsymbol{x}^{(k)}_{ij}]\right)\boldsymbol{Z}^\tau_{n,p}\nnl
&\ph{=}+\sum_{k\geq 1 }\left(\boldsymbol{x}^{(k)}_{ij}+(-1)^k\boldsymbol{x}^{(k)}_{ji}\right)\partial_jg^{(k)}_{ij}\boldsymbol{Z}^\tau_{n,p}+\sum_{k,l\geq 1 }[\boldsymbol{x}^{(k)}_{ij},\boldsymbol{x}^{(l)}_{ji}]g_{ij}^{(k)}g_{ji}^{(l)}\boldsymbol{Z}^\tau_{n,p}\nnl
&\ph{=}+\sum_{k\geq 1 }\sum_{\begin{smallmatrix}
r\in \{1,p+2,\dots,n\}\\r\neq i,j
\end{smallmatrix}}g^{(k)}_{jr}[\boldsymbol{x}^{(0)}_{i},\boldsymbol{x}^{(k)}_{jr}]\boldsymbol{Z}^\tau_{n,p}+\sum_{k\geq 1 }g^{(k)}_{ir}\sum_{\begin{smallmatrix}
r\in \{1,p+2,\dots,n\}\\r\neq i,j
\end{smallmatrix}}[\boldsymbol{x}^{(k)}_{ir},\boldsymbol{x}^{(0)}_{j}]\boldsymbol{Z}^\tau_{n,p}\nnl
&\ph{=}+\sum_{k,l\geq 1 }\sum_{\begin{smallmatrix}
r,s\in \{1,p+2,\dots,n\}\\r,s\neq i,j\\r\neq s
\end{smallmatrix}}[\boldsymbol{x}^{(k)}_{ir},\boldsymbol{x}^{(l)}_{js}]g^{(k)}_{ir} g^{(l)}_{js}\boldsymbol{Z}^\tau_{n,p}\nnl
&\ph{=}+\sum_{k,l\geq 1 }\sum_{\begin{smallmatrix}
r\in \{1,p+2,\dots,n\}\\r\neq i,j
\end{smallmatrix}}\left([\boldsymbol{x}^{(k)}_{ir},\boldsymbol{x}^{(l)}_{jr}]g^{(k)}_{ir} g^{(l)}_{jr}+[\boldsymbol{x}^{(k)}_{ir},\boldsymbol{x}^{(l)}_{ji}]g^{(k)}_{ir} g^{(l)}_{ji}+[\boldsymbol{x}^{(k)}_{ij},\boldsymbol{x}^{(l)}_{jr}]g^{(k)}_{ij} g^{(l)}_{jr}\right)\boldsymbol{Z}^\tau_{n,p} \, ,
\end{align}

where we have used that $\partial_i g_{ji}^{(k)}=-\partial_j g_{ji}^{(k)}=-(-1)^k \partial_j g_{ij}^{(k)}$. Since for $k\geq 1$ the derivative $\partial_j g_{ij}^{(k)}$ has terms proportional to the Eisenstein series of weight two, which does not appear in $g_{ij}^{(k)}$, the above calculation implies 

\begin{align}
\boldsymbol{x}^{(k)}_{ji}&=(-1)^{k+1}\boldsymbol{x}^{(k)}_{ij}\,,\quad k\geq 1\,.
\end{align}

This last  parity relation implies that $\sum_{k,l > 0 }[\boldsymbol{x}^{(k)}_{ij},\boldsymbol{x}^{(l)}_{ji}]g_{ij}^{(k)}g_{ji}^{(l)}=0$, so there is no constraint on $[\boldsymbol{x}^{(k)}_{ij},\boldsymbol{x}^{(l)}_{ij}]$  for $k,l\geq 1$ from the above equation. This can be contrasted with \eqn{eqn:bonusRelationRepresentation}, which holds for the particular matrices found in this work.

Moreover for disjoint $i,j,r,s$ and $k,l\geq 1$, the products $g_{ir}^{(k)}g^{(l)}_{js}$ are independent, such that
\begin{align}\label{eqn:comRelijkl}
[\boldsymbol{x}^{(k)}_{ir},\boldsymbol{x}^{(l)}_{js}]&=0\,,\quad |\{i,j,r,s\}|=4\,.
\end{align}

Therefore, the leftover constraint is 
\begin{align}
0
&=[\boldsymbol{x}^{(0)}_{i},\boldsymbol{x}^{(0)}_{j}]\boldsymbol{Z}^\tau_{n,p}+\sum_{k\geq 1 }g^{(k)}_{ji}[\boldsymbol{x}^{(0)}_{i}+\boldsymbol{x}^{(0)}_{j},\boldsymbol{x}^{(k)}_{ji}]\boldsymbol{Z}^\tau_{n,p}\nnl
&\ph{=}+\sum_{k\geq 1 }\sum_{\begin{smallmatrix}
	r\in \{1,p+2,\dots,n\}\\r\neq i,j
	\end{smallmatrix}}g^{(k)}_{jr}[\boldsymbol{x}^{(0)}_{i},\boldsymbol{x}^{(k)}_{jr}]\boldsymbol{Z}^\tau_{n,p}+\sum_{k\geq 1 }g^{(k)}_{ir}\sum_{\begin{smallmatrix}
	r\in \{1,p+2,\dots,n\}\\r\neq i,j
	\end{smallmatrix}}[\boldsymbol{x}^{(k)}_{ir},\boldsymbol{x}^{(0)}_{j}]\boldsymbol{Z}^\tau_{n,p}\nnl
&\ph{=}+\sum_{k,l\geq 1 }\sum_{\begin{smallmatrix}
	r\in \{1,p+2,\dots,n\}\\r\neq i,j
	\end{smallmatrix}}\left([\boldsymbol{x}^{(k)}_{ir},\boldsymbol{x}^{(l)}_{ji}]g^{(k)}_{ir} g^{(l)}_{ji}+[\boldsymbol{x}^{(k)}_{ij},\boldsymbol{x}^{(l)}_{jr}]g^{(k)}_{ij} g^{(l)}_{jr}\right)\boldsymbol{Z}^\tau_{n,p}
\nnl
&\ph{=}+\sum_{
	k,l\geq 1}\sum_{\begin{smallmatrix}
	r\in \{1,p+2,\dots,n\}\\r\neq i,j
	\end{smallmatrix}}[\boldsymbol{x}^{(k)}_{ir},\boldsymbol{x}^{(l)}_{jr}](-1)^{k+1}g_{ji}^{(k+l)}\boldsymbol{Z}^\tau_{n,p}
\nnl
&\ph{=}+\sum_{
	k,l\geq 1}\sum_{\begin{smallmatrix}
	r\in \{1,p+2,\dots,n\}\\r\neq i,j
	\end{smallmatrix}}[\boldsymbol{x}^{(k)}_{ir},\boldsymbol{x}^{(l)}_{jr}]\sum_{a=0}^l\binom{k+a-1}{k-1}g_{ji}^{(l-a)}g_{ir}^{(k+a)}\boldsymbol{Z}^\tau_{n,p}
\nnl
&\ph{=}+\sum_{
	k,l\geq 1}\sum_{\begin{smallmatrix}
	r\in \{1,p+2,\dots,n\}\\r\neq i,j
	\end{smallmatrix}}[\boldsymbol{x}^{(k)}_{ir},\boldsymbol{x}^{(l)}_{jr}]\sum_{a=0}^k\binom{l+a-1}{l-1}g_{ij}^{(k-a)}g_{jr}^{(l+a)}\boldsymbol{Z}^\tau_{n,p}\nnl
&=[\boldsymbol{x}^{(0)}_{i},\boldsymbol{x}^{(0)}_{j}]\boldsymbol{Z}^\tau_{n,p}\nnl
&\ph{=}+\sum_{k\geq 1 }g^{(k)}_{ji}\left([\boldsymbol{x}^{(0)}_{i}+\boldsymbol{x}^{(0)}_{j},\boldsymbol{x}^{(k)}_{ji}]+\sum_{
	l=1}^{k-1}\sum_{\begin{smallmatrix}
	r\in \{1,p+2,\dots,n\}\\r\neq i,j
	\end{smallmatrix}}[\boldsymbol{x}^{(k-l)}_{ri},\boldsymbol{x}^{(l)}_{jr}]\right)\boldsymbol{Z}^\tau_{n,p}\nnl
&\ph{=}+\sum_{k\geq 1 }\sum_{\begin{smallmatrix}
	r\in \{1,p+2,\dots,n\}\\r\neq i,j
	\end{smallmatrix}}g^{(k)}_{jr}\left([\boldsymbol{x}^{(0)}_{i},\boldsymbol{x}^{(k)}_{jr}]+\sum_{a=1}^{k-1}\binom{k-1}{a}[\boldsymbol{x}^{(a)}_{ir},\boldsymbol{x}^{(k-a)}_{jr}]\right)\boldsymbol{Z}^\tau_{n,p}\nnl
&\ph{=}+\sum_{k\geq 1 }\sum_{\begin{smallmatrix}
	r\in \{1,p+2,\dots,n\}\\r\neq i,j
	\end{smallmatrix}}g^{(k)}_{ir}\left([\boldsymbol{x}^{(k)}_{ir},\boldsymbol{x}^{(0)}_{j}]+\sum_{a=1}^{k-1}\binom{k-1}{a}[\boldsymbol{x}^{(k-a)}_{ir},\boldsymbol{x}^{(a)}_{jr}]\right)\boldsymbol{Z}^\tau_{n,p}\nnl
&\ph{=}+\sum_{k,l\geq 1 }\sum_{\begin{smallmatrix}
r\in \{1,p+2,\dots,n\}\\r\neq i,j
\end{smallmatrix}}g^{(k)}_{ir} g^{(l)}_{ji}\left([\boldsymbol{x}^{(k)}_{ir},\boldsymbol{x}^{(l)}_{ji}]+\sum_{a=0}^{k-1}\binom{k-1}{a}[\boldsymbol{x}^{(k-a)}_{ir},\boldsymbol{x}^{(l+a)}_{jr}]\right)\boldsymbol{Z}^\tau_{n,p}\nnl
&\ph{=}+\sum_{k,l\geq 1 }\sum_{\begin{smallmatrix}
r\in \{1,p+2,\dots,n\}\\r\neq i,j
\end{smallmatrix}}g^{(k)}_{ij} g^{(l)}_{jr}\left([\boldsymbol{x}^{(k)}_{ij},\boldsymbol{x}^{(l)}_{jr}]+\sum_{a=0}^{l-1}\binom{l-1}{a}[\boldsymbol{x}^{(k+a)}_{ir},\boldsymbol{x}^{(l-a)}_{jr}]\right)\boldsymbol{Z}^\tau_{n,p}\,,
\end{align}
where we have used the Fay identity, followed by a change of summation variables. All the lines have to vanish separately due to the functional independence of the factors $g^{(k)}_{jr},g^{(k)}_{ir},g^{(k)}_{ir}g^{(l)}_{ji},g^{(k)}_{ij}g^{(l)}_{jr}$, which yields the following commutators for distinct labels $i,j,r\in \{1,p+2,\dots,n\}$ and $k,l\geq 0$:
\begin{align}\label{eqn:comXrij}
[\boldsymbol{x}^{(0)}_{i},\boldsymbol{x}^{(0)}_{j}]&=0\,,\nonumber\\
[\boldsymbol{x}^{(0)}_{i}+\boldsymbol{x}^{(0)}_{j},\boldsymbol{x}^{(k)}_{ji}]+\sum_{
	l=1}^{k-1}\sum_{\begin{smallmatrix}
	r\in \{1,p+2,\dots,n\}\\r\neq i,j
	\end{smallmatrix}}[\boldsymbol{x}^{(k-l)}_{ri},\boldsymbol{x}^{(l)}_{jr}]&=0\,,\nonumber\\
[\boldsymbol{x}^{(k)}_{ir},\boldsymbol{x}^{(0)}_{j}]+\sum_{a=1}^{k-1}\binom{k-1}{a}[\boldsymbol{x}^{(k-a)}_{ir},\boldsymbol{x}^{(a)}_{jr}]&=0\,,\nonumber\\
[\boldsymbol{x}^{(k)}_{ir},\boldsymbol{x}^{(l)}_{ji}+\boldsymbol{x}^{(l)}_{jr}]+\sum_{a=1}^{k-1}\binom{k-1}{a}[\boldsymbol{x}^{(k-a)}_{ir},\boldsymbol{x}^{(l+a)}_{jr}]&=0\,.
\end{align}

\subsubsection{Commutators from $[ \partial_{\tau},\partial_{z_i}]=0$}
Now, let us investigate the consequence of the vanishing of the commutator
\begin{align}
[2\pi i \partial_{\tau},\partial_{z_i}]\boldsymbol{Z}^\tau_{n,p}&=0\,.
\end{align}

Using the identity $2\pi i  \partial_{\tau} g^{(k)}_{ir}=k \partial_{z_i}g_{ir}^{(k+1)}$ implied by the mixed heat equation, we can calculate 
{\small
\begin{align}
&2\pi i \partial_{\tau}\partial_{z_i}\boldsymbol{Z}^\tau_{n,p}\nnl
&=\left(\sum_{k\geq 1 }\sum_{\begin{smallmatrix}
	r\in \{1,p+2,\dots,n\}\\r\neq i
	\end{smallmatrix}}\boldsymbol{x}^{(k)}_{ir}(2\pi i \partial_{\tau}g^{(k)}_{ir})\right)\boldsymbol{Z}^\tau_{n,p}\nnl
&\ph{=}+\left(\boldsymbol{x}^{(0)}_{i}+\sum_{l\geq 1 }\sum_{\begin{smallmatrix}
	s\in \{1,p+2,\dots,n\}\\s\neq i
	\end{smallmatrix}}\boldsymbol{x}^{(l)}_{is}g^{(l)}_{is}\right)\left(-\boldsymbol{\epsilon}^{(0)}+\sum_{k\geq 4}(1-k)G_k \boldsymbol{\epsilon}^{(k)}+ \! \! \! \! \!\! \! \! \! \! \sum_{\begin{smallmatrix}
	r,q\in \{1,p+2,\dots,n\}\\
	q< r
	\end{smallmatrix}}\sum_{k\geq 2 }(k-1)\boldsymbol{b}_{qr}^{(k)}g^{(k)}_{qr}\right)\boldsymbol{Z}^\tau_{n,p}\nnl
&=\sum_{k\geq 2 }\left(\sum_{\begin{smallmatrix}
	r\in \{1,p+2,\dots,n\}\\r\neq i
	\end{smallmatrix}}\boldsymbol{x}^{(k-1)}_{ir}(k-1) (\partial_{z_i}g_{ir}^{(k)})\right)\boldsymbol{Z}^\tau_{n,p}\nnl
&\ph{=}+\left(\boldsymbol{x}^{(0)}_{i}+\sum_{l\geq 1 }\sum_{\begin{smallmatrix}
	s\in \{1,p+2,\dots,n\}\\s\neq i
	\end{smallmatrix}}\boldsymbol{x}^{(l)}_{is}g^{(l)}_{is}\right)\left(-\boldsymbol{\epsilon}^{(0)}+\sum_{k\geq 4}(1-k)G_k \boldsymbol{\epsilon}^{(k)}+ \! \! \! \! \!\! \! \! \! \! \sum_{\begin{smallmatrix}
	r,q\in \{1,p+2,\dots,n\}\\
	q< r
	\end{smallmatrix}}\sum_{k\geq 2 }(k-1)\boldsymbol{b}_{qr}^{(k)}g^{(k)}_{qr}\right)\boldsymbol{Z}^\tau_{n,p}
\end{align}
}
and
{\small
\begin{align}
&\partial_{z_i}2\pi i \partial_{\tau}\boldsymbol{Z}^\tau_{n,p}\nnl
&=\left(\sum_{\begin{smallmatrix}
	r,q\in \{1,p+2,\dots,n\}\\
	q< r
	\end{smallmatrix}}\sum_{k\geq 2 }(k-1)\boldsymbol{b}_{qr}^{(k)}(\partial_{z_i}g^{(k)}_{qr})\right)\boldsymbol{Z}^\tau_{n,p}\nnl
&\ph{=}+\left(-\boldsymbol{\epsilon}^{(0)}+\sum_{k\geq 4}(1-k)G_k \boldsymbol{\epsilon}^{(k)}+ \! \! \! \! \!\! \! \! \! \! \sum_{\begin{smallmatrix}
	r,q\in \{1,p+2,\dots,n\}\\
	q< r
	\end{smallmatrix}}\sum_{k\geq 2 }(k-1)\boldsymbol{b}_{qr}^{(k)}g^{(k)}_{qr}\right)\left(\boldsymbol{x}^{(0)}_{i}+\sum_{l\geq 1 }\sum_{\begin{smallmatrix}
	s\in \{1,p+2,\dots,n\}\\s\neq i
	\end{smallmatrix}}\boldsymbol{x}^{(l)}_{is}g^{(l)}_{is}\right)\boldsymbol{Z}^\tau_{n,p}\nnl
&=\left(\sum_{\begin{smallmatrix}
r\in \{1,p+2,\dots,n\}\\
r< i
\end{smallmatrix}}\sum_{k\geq 2 }(k-1)\boldsymbol{b}_{ri}^{(k)}(-1)^k(\partial_{z_i}g^{(k)}_{ir})\right)\boldsymbol{Z}^\tau_{n,p}+\left(\sum_{\begin{smallmatrix}
r\in \{1,p+2,\dots,n\}\\
i< r
\end{smallmatrix}}\sum_{k\geq 2 }(k-1)\boldsymbol{b}_{ir}^{(k)}(\partial_{z_i}g^{(k)}_{ir})\right)\boldsymbol{Z}^\tau_{n,p}\nnl
&\ph{=}+\left(-\boldsymbol{\epsilon}^{(0)}+\sum_{k\geq 4}(1-k)G_k \boldsymbol{\epsilon}^{(k)}+ \! \! \! \! \!\! \! \! \! \!  \sum_{\begin{smallmatrix}
r,q\in \{1,p+2,\dots,n\}\\
q< r
\end{smallmatrix}}\sum_{k\geq 2 }(k-1)\boldsymbol{b}_{qr}^{(k)}g^{(k)}_{qr}\right)\left(\boldsymbol{x}^{(0)}_{i}+\sum_{l\geq 1 }\sum_{\begin{smallmatrix}
s\in \{1,p+2,\dots,n\}\\s\neq i
\end{smallmatrix}}\boldsymbol{x}^{(l)}_{is}g^{(l)}_{is}\right)\boldsymbol{Z}^\tau_{n,p}\,.
\end{align}
}
Therefore
{\small
\begin{align}
0&=[2\pi i \partial_{\tau},\partial_{z_i}]\boldsymbol{Z}^\tau_{n,p}\nnl
&=\sum_{k\geq 2 }(k-1)\left(\sum_{\begin{smallmatrix}
	r\in \{1,p+2,\dots,n\}\\r< i
	\end{smallmatrix}} (\partial_{z_i}g_{ir}^{(k)})(\boldsymbol{x}^{(k-1)}_{ri}-(-1)^k\boldsymbol{b}^{(k)}_{ri})+ \! \! \! \! \!\! \! \! \! \! \sum_{\begin{smallmatrix}
	r\in \{1,p+2,\dots,n\}\\i< r
	\end{smallmatrix}} (\partial_{z_i}g_{ir}^{(k)})(\boldsymbol{x}^{(k-1)}_{ir}-\boldsymbol{b}^{(k)}_{ir})\right)\boldsymbol{Z}^\tau_{n,p}\nnl
&\ph{=}+\left(\boldsymbol{x}^{(0)}_{i}+\sum_{l\geq 1 }\sum_{\begin{smallmatrix}
	s\in \{1,p+2,\dots,n\}\\s\neq i
	\end{smallmatrix}}\boldsymbol{x}^{(l)}_{is}g^{(l)}_{is}\right)\left(-\boldsymbol{\epsilon}^{(0)}+\sum_{k\geq 4}(1-k)G_k \boldsymbol{\epsilon}^{(k)}+ \! \! \! \! \!\! \! \! \! \! \sum_{\begin{smallmatrix}
	r,q\in \{1,p+2,\dots,n\}\\
	q< r
	\end{smallmatrix}}\sum_{k\geq 2 }(k-1)\boldsymbol{b}_{qr}^{(k)}g^{(k)}_{qr}\right)\boldsymbol{Z}^\tau_{n,p}\nnl
&\ph{=}-\left(-\boldsymbol{\epsilon}^{(0)}+\sum_{k\geq 4}(1-k)G_k \boldsymbol{\epsilon}^{(k)}+ \! \! \! \! \! \! \! \! \! \! \sum_{\begin{smallmatrix}
	r,q\in \{1,p+2,\dots,n\}\\
	q< r
	\end{smallmatrix}}\sum_{k\geq 2 }(k-1)\boldsymbol{b}_{qr}^{(k)}g^{(k)}_{qr}\right)\left(\boldsymbol{x}^{(0)}_{i}+\sum_{l\geq 1 }\sum_{\begin{smallmatrix}
	s\in \{1,p+2,\dots,n\}\\s\neq i
	\end{smallmatrix}}\boldsymbol{x}^{(l)}_{is}g^{(l)}_{is}\right)\boldsymbol{Z}^\tau_{n,p}\,.
\end{align}
}
The first line has to vanish due to the appearance of derivatives $\partial_{z_i}g_{ir}^{(k)}$, leading to
\begin{align}\label{eqn:bx}
\boldsymbol{b}^{(k)}_{ir}&=\boldsymbol{x}^{(k-1)}_{ir}\,,
\end{align}
which also serves as a definition for $\boldsymbol{b}^{(k)}_{ir}$ in the case $r>i$. Hence
\begin{align}
0&=\left(-[\boldsymbol{x}^{(0)}_{i},\boldsymbol{\epsilon}^{(0)}]+\sum_{k\geq 4}(1-k)G_k [\boldsymbol{x}^{(0)}_{i},\boldsymbol{\epsilon}^{(k)}]\right)\boldsymbol{Z}^\tau_{n,p}\nnl
&\ph{=}+\sum_{l\geq 1}\sum_{\begin{smallmatrix}
	s\in \{1,p+2,\dots,n\}\\s\neq i
	\end{smallmatrix}}g^{(l)}_{is}\left(-[\boldsymbol{x}^{(l)}_{is},\boldsymbol{\epsilon}^{(0)}]+\sum_{k\geq 4}(1-k)G_k [\boldsymbol{x}^{(l)}_{is},\boldsymbol{\epsilon}^{(k)}]\right)\boldsymbol{Z}^\tau_{n,p}\nnl
&\ph{=}+\left(\sum_{\begin{smallmatrix}
r,q\in \{1,p+2,\dots,n\}\\
q< r
\end{smallmatrix}}\sum_{k\geq 2 }(k-1)[\boldsymbol{x}^{(0)}_{i},\boldsymbol{x}_{qr}^{(k-1)}]g^{(k)}_{qr}\right)\boldsymbol{Z}^\tau_{n,p}\nnl
&\ph{=}+\sum_{l\geq 1 }\sum_{\begin{smallmatrix}
	s\in \{1,p+2,\dots,n\}\\s\neq i
	\end{smallmatrix}}g^{(l)}_{is}\left(\sum_{\begin{smallmatrix}
	r,q\in \{1,p+2,\dots,n\}\\
	q< r
	\end{smallmatrix}}\sum_{k\geq 2 }(k-1)[\boldsymbol{x}^{(l)}_{is},\boldsymbol{x}_{qr}^{(k-1)}]g^{(k)}_{qr}\right)\boldsymbol{Z}^\tau_{n,p}\,.
\end{align}
Note that each term in the sum over $q,r$ on the last two lines is symmetric in $q,r$, due to
\begin{align}
\boldsymbol{x}_{qr}^{(k-1)}g^{(k)}_{qr}&=\boldsymbol{x}_{rq}^{(k-1)}g^{(k)}_{rq}\,,
\end{align}
such that the first one can be decomposed 
as 
\begin{align}
&\sum_{\begin{smallmatrix}
	r,q\in \{1,p+2,\dots,n\}\\
	q< r
	\end{smallmatrix}}\nnl
&=\frac{1}{2}\sum_{\begin{smallmatrix}
	r,q\in \{1,p+2,\dots,n\}\\
	q\neq r
	\end{smallmatrix}}\nnl
&=\frac{1}{2}\Big(\sum_{\begin{smallmatrix}
	r,q\in \{1,p+2,\dots,n\}\setminus\{i\}\\
	q\neq r\\
	\end{smallmatrix}}+\sum_{\begin{smallmatrix}
	q\in \{1,p+2,\dots,n\}\setminus\{i\}\\
	r=i\\
	\end{smallmatrix}}+\sum_{\begin{smallmatrix}
	r\in \{1,p+2,\dots,n\}\setminus\{i\}\\
	q=i\\
	\end{smallmatrix}}\Big)\nnl
&=\sum_{\begin{smallmatrix}
	r,q\in \{1,p+2,\dots,n\}\setminus\{i\}\\
	q< r\\
	\end{smallmatrix}}+\sum_{\begin{smallmatrix}
	r\in \{1,p+2,\dots,n\}\setminus\{i\}\\
	q=i\\
	\end{smallmatrix}}\,.
\end{align}
and the second as follows
\begin{align}
&\sum_{\begin{smallmatrix}
	r,q\in \{1,p+2,\dots,n\}\\
	q< r
	\end{smallmatrix}}\nnl
&=\frac{1}{2}\sum_{\begin{smallmatrix}
	r,q\in \{1,p+2,\dots,n\}\\
	q\neq r
	\end{smallmatrix}}\nnl
&=\frac{1}{2}\Big(\sum_{\begin{smallmatrix}
	r,q\in \{1,p+2,\dots,n\}\setminus\{i,s\}\\
	q\neq r\\
	\end{smallmatrix}}+\sum_{\begin{smallmatrix}
	q\in \{1,p+2,\dots,n\}\setminus\{i,s\}\\
	r=i\\
	\end{smallmatrix}}+\sum_{\begin{smallmatrix}
	q\in \{1,p+2,\dots,n\}\setminus\{i,s\}\\
	r=s\\
	\end{smallmatrix}}\nnl
&\ph{=\frac{1}{2}\Big(}+\sum_{\begin{smallmatrix}
	r\in \{1,p+2,\dots,n\}\setminus\{i,s\}\\
	q=i\\
	\end{smallmatrix}}+\sum_{\begin{smallmatrix}
	r\in \{1,p+2,\dots,n\}\setminus\{i,s\}\\
	q=s\\
	\end{smallmatrix}}+\sum_{\begin{smallmatrix}
	q=i\\r=s
	\end{smallmatrix}}+\sum_{\begin{smallmatrix}
	q=s\\r=i
	\end{smallmatrix}}\Big)\nnl
&=\sum_{\begin{smallmatrix}
	r,q\in \{1,p+2,\dots,n\}\setminus\{i,s\}\\
	q< r\\
	\end{smallmatrix}}+\sum_{\begin{smallmatrix}
	q\in \{1,p+2,\dots,n\}\setminus\{i,s\}\\
	r=i\\
	\end{smallmatrix}}+\sum_{\begin{smallmatrix}
	q\in \{1,p+2,\dots,n\}\setminus\{i,s\}\\
	r=s\\
	\end{smallmatrix}}+\sum_{\begin{smallmatrix}
	q=s\\r=i
	\end{smallmatrix}}\,.
\end{align}
Thus, we can proceed with 
\begin{align}\label{eqn:tauDeriv}
0&=\left(-[\boldsymbol{x}^{(0)}_{i},\boldsymbol{\epsilon}^{(0)}]+\sum_{k\geq 4}(1-k)G_k [\boldsymbol{x}^{(0)}_{i},\boldsymbol{\epsilon}^{(k)}]+\! \! \! \! \!\! \! \! \! \! \sum_{\begin{smallmatrix}
	s\in \{1,p+2,\dots,n\}\setminus\{i\}
	\end{smallmatrix}}\sum_{k\geq 2 }(k-1)[\boldsymbol{x}^{(0)}_{i},\boldsymbol{x}_{is}^{(k-1)}]g^{(k)}_{is}\right)\boldsymbol{Z}^\tau_{n,p}\nnl
&\ph{=}+\sum_{l\geq 1}\sum_{\begin{smallmatrix}
	s\in \{1,p+2,\dots,n\}\\s\neq i
	\end{smallmatrix}} \! \! \! \! \! g^{(l)}_{is}\left(-[\boldsymbol{x}^{(l)}_{is},\boldsymbol{\epsilon}^{(0)}]+\sum_{k\geq 4}(1-k)G_k [\boldsymbol{x}^{(l)}_{is},\boldsymbol{\epsilon}^{(k)}]+\sum_{k\geq 2 }(k-1)[\boldsymbol{x}^{(l)}_{is},\boldsymbol{x}_{is}^{(k-1)}]g^{(k)}_{is}\right)\boldsymbol{Z}^\tau_{n,p}\nnl
&\ph{=}+\left(\sum_{\begin{smallmatrix}
	r,q\in \{1,p+2,\dots,n\}\setminus\{i\}\\
	q< r
	\end{smallmatrix}}\sum_{k\geq 2 }(k-1)[\boldsymbol{x}^{(0)}_{i},\boldsymbol{x}_{qr}^{(k-1)}]g^{(k)}_{qr}\right)\boldsymbol{Z}^\tau_{n,p}\nnl
&\ph{=}-\sum_{l\geq 1 }\sum_{\begin{smallmatrix}
s\in \{1,p+2,\dots,n\}\\s\neq i
\end{smallmatrix}}g^{(l)}_{si}\left(\sum_{\begin{smallmatrix}
q\in \{1,p+2,\dots,n\}\setminus\{i,s\}
\end{smallmatrix}}\sum_{k\geq 2 }(k-1)[\boldsymbol{x}^{(l)}_{si},\boldsymbol{x}_{qi}^{(k-1)}]g^{(k)}_{qi}\right)\boldsymbol{Z}^\tau_{n,p}\nnl
&\ph{=}-\sum_{l\geq 1 }\sum_{\begin{smallmatrix}
s\in \{1,p+2,\dots,n\}\\s\neq i
\end{smallmatrix}}g^{(l)}_{si}\left(\sum_{\begin{smallmatrix}
q\in \{1,p+2,\dots,n\}\setminus\{i,s\}
\end{smallmatrix}}\sum_{k\geq 2 }(k-1)[\boldsymbol{x}^{(l)}_{si},\boldsymbol{x}_{qs}^{(k-1)}]g^{(k)}_{qs}\right)\boldsymbol{Z}^\tau_{n,p}\nnl
&\ph{=}-\sum_{l\geq 1 }\sum_{\begin{smallmatrix}
	s\in \{1,p+2,\dots,n\}\\s\neq i
	\end{smallmatrix}}g^{(l)}_{si}\left(\sum_{\begin{smallmatrix}
	r,q\in \{1,p+2,\dots,n\}\setminus\{i,s\}\\
	q< r
	\end{smallmatrix}}\sum_{k\geq 2 }(k-1)[\boldsymbol{x}^{(l)}_{si},\boldsymbol{x}_{qr}^{(k-1)}]g^{(k)}_{qr}\right)\boldsymbol{Z}^\tau_{n,p}\,.
\end{align}
The last line of \eqn{eqn:tauDeriv} has to vanish since it depends on four distinct punctures, reproducing \eqn{eqn:comRelijkl}. The products $g^{(l)}_{si}g^{(k)}_{si}$ appearing in the second line can be rewritten using for $a+b\geq 2$
\begin{align}
&b g_{is}^{(a)} g_{is}^{(b+1)} - a g_{is}^{(a+1)} g_{is}^{(b)} \nonumber\\
&= \frac{ (b-a) (a+b+1)! }{(a+1)!(b+1)!} g_{is}^{(a+b+1)}  - (-1)^b (a+b) G_{a+b+1} 
 \nonumber\\
&\phantom{=}
+ \sum_{k=4}^{a+1} { a+b-k \choose b-1 } (k-1) G_{k} g_{is}^{(a+b+1-k)}
- \sum_{k=4}^{b+1} { a+b-k \choose a-1 } (k-1) G_{k} g_{is}^{(a+b+1-k)}\, .
\end{align}
This yields 
{\small
\begin{align}\label{eqn:firstLines}
0&=\left(-[\boldsymbol{x}^{(0)}_{i},\boldsymbol{\epsilon}^{(0)}]+\sum_{k\geq 4}(1-k)G_k [\boldsymbol{x}^{(0)}_{i},\boldsymbol{\epsilon}^{(k)}]+\sum_{\begin{smallmatrix}
	s\in \{1,p+2,\dots,n\}\setminus\{i\}
	\end{smallmatrix}}\sum_{k\geq 2 }(k-1)[\boldsymbol{x}^{(0)}_{i},\boldsymbol{x}_{is}^{(k-1)}]g^{(k)}_{is}\right)\boldsymbol{Z}^\tau_{n,p}\nnl
&\ph{=}+\sum_{l\geq 1}\sum_{\begin{smallmatrix}
	s\in \{1,p+2,\dots,n\}\\s\neq i
	\end{smallmatrix}}g^{(l)}_{is}\left(-[\boldsymbol{x}^{(l)}_{is},\boldsymbol{\epsilon}^{(0)}]+\sum_{k\geq 4}(1-k)G_k [\boldsymbol{x}^{(l)}_{is},\boldsymbol{\epsilon}^{(k)}]+\sum_{k\geq 2 }(k-1)[\boldsymbol{x}^{(l)}_{is},\boldsymbol{x}_{is}^{(k-1)}]g^{(k)}_{is}\right)\boldsymbol{Z}^\tau_{n,p}\nnl
&\ph{=}+\left(\sum_{\begin{smallmatrix}
r,q\in \{1,p+2,\dots,n\}\setminus\{i\}\\
q< r
\end{smallmatrix}}\sum_{k\geq 2 }(k-1)[\boldsymbol{x}^{(0)}_{i},\boldsymbol{x}_{qr}^{(k-1)}]g^{(k)}_{qr}\right)\boldsymbol{Z}^\tau_{n,p}\nnl
&\ph{=}-\sum_{l\geq 1 }\sum_{\begin{smallmatrix}
s\in \{1,p+2,\dots,n\}\\s\neq i
\end{smallmatrix}}g^{(l)}_{si}\left(\sum_{\begin{smallmatrix}
q\in \{1,p+2,\dots,n\}\setminus\{i,s\}
\end{smallmatrix}}\sum_{k\geq 2 }(k-1)[\boldsymbol{x}^{(l)}_{si},\boldsymbol{x}_{qi}^{(k-1)}]g^{(k)}_{qi}\right)\boldsymbol{Z}^\tau_{n,p}\nnl
&\ph{=}-\sum_{l\geq 1 }\sum_{\begin{smallmatrix}
s\in \{1,p+2,\dots,n\}\\s\neq i
\end{smallmatrix}}g^{(l)}_{si}\left(\sum_{\begin{smallmatrix}
q\in \{1,p+2,\dots,n\}\setminus\{i,s\}
\end{smallmatrix}}\sum_{k\geq 2 }(k-1)[\boldsymbol{x}^{(l)}_{si},\boldsymbol{x}_{qs}^{(k-1)}]g^{(k)}_{qs}\right)\boldsymbol{Z}^\tau_{n,p}\nnl
&\ph{=}-\sum_{l\geq 1 }\sum_{\begin{smallmatrix}
s\in \{1,p+2,\dots,n\}\\s\neq i
\end{smallmatrix}}g^{(l)}_{si}\left(\sum_{\begin{smallmatrix}
r,q\in \{1,p+2,\dots,n\}\setminus\{i,s\}\\
q< r
\end{smallmatrix}}\sum_{k\geq 2 }(k-1)[\boldsymbol{x}^{(l)}_{si},\boldsymbol{x}_{qr}^{(k-1)}]g^{(k)}_{qr}\right)\boldsymbol{Z}^\tau_{n,p}\nnl
&=\left(-[\boldsymbol{x}^{(0)}_{i},\boldsymbol{\epsilon}^{(0)}]+\sum_{k\geq 4}(1-k)G_k [\boldsymbol{x}^{(0)}_{i},\boldsymbol{\epsilon}^{(k)}]\right)\boldsymbol{Z}^\tau_{n,p}\nnl
&\ph{=}+\sum_{k\geq 1 }\sum_{\begin{smallmatrix}
	s\in \{1,p+2,\dots,n\}\setminus\{i\}
	\end{smallmatrix}}g^{(k)}_{is}\left((k-1)[\boldsymbol{x}^{(0)}_{i},\boldsymbol{x}_{is}^{(k-1)}]-[\boldsymbol{x}^{(k)}_{is},\boldsymbol{\epsilon}^{(0)}]\right)\boldsymbol{Z}^\tau_{n,p}\nnl
&\ph{=}+\sum_{l\geq 1}\sum_{k\geq 4}\sum_{\begin{smallmatrix}
	s\in \{1,p+2,\dots,n\}\setminus\{i\}
	\end{smallmatrix}}g^{(l)}_{is}G_k(1-k) [\boldsymbol{x}^{(l)}_{is},\boldsymbol{\epsilon}^{(k)}]\boldsymbol{Z}^\tau_{n,p}\nnl
&\ph{=}+\sum_{1\leq a<b}\sum_{\begin{smallmatrix}
	s\in \{1,p+2,\dots,n\}\setminus\{i\}
	\end{smallmatrix}}\Bigg(\frac{ (b-a) (a+b+1)! }{(a+1)!(b+1)!} g_{is}^{(a+b+1)}  - (-1)^b (a{+}b) G_{a+b+1} 
\nonumber\\
&\phantom{=}
+ \sum_{k=4}^{a+1} { a+b-k \choose b-1 } (k-1) G_{k} g_{is}^{(a+b+1-k)}
- \sum_{k=4}^{b+1} { a+b-k \choose a-1 } (k-1) G_{k} g_{is}^{(a+b+1-k)} \Bigg)[\boldsymbol{x}^{(a)}_{is},\boldsymbol{x}_{is}^{(b)}]\boldsymbol{Z}^\tau_{n,p}\nnl
&\ph{=}+\left(\sum_{\begin{smallmatrix}
	r,q\in \{1,p+2,\dots,n\}\setminus\{i\}\\
	q< r
	\end{smallmatrix}}\sum_{k\geq 2 }(k-1)[\boldsymbol{x}^{(0)}_{i},\boldsymbol{x}_{qr}^{(k-1)}]g^{(k)}_{qr}\right)\boldsymbol{Z}^\tau_{n,p}\nnl
&\ph{=}-\sum_{l\geq 1 }\sum_{\begin{smallmatrix}
	s\in \{1,p+2,\dots,n\}\\s\neq i
	\end{smallmatrix}}g^{(l)}_{si}\left(\sum_{\begin{smallmatrix}
	q\in \{1,p+2,\dots,n\}\setminus\{i,s\}
	\end{smallmatrix}}\sum_{k\geq 2 }(k-1)[\boldsymbol{x}^{(l)}_{si},\boldsymbol{x}_{qi}^{(k-1)}]g^{(k)}_{qi}\right)\boldsymbol{Z}^\tau_{n,p}\nnl
&\ph{=}-\sum_{l\geq 1 }\sum_{\begin{smallmatrix}
	s\in \{1,p+2,\dots,n\}\\s\neq i
	\end{smallmatrix}}g^{(l)}_{si}\left(\sum_{\begin{smallmatrix}
	q\in \{1,p+2,\dots,n\}\setminus\{i,s\}
	\end{smallmatrix}}\sum_{k\geq 2 }(k-1)[\boldsymbol{x}^{(l)}_{si},\boldsymbol{x}_{qs}^{(k-1)}]g^{(k)}_{qs}\right)\boldsymbol{Z}^\tau_{n,p}\nnl
&\ph{=}-\sum_{l\geq 1 }\sum_{\begin{smallmatrix}
	s\in \{1,p+2,\dots,n\}\\s\neq i
	\end{smallmatrix}}g^{(l)}_{si}\left(\sum_{\begin{smallmatrix}
	r,q\in \{1,p+2,\dots,n\}\setminus\{i,s\}\\
	q< r
	\end{smallmatrix}}\sum_{k\geq 2 }(k-1)[\boldsymbol{x}^{(l)}_{si},\boldsymbol{x}_{qr}^{(k-1)}]g^{(k)}_{qr}\right)\boldsymbol{Z}^\tau_{n,p}\nnl
&=\left(-[\boldsymbol{x}^{(0)}_{i},\boldsymbol{\epsilon}^{(0)}]+\sum_{k\geq 4}(1-k)G_k \left([\boldsymbol{x}^{(0)}_{i},\boldsymbol{\epsilon}^{(k)}]+\sum_{l= k/2}^{k-2}\sum_{\begin{smallmatrix}
	s\in \{1,p+2,\dots,n\}\setminus\{i\}
	\end{smallmatrix}}(-1)^l [\boldsymbol{x}^{(k-l-1)}_{is},\boldsymbol{x}_{is}^{(l)}]\right)\right)\boldsymbol{Z}^\tau_{n,p}\nnl
&\ph{=}-\sum_{k\geq 1 }\sum_{\begin{smallmatrix}
	s\in \{1,p+2,\dots,n\}\setminus\{i\}
	\end{smallmatrix}}g^{(k)}_{is}\left([\boldsymbol{x}^{(k)}_{is},\boldsymbol{\epsilon}^{(0)}]-(k-1)[\boldsymbol{x}^{(0)}_{i},\boldsymbol{x}_{is}^{(k-1)}]-\sum_{l=1}^{k-2}\binom{k-1}{l-1}
[\boldsymbol{x}^{(k-l-1)}_{is},\boldsymbol{x}_{is}^{(l)}]\right)\boldsymbol{Z}^\tau_{n,p}\nnl
&\ph{=}+\sum_{l\geq 1}\sum_{k\geq 4}\sum_{\begin{smallmatrix}
	s\in \{1,p+2,\dots,n\}\setminus\{i\}
	\end{smallmatrix}}g^{(l)}_{is}G_k(1-k) \left([\boldsymbol{x}^{(l)}_{is},\boldsymbol{\epsilon}^{(k)}]-\sum_{b=1}^l{ l-1 \choose b-1 }
[\boldsymbol{x}^{(l+k-b-1)}_{is},\boldsymbol{x}_{is}^{(b)}]\right)\boldsymbol{Z}^\tau_{n,p}\nnl
&\ph{=}+\left(\sum_{\begin{smallmatrix}
	r,q\in \{1,p+2,\dots,n\}\setminus\{i\}\\
	q< r
	\end{smallmatrix}}\sum_{k\geq 2 }(k-1)[\boldsymbol{x}^{(0)}_{i},\boldsymbol{x}_{qr}^{(k-1)}]g^{(k)}_{qr}\right)\boldsymbol{Z}^\tau_{n,p}\nnl
&\ph{=}-\sum_{l\geq 1 }\sum_{\begin{smallmatrix}
	s\in \{1,p+2,\dots,n\}\\s\neq i
	\end{smallmatrix}}g^{(l)}_{si}\left(\sum_{\begin{smallmatrix}
	q\in \{1,p+2,\dots,n\}\setminus\{i,s\}
	\end{smallmatrix}}\sum_{k\geq 2 }(k-1)[\boldsymbol{x}^{(l)}_{si},\boldsymbol{x}_{qi}^{(k-1)}]g^{(k)}_{qi}\right)\boldsymbol{Z}^\tau_{n,p}\nnl
&\ph{=}-\sum_{l\geq 1 }\sum_{\begin{smallmatrix}
	s\in \{1,p+2,\dots,n\}\\s\neq i
	\end{smallmatrix}}g^{(l)}_{si}\left(\sum_{\begin{smallmatrix}
	q\in \{1,p+2,\dots,n\}\setminus\{i,s\}
	\end{smallmatrix}}\sum_{k\geq 2 }(k-1)[\boldsymbol{x}^{(l)}_{si},\boldsymbol{x}_{qs}^{(k-1)}]g^{(k)}_{qs}\right)\boldsymbol{Z}^\tau_{n,p}\nnl
&\ph{=}-\sum_{l\geq 1 }\sum_{\begin{smallmatrix}
	s\in \{1,p+2,\dots,n\}\\s\neq i
	\end{smallmatrix}}g^{(l)}_{si}\left(\sum_{\begin{smallmatrix}
	r,q\in \{1,p+2,\dots,n\}\setminus\{i,s\}\\
	q< r
	\end{smallmatrix}}\sum_{k\geq 2 }(k-1)[\boldsymbol{x}^{(l)}_{si},\boldsymbol{x}_{qr}^{(k-1)}]g^{(k)}_{qr}\right)\boldsymbol{Z}^\tau_{n,p}\,.
\end{align}
}
The terms proportional to $G_k$ and $g^{(l)}_{is}G_k$ on the first and third line have to vanish separately, which leads to the commutation relations 
\begin{align}
[\boldsymbol{x}^{(0)}_{i},\boldsymbol{\epsilon}^{(0)}]&=0\,,\nnl
\ph{=}[ \boldsymbol{x}^{(0)}_{i},\boldsymbol{\epsilon}^{(k)} ]&=\sum_{l= k/2}^{k-2}\sum_{\begin{smallmatrix}
	s\in \{1,p+2,\dots,n\}\setminus\{i\}
	\end{smallmatrix}}(-1)^l [\boldsymbol{x}_{is}^{(l)},\boldsymbol{x}^{(k-l-1)}_{is}]\,,\nnl
\ph{=}[\boldsymbol{x}^{(l)}_{is},\boldsymbol{\epsilon}^{(k)}]&=\sum_{b=1}^l{ l-1 \choose b-1 }
[\boldsymbol{x}^{(l+k-b-1)}_{is},\boldsymbol{x}_{is}^{(b)}]
\end{align}
for $l\geq 1$, $k\geq 4$. 

The products $g^{(l)}_{si}g^{(k)}_{qi}$ on the third last line of \eqn{eqn:firstLines} can be rewritten using the Fay identity
\begin{align}
g^{(l)}_{si}g^{(k)}_{qi}&=(-1)^{l+1}g^{(l+k)}_{qs}+\sum_{r=0}^k \binom{l+r-1}{l-1}g^{(k-r)}_{qs}g^{(l+r)}_{si}+\sum_{r=0}^l \binom{k+r-1}{k-1}g^{(l-r)}_{sq}g^{(k+r)}_{qi}\,.
\end{align}
Thus, the remaining terms of \eqn{eqn:firstLines} are
{\small
\begin{align}\label{eqn:tauDeriv2}
0&=-\sum_{k\geq 1 }\sum_{\begin{smallmatrix}
	s\in \{1,p+2,\dots,n\}\setminus\{i\}
	\end{smallmatrix}}g^{(k)}_{is}\left([\boldsymbol{x}^{(k)}_{is},\boldsymbol{\epsilon}^{(0)}]-(k-1)[\boldsymbol{x}^{(0)}_{i},\boldsymbol{x}_{is}^{(k-1)}]-\sum_{l=1}^{k-2}\binom{k-1}{l-1}
[\boldsymbol{x}^{(k-l-1)}_{is},\boldsymbol{x}_{is}^{(l)}]\right)\boldsymbol{Z}^\tau_{n,p}\nnl
&\ph{=}+\sum_{\begin{smallmatrix}
	r,q\in \{1,p+2,\dots,n\}\setminus\{i\}\\
	q< r
	\end{smallmatrix}}\sum_{k\geq 2 }(k-1)[\boldsymbol{x}^{(0)}_{i},\boldsymbol{x}_{qr}^{(k-1)}]g^{(k)}_{qr}\boldsymbol{Z}^\tau_{n,p}\nnl
&\ph{=}-\sum_{l,k\geq 1 }\sum_{\begin{smallmatrix}
	s,q\in \{1,p+2,\dots,n\}\setminus\{i\}\\s\neq q
	\end{smallmatrix}}(-1)^{l+1}g^{(l+k)}_{qs}(k-1)[\boldsymbol{x}^{(l)}_{si},\boldsymbol{x}_{qi}^{(k-1)}]\boldsymbol{Z}^\tau_{n,p}\nnl
&\ph{=}-\sum_{l,k\geq 1 }\sum_{\begin{smallmatrix}
	s,q\in \{1,p+2,\dots,n\}\setminus\{i\}\\s\neq q
	\end{smallmatrix}}\sum_{r=0}^k \binom{l+r-1}{l-1}g^{(k-r)}_{qs}g^{(l+r)}_{si}(k-1)[\boldsymbol{x}^{(l)}_{si},\boldsymbol{x}_{qi}^{(k-1)}]\boldsymbol{Z}^\tau_{n,p}\nnl
&\ph{=}-\sum_{l,k\geq 1 }\sum_{\begin{smallmatrix}
	s,q\in \{1,p+2,\dots,n\}\setminus\{i\}\\s\neq q
	\end{smallmatrix}}\sum_{r=0}^l \binom{k+r-1}{k-1}g^{(l-r)}_{sq}g^{(k+r)}_{qi}(k-1)[\boldsymbol{x}^{(l)}_{si},\boldsymbol{x}_{qi}^{(k-1)}]\boldsymbol{Z}^\tau_{n,p}\nnl
&\ph{=}-\sum_{l,k\geq 1 }\sum_{\begin{smallmatrix}
	s\in \{1,p+2,\dots,n\}\setminus\{i\}\\s\neq q
	\end{smallmatrix}}g^{(l)}_{si}g^{(k)}_{qs}(k-1)[\boldsymbol{x}^{(l)}_{si},\boldsymbol{x}_{qs}^{(k-1)}]\boldsymbol{Z}^\tau_{n,p}\,.
\end{align}
}
We can interchange the role of the summation indices $s$ and $q$ in the second last line, such that all products of integration kernels are of the form $g^{(l)}_{qs}g^{(k)}_{si}$. Additionally collecting terms $g^{(k)}_{si}$, $g^{(l)}_{qs}$ and $g^{(l)}_{qs}g^{(k)}_{si}$ leads to 
{\small
\begin{align}\label{eqn:tauDeriv3}
0
&=-\sum_{k\geq 1 }\sum_{\begin{smallmatrix}
	s\in \{1,p+2,\dots,n\}\setminus\{i\}
	\end{smallmatrix}}g^{(k)}_{is}\Bigg([\boldsymbol{x}^{(k)}_{is},\boldsymbol{\epsilon}^{(0)}]-(k-1)[\boldsymbol{x}^{(0)}_{i},\boldsymbol{x}_{is}^{(k-1)}]-\sum_{l=1}^{k-2}\binom{k-1}{l-1}
[\boldsymbol{x}^{(k-l-1)}_{is},\boldsymbol{x}_{is}^{(l)}]\nnl
&\ph{=}- \!\!\!\!\!\!\!\!\! \sum_{\begin{smallmatrix}
	q\in \{1,p+2,\dots,n\}\\
	q\neq i,s
	\end{smallmatrix}}\left(\sum_{l=1}^{k-2} \binom{k-1}{l-1}(k-l-1)[\boldsymbol{x}^{(l)}_{is},\boldsymbol{x}_{iq}^{(k-l-1)}]+\sum_{l=1}^{k-2} \binom{k-1}{k-l-1}(k-l-1)[\boldsymbol{x}^{(l)}_{iq},\boldsymbol{x}_{is}^{(k-l-1)}]\right)\Bigg)\boldsymbol{Z}^\tau_{n,p}\nnl
&\ph{=}+\sum_{k\geq 2 }\sum_{\begin{smallmatrix}
	s,q\in \{1,p+2,\dots,n\}\setminus\{i\}\\
	q\neq s
	\end{smallmatrix}}g^{(k)}_{qs}\left(\frac{k-1}{2}[\boldsymbol{x}^{(0)}_{i},\boldsymbol{x}_{qs}^{(k-1)}]-\sum_{l=1}^{k-2}(-1)^{l+1}(k-l-1)[\boldsymbol{x}^{(l)}_{si},\boldsymbol{x}_{qi}^{(k-l-1)}]\right)\boldsymbol{Z}^\tau_{n,p}\nnl
&\ph{=}-\sum_{l,k\geq 1 }\sum_{\begin{smallmatrix}
	s,q\in \{1,p+2,\dots,n\}\setminus\{i\}\\s\neq q
	\end{smallmatrix}}\sum_{r=0}^{k-1} \binom{l+r-1}{l-1}g^{(k-r)}_{qs}g^{(l+r)}_{si}(k-1)[\boldsymbol{x}^{(l)}_{si},\boldsymbol{x}_{qi}^{(k-1)}]\boldsymbol{Z}^\tau_{n,p}\nnl
&\ph{=}-\sum_{l,k\geq 1 }\sum_{\begin{smallmatrix}
	s,q\in \{1,p+2,\dots,n\}\setminus\{i\}\\s\neq q
	\end{smallmatrix}}\sum_{r=0}^{l-1} \binom{k+r-1}{k-1}g^{(l-r)}_{qs}g^{(k+r)}_{si}(k-1)[\boldsymbol{x}^{(l)}_{qi},\boldsymbol{x}_{si}^{(k-1)}]\boldsymbol{Z}^\tau_{n,p}\nnl
&\ph{=}-\sum_{l,k\geq 1 }\sum_{\begin{smallmatrix}
	s\in \{1,p+2,\dots,n\}\setminus\{i\}\\s\neq q
	\end{smallmatrix}}g^{(l)}_{si}g^{(k)}_{qs}(k-1)[\boldsymbol{x}^{(l)}_{si},\boldsymbol{x}_{qs}^{(k-1)}]\boldsymbol{Z}^\tau_{n,p}\,.
\end{align}
}
The terms proportional to products $g^{(l)}_{si}g^{(k)}_{qs}$ for $l,k\geq 1$ on the three last lines have to vanish separately, leading to
\begin{align}\label{eqn:tauDeriv4}
0
&=\sum_{l,k\geq 1 }\sum_{\begin{smallmatrix}
	s,q\in \{1,p+2,\dots,n\}\setminus\{i\}\\s\neq q
	\end{smallmatrix}}\sum_{r=0}^{k-1} \binom{l+r-1}{l-1}g^{(k-r)}_{qs}g^{(l+r)}_{si}(k-1)[\boldsymbol{x}^{(l)}_{si},\boldsymbol{x}_{qi}^{(k-1)}]\boldsymbol{Z}^\tau_{n,p}\nnl
&\ph{=}+\sum_{l,k\geq 1 }\sum_{\begin{smallmatrix}
	s,q\in \{1,p+2,\dots,n\}\setminus\{i\}\\s\neq q
	\end{smallmatrix}}\sum_{r=0}^{l-1} \binom{k+r-1}{k-1}g^{(l-r)}_{qs}g^{(k+r)}_{si}(k-1)[\boldsymbol{x}^{(l)}_{qi},\boldsymbol{x}_{si}^{(k-1)}]\boldsymbol{Z}^\tau_{n,p}\nnl
&\ph{=}+\sum_{l,k\geq 1 }\sum_{\begin{smallmatrix}
	s\in \{1,p+2,\dots,n\}\setminus\{i\}\\s\neq q
	\end{smallmatrix}}g^{(l)}_{si}g^{(k)}_{qs}(k-1)[\boldsymbol{x}^{(l)}_{si},\boldsymbol{x}_{qs}^{(k-1)}]\boldsymbol{Z}^\tau_{n,p}\nnl
&=\sum_{l,k\geq 1 }\sum_{\begin{smallmatrix}
	s,q\in \{1,p+2,\dots,n\}\setminus\{i\}\\s\neq q
	\end{smallmatrix}}g^{(l)}_{si}g^{(k)}_{qs}\sum_{r=0}^{l-1} \binom{l-1}{r}(k+r-1)[\boldsymbol{x}^{(l-r)}_{si},\boldsymbol{x}_{qi}^{(k+r-1)}]\boldsymbol{Z}^\tau_{n,p}\nnl
&\ph{=}+\sum_{l,k\geq 1 }\sum_{\begin{smallmatrix}
	s,q\in \{1,p+2,\dots,n\}\setminus\{i\}\\s\neq q
	\end{smallmatrix}}g^{(l)}_{si}g^{(k)}_{qs}\sum_{r=0}^{l-1} \binom{l-1}{r}(l-r-1)[\boldsymbol{x}^{(k+r)}_{qi},\boldsymbol{x}_{si}^{(l-r-1)}]\boldsymbol{Z}^\tau_{n,p}\nnl
&\ph{=}+\sum_{l,k\geq 1 }\sum_{\begin{smallmatrix}
	s\in \{1,p+2,\dots,n\}\setminus\{i\}\\s\neq q
	\end{smallmatrix}}g^{(l)}_{si}g^{(k)}_{qs}(k-1)[\boldsymbol{x}^{(l)}_{si},\boldsymbol{x}_{qs}^{(k-1)}]\boldsymbol{Z}^\tau_{n,p}\nnl
&=\sum_{l,k\geq 1 }\sum_{\begin{smallmatrix}
	s,q\in \{1,p+2,\dots,n\}\setminus\{i\}\\s\neq q
	\end{smallmatrix}}g^{(l)}_{si}g^{(k)}_{qs}\Bigg((k-1)[\boldsymbol{x}^{(l)}_{si},\boldsymbol{x}_{qs}^{(k-1)}]+(k-1)[\boldsymbol{x}^{(l)}_{si},\boldsymbol{x}_{qi}^{(k-1)}]\nnl
&\ph{=}+\sum_{r=1}^{l-1}\left( \binom{l-1}{r}(k+r-1)-\binom{l-1}{r-1}(l-r)\right)[\boldsymbol{x}^{(l-r)}_{si},\boldsymbol{x}_{qi}^{(k+r-1)}]\Bigg)\boldsymbol{Z}^\tau_{n,p}
\,.
\end{align}
The identity 
\begin{align}
\binom{l-1}{r}(k+r-1)-\binom{l-1}{r-1}(l-r)&= \binom{l-1}{r}(k-1)
\end{align}
finally leads to 
\begin{align}\label{eqn:tauDeriv5}
0
&=\sum_{l\geq 1 }\sum_{k\geq 2 }\sum_{\begin{smallmatrix}
	s,q\in \{1,p+2,\dots,n\}\setminus\{i\}\\s\neq q
	\end{smallmatrix}}g^{(l)}_{si}g^{(k)}_{qs}(k-1)\Bigg([\boldsymbol{x}^{(l)}_{si},\boldsymbol{x}_{qs}^{(k-1)}]+[\boldsymbol{x}^{(l)}_{si},\boldsymbol{x}_{qi}^{(k-1)}]\nnl
&\ph{=}+\sum_{r=1}^{l-1} \binom{l-1}{r}[\boldsymbol{x}^{(l-r)}_{si},\boldsymbol{x}_{qi}^{(k+r-1)}]\Bigg)\boldsymbol{Z}^\tau_{n,p}
\,.
\end{align}
However, the bracket on the right-hand side of \eqn{eqn:tauDeriv5} vanishes trivially due to the commutation relation \eqref{eqn:comXrij}, i.e.
\begin{align}
[\boldsymbol{x}^{(l)}_{si},\boldsymbol{x}^{(k)}_{qs}]+[\boldsymbol{x}^{(l)}_{si},\boldsymbol{x}^{(k)}_{qi}]+\sum_{r=1}^{l-1}\binom{l-1}{r}[\boldsymbol{x}^{(l-r)}_{si},\boldsymbol{x}^{(k+r)}_{qi}]&=0\,,\quad k,l\geq 1\,.
\end{align}

Thus, we are left with the first three lines of \eqn{eqn:tauDeriv3}. The terms on the first two lines proportional to $g_{si}^{(k)}$ and on the third line proportional to $g_{qs}^{(k)}$ have to vanish separately, since the latter are independent of the puncture $z_i$. Using
\begin{align}
\binom{k-1}{l}l-\binom{k-1}{l-1}(k-l-1)&= \binom{k-1}{l-1}\,,
\end{align}
the former term leads to
{\small
\begin{align}
0&=[\boldsymbol{x}^{(k)}_{is},\boldsymbol{\epsilon}^{(0)}]-(k-1)[\boldsymbol{x}^{(0)}_{i},\boldsymbol{x}_{is}^{(k-1)}]-\sum_{l=1}^{k-2}\binom{k-1}{l-1}
[\boldsymbol{x}^{(k-l-1)}_{is},\boldsymbol{x}_{is}^{(l)}]\nnl
&\ph{=}- \sum_{\begin{smallmatrix}
	q\in \{1,p+2,\dots,n\}\\
	q\neq i,s
	\end{smallmatrix}}\left(\sum_{l=1}^{k-2} \binom{k-1}{l-1}(k-l-1)[\boldsymbol{x}^{(l)}_{is},\boldsymbol{x}_{iq}^{(k-l-1)}]+\sum_{l=1}^{k-2} \binom{k-1}{k-l-1}(k-l-1)[\boldsymbol{x}^{(l)}_{iq},\boldsymbol{x}_{is}^{(k-l-1)}]\right)\nnl
&=[\boldsymbol{x}^{(k)}_{is},\boldsymbol{\epsilon}^{(0)}]-(k-1)[\boldsymbol{x}^{(0)}_{i},\boldsymbol{x}_{is}^{(k-1)}]-\sum_{l=1}^{k-2}\binom{k-1}{l-1}
[\boldsymbol{x}^{(k-l-1)}_{is},\boldsymbol{x}_{is}^{(l)}]\nnl
&\ph{=}-\sum_{\begin{smallmatrix}
	q\in \{1,p+2,\dots,n\}\\
	q\neq i,s
	\end{smallmatrix}}\sum_{l=1}^{k-2}\left( \binom{k-1}{l-1}(k-l-1)+ \binom{k-1}{l}l\right)[\boldsymbol{x}^{(l)}_{is},\boldsymbol{x}_{iq}^{(k-l-1)}]\nnl
&=[\boldsymbol{x}^{(k)}_{is},\boldsymbol{\epsilon}^{(0)}]-(k-1)[\boldsymbol{x}^{(0)}_{i},\boldsymbol{x}_{is}^{(k-1)}]-\sum_{l=1}^{k-2}\binom{k-1}{l-1}
[\boldsymbol{x}^{(k-l-1)}_{is},\boldsymbol{x}_{is}^{(l)}]\nnl
&\ph{=}-\sum_{\begin{smallmatrix}
q\in \{1,p+2,\dots,n\}\\
q\neq i,s
\end{smallmatrix}}\sum_{l=1}^{k-2}\left( \binom{k-1}{l-1}(k-l-1)- \binom{k-1}{l}l\right)[\boldsymbol{x}^{(l)}_{is},\boldsymbol{x}_{iq}^{(k-l-1)}]\nnl
&=[\boldsymbol{x}^{(k)}_{is},\boldsymbol{\epsilon}^{(0)}]-(k-1)[\boldsymbol{x}^{(0)}_{i},\boldsymbol{x}_{is}^{(k-1)}]-\sum_{l=1}^{k-2}\binom{k-1}{l-1}
[\boldsymbol{x}^{(k-l-1)}_{is},\boldsymbol{x}_{is}^{(l)}]\nnl
&\ph{=}+\sum_{\begin{smallmatrix}
q\in \{1,p+2,\dots,n\}\\
q\neq i,s
\end{smallmatrix}}\sum_{l=1}^{k-2} \binom{k-1}{l-1}[\boldsymbol{x}^{(l)}_{is},\boldsymbol{x}_{iq}^{(k-l-1)}]\,,
\end{align}
}
such that
\begin{align}
[\boldsymbol{x}^{(k)}_{is},\boldsymbol{\epsilon}^{(0)}]&=(k-1)[\boldsymbol{x}_{is}^{(k-1)},\boldsymbol{x}^{(0)}_{i}]+\sum_{\begin{smallmatrix}
	q\in \{1,p+2,\dots,n\}\\
	q\neq i
	\end{smallmatrix}}\sum_{l=1}^{k-2} \binom{k-1}{l-1}[\boldsymbol{x}_{iq}^{(k-l-1)},\boldsymbol{x}^{(l)}_{is}]
\end{align}
for $k\geq 1$. Therefore, we are left with the terms proportional to  $g_{qs}^{(k)}$ in \eqn{eqn:tauDeriv3}, where various terms cancel pairwise as follows
\begin{align}
0&=\sum_{k\geq 2 }\sum_{\begin{smallmatrix}
	s,q\in \{1,p+2,\dots,n\}\setminus\{i\}\\
	q\neq s
	\end{smallmatrix}}g^{(k)}_{qs}\left(\frac{k-1}{2}[\boldsymbol{x}^{(0)}_{i},\boldsymbol{x}_{qs}^{(k-1)}]-\sum_{l=1}^{k-2}(-1)^{l+1}(k-l-1)[\boldsymbol{x}^{(l)}_{si},\boldsymbol{x}_{qi}^{(k-l-1)}]\right)\boldsymbol{Z}^\tau_{n,p}\nnl
&=\sum_{k\geq 2 }\sum_{\begin{smallmatrix}
	s,q\in \{1,p+2,\dots,n\}\setminus\{i\}\\
	q\neq s
	\end{smallmatrix}}g^{(k)}_{qs}\frac{k-1}{2}[\boldsymbol{x}^{(0)}_{i},\boldsymbol{x}_{qs}^{(k-1)}]\boldsymbol{Z}^\tau_{n,p}\nnl
&\ph{=}-\frac{1}{2}\sum_{k\geq 2 }\sum_{\begin{smallmatrix}
	s,q\in \{1,p+2,\dots,n\}\setminus\{i\}\\
	q\neq s
	\end{smallmatrix}}g^{(k)}_{qs}\sum_{l=1}^{k-2}(-1)^{l+1}(k-l-1)[\boldsymbol{x}^{(l)}_{si},\boldsymbol{x}_{qi}^{(k-l-1)}]\boldsymbol{Z}^\tau_{n,p}\nnl
&\ph{=}-\frac{1}{2}\sum_{k\geq 2 }\sum_{\begin{smallmatrix}
s,q\in \{1,p+2,\dots,n\}\setminus\{i\}\\
q\neq s
\end{smallmatrix}}g^{(k)}_{sq}\sum_{l=1}^{k-2}(-1)^{k-l}l[\boldsymbol{x}^{(k-l-1)}_{qi},\boldsymbol{x}_{si}^{(l)}]\boldsymbol{Z}^\tau_{n,p}\nnl
&=\sum_{k\geq 2 }\sum_{\begin{smallmatrix}
	s,q\in \{1,p+2,\dots,n\}\setminus\{i\}\\
	q\neq s
	\end{smallmatrix}}g^{(k)}_{qs}\frac{k-1}{2}\left([\boldsymbol{x}^{(0)}_{i},\boldsymbol{x}_{qs}^{(k-1)}]-\sum_{l=1}^{k-2}(-1)^{l+1}[\boldsymbol{x}^{(l)}_{si},\boldsymbol{x}_{qi}^{(k-l-1)}]\right)\boldsymbol{Z}^\tau_{n,p}\nnl
&\ph{=}-\frac{1}{2}\sum_{k\geq 2 }\sum_{\begin{smallmatrix}
	s,q\in \{1,p+2,\dots,n\}\setminus\{i\}\\
	q\neq s
	\end{smallmatrix}}g^{(k)}_{qs}\sum_{l=1}^{k-2}(-1)^{l}l\left([\boldsymbol{x}^{(l)}_{si},\boldsymbol{x}_{qi}^{(k-l-1)}]+[\boldsymbol{x}^{(k-l-1)}_{qi},\boldsymbol{x}_{si}^{(l)}]\right)\boldsymbol{Z}^\tau_{n,p}\nnl
&=\sum_{k\geq 2 }\sum_{\begin{smallmatrix}
s,q\in \{1,p+2,\dots,n\}\setminus\{i\}\\
q\neq s
\end{smallmatrix}}g^{(k)}_{qs}\frac{k-1}{2}\left([\boldsymbol{x}^{(0)}_{i},\boldsymbol{x}_{qs}^{(k-1)}]-\sum_{l=1}^{k-2}(-1)^{l+1}[\boldsymbol{x}^{(l)}_{si},\boldsymbol{x}_{qi}^{(k-l-1)}]\right)\boldsymbol{Z}^\tau_{n,p}\,,
\end{align}
such that the remaining commutator relation is given by 
\begin{align}
[\boldsymbol{x}^{(0)}_{i},\boldsymbol{x}_{qs}^{(k)}]&=\sum_{l=1}^{k-1}[\boldsymbol{x}^{(l)}_{is},\boldsymbol{x}_{qi}^{(k-l)}]\,,\quad |\{i,s,q\}|=3\
\end{align}
for $k\geq 1$. 

\section{Eigenvalue equations}\label{app:evalEq}
In this appendix, the eigenvalue equations for the matrix $\boldsymbol{U}_{p+3,p+1}$ from \eqn{eqn:limitZTaunp} are determined starting from the following condition
\begin{align}
[\lim_{n,p,k},2\pi i \partial_{\tau}]\boldsymbol{Z}^\tau_{n,p}&=0
\end{align}
for $k=2,\dots, p-n$, where
\begin{align}
\lim_{n,p,k}&=\lim_{z_{p+k}\to 0}(-2 \pi i z_{p+k})^{s_{(12\dots p+k-1),p+k}}\dots \lim_{z_{p+3\to 0}}(-2 \pi i z_{p+3})^{s_{(12\dots p+2),p+3}}\lim_{z_{p+2\to 0}} (-2 \pi i z_{p+2})^{s_{12\dots p+2}}\,.
\end{align}

First evaluating the derivative, then the limit yields 
\begin{align}
&\lim_{n,p,k} 2\pi i \partial_{\tau} \boldsymbol{Z}^\tau_{n,p}\nnl
&=\lim_{n,p,k} \Bigg(-\boldsymbol{\epsilon}^{(0)}+\sum_{k\geq 4}(1-k)G_k \boldsymbol{\epsilon}^{(k)}+\sum_{\begin{smallmatrix}
	r,q\in \{1,p+2,\dots,n\}\\
	q< r
	\end{smallmatrix}}\sum_{k\geq 2 }(k-1)\boldsymbol{x}_{qr}^{(k-1)}g^{(k)}_{qr}\Bigg)\boldsymbol{Z}^\tau_{n,p}\nnl
&= \Bigg(-\boldsymbol{\epsilon}^{(0)}+\sum_{k\geq 4}(1-k)G_k\boldsymbol{\epsilon}^{(k)}\nnl
&\ph{=}+\sum_{k\geq 2 }(k-1)\Big(\sum_{\begin{smallmatrix}
	r,q\in \{1,p+2,\dots,p{+}k\}\\
	q< r
	\end{smallmatrix}}+\sum_{\begin{smallmatrix}
	q\in \{1,p+2,\dots,p{+}k\}\\
	r\in \{p{+}k{+}1,\dots,n\}
	\end{smallmatrix}}+\sum_{\begin{smallmatrix}
	r,q\in \{p{+}k{+}1,\dots,n\}\\
	q< r
	\end{smallmatrix}}\Big)\boldsymbol{x}_{qr}^{(k-1)}\lim_{n,p,k}g^{(k)}_{qr}\Bigg)\boldsymbol{Z}^\tau_{n,p}\nnl
&= \Bigg(-\boldsymbol{\epsilon}^{(0)}+\sum_{k\geq 4}(1-k)G_k\Big(\boldsymbol{\epsilon}^{(k)}+\sum_{\begin{smallmatrix}
	r,q\in \{1,p+2,\dots,p{+}k\}\\
	q< r
	\end{smallmatrix}}\boldsymbol{x}_{qr}^{(k-1)}\Big)-G_2\sum_{\begin{smallmatrix}
	r,q\in \{1,p+2,\dots,p{+}k\}\\
	q< r
	\end{smallmatrix}}\boldsymbol{x}_{qr}^{(1)}\nnl
&\ph{=}+\sum_{k\geq 2 }(k-1)\Big(\sum_{\begin{smallmatrix}
q\in \{1,p+2,\dots,p{+}k\}\\
r\in \{p{+}k{+}1,\dots,n\}
\end{smallmatrix}}+\sum_{\begin{smallmatrix}
r,q\in \{p{+}k{+}1,\dots,n\}\\
q< r
\end{smallmatrix}}\Big)\boldsymbol{x}_{qr}^{(k-1)}\lim_{n,p,k}g^{(k)}_{qr}\Bigg)\boldsymbol{Z}^\tau_{n,p}\nnl
&= \Bigg(-\boldsymbol{\epsilon}^{(0)}+\sum_{k\geq 4}(1-k)G_k\Big(\boldsymbol{\epsilon}^{(k)}+\sum_{r=p+2}^{p+k}\boldsymbol{X}_{r1}^{(k-1)}\Big)-G_2\sum_{r=p+2}^{p+k}\boldsymbol{X}_{r1}^{(1)}\nnl
&\ph{=}+\sum_{k\geq 2 }(k-1)\Big(\sum_{\begin{smallmatrix}
q\in \{1,p+2,\dots,p{+}k\}\\
r\in \{p{+}k{+}1,\dots,n\}
\end{smallmatrix}}+\sum_{\begin{smallmatrix}
r,q\in \{p{+}k{+}1,\dots,n\}\\
q< r
\end{smallmatrix}}\Big)\boldsymbol{x}_{rq}^{(k-1)}\lim_{n,p,k}g^{(k)}_{rq}\Bigg)\nnl
&\ph{=}e^{s_{12\dots p+k}\omega(1,0)}\prod_{
	p+k<j\leq n}e^{-s_{(12\dots p+k),j}\CG_{j1}^{\tau}}\KN^{\tau}_{p+k+1\dots n}\begin{pmatrix}\boldsymbol{U}_{p+3,p+1}
\\
0
\end{pmatrix}\boldsymbol{Z}^\textrm{tree}_{p+3,p+2}\,,
\end{align}
where we have used that 
\begin{align}
\lim_{z_i\to 0} g^{(k)}_{i1}&=-G_k\,.
\end{align}

On the other hand, changing the order of the limit and derivative leads to
\begin{align}
& 2\pi i \partial_{\tau} \lim_{n,p,k} \boldsymbol{Z}^\tau_{n,p}\nnl
&=2\pi i \partial_{\tau}\Bigg( e^{s_{12\dots p+k}\omega(1,0)}\prod_{
	p+k<j\leq n}e^{-s_{(12\dots p+k),j}\CG_{j1}^{\tau}}\KN^{\tau}_{p+k+1\dots n}\Bigg)\begin{pmatrix}\boldsymbol{U}_{p+3,p+1}
\\
0
\end{pmatrix}\boldsymbol{Z}^\textrm{tree}_{p+3,p+2}\nnl
&=\Bigg( (G_2-2\zeta_2)s_{12\dots p+k}+\sum_{
	p+k<j\leq n}s_{(12\dots p+k),j}(g^{(2)}_{1j}-2\zeta_2) -\sum_{\begin{smallmatrix}
	q,r\in \{p+k+1\dots n\}\\
	q<r
	\end{smallmatrix}}s_{qr}(g^{(2)}_{qr}+2\zeta_2)\Bigg)\nnl
&\ph{=}e^{s_{12\dots p+k}\omega(1,0)}\prod_{
	p+k<j\leq n}e^{-s_{(12\dots p+k),j}\CG_{j1}^{\tau}}\KN^{\tau}_{p+k+1\dots n}\begin{pmatrix}\boldsymbol{U}_{p+3,p+1}
\\
0
\end{pmatrix}\boldsymbol{Z}^\textrm{tree}_{p+3,p+2}\,,
\end{align}
where we have used that 
\begin{align}
2\pi i \partial_{\tau} \omega(1,0)&=G_2-2\zeta_2
\end{align}
and
\begin{align}
2\pi i \partial_{\tau}\CG_{j1}^{\tau}&=-(g^{(2)}_{1j}+2\zeta_2)\,,\nnl
2\pi i \partial_{\tau} \KN^{\tau}_{p+k+1\dots n}&=-\sum_{\begin{smallmatrix}
	q,r\in \{p+k+1\dots n\}\\
	q<r
	\end{smallmatrix}}s_{qr}(g^{(2)}_{qr}+2\zeta_2)\KN^{\tau}_{p+k+1\dots n}\,.
\end{align}

Comparing the coefficients of $G_2$ in the above calculations leads to the eigenvalue equations 
\begin{align}\label{eqn:app:eigenvalueEq}
\boldsymbol{X}_{p{+}2,1}^{(1)}\begin{pmatrix}\boldsymbol{U}_{p+3,p+1}\\0\end{pmatrix}&=-s_{12\dots p{+}2}\begin{pmatrix}\boldsymbol{U}_{p+3,p+1}\\0\end{pmatrix}\,,\nnl
\boldsymbol{X}_{p{+}k,1}^{(1)}\begin{pmatrix}\boldsymbol{U}_{p+3,p+1}\\0\end{pmatrix}&=-s_{12\dots p{+}k{-}1,p{+}k}\begin{pmatrix}\boldsymbol{U}_{p+3,p+1}\\0\end{pmatrix}\,,\quad 2<k\leq n-p\,.
\end{align}
\section{Pole subtraction for $Z^{\tau}_{4,1}$-integrals.}\label{app:PoleSubraction}

Here we summarize the computation of the $O(\eta^{0}_2)$ coefficients of $	V^\tau_1$ and 	$V^\tau_2$ defined in \eqn{eqn:mu1mu2mu3Defn}. For future convenience, we define

\begin{align}
h_1(z_2)=\exp[ -s_{23}\tilde{\Gamma}\left(\begin{smallmatrix}
	1\\
	z_3
\end{smallmatrix}; z_2,\tau\right) -s_{24} \tilde{\Gamma}\left(\begin{smallmatrix}
	1\\
	z_4
\end{smallmatrix}; z_2,\tau\right)] \, ,
\end{align}
and
\begin{align}
	h_2(z_2)=\exp[-s_{12}\tilde{\Gamma}\left(\begin{smallmatrix}
		1\\
		0
	\end{smallmatrix}; z_2,\tau\right) -s_{24} \tilde{\Gamma}\left(\begin{smallmatrix}
		1\\
		z_4
	\end{smallmatrix}; z_2,\tau\right)] \, .
\end{align}
Note that $h_1(z_2)$  behaves like  $\mathcal{O}(1)$ as $z_2\rightarrow 0$, and  $h_1(z_2)$ behaves like $\mathcal{O}(1)$ when  $z_2 \rightarrow z_3$. Now, we  notice that  the simple poles of $g^{(1)}(z_j-z_2)$, for $j=1,3$ appear in the endpoints of the integration contour need to be addressed. Following \rcite{Broedel:2014vla}, we can write for the $eta^0_2$- component of   $V^\tau_1$:

\begin{align} \label{eq:V1tauComputation}
	V^\tau_1\big|_{\eta^0_2}= &\int_0^{z_3}\dd z_2\, e^{-s_{12}\tilde{\Gamma}\left(\begin{smallmatrix}
			1\\
			0
		\end{smallmatrix}; z_2,\tau\right)}  h_1(z_2) g^{(1)}(-z_2) \nnl  
	=& \int_0^{z_3}\dd z_2\,  e^{-s_{12}\tilde{\Gamma}\left(\begin{smallmatrix}
		1\\
		0
	\end{smallmatrix}; z_2,\tau\right)}  [ h_1(z_2) - h_1(0)] g^{(1)}(-z_2)  + \int_0^{z_3}\dd z_2\,   e^{-s_{12}\tilde{\Gamma}\left(\begin{smallmatrix}
	1\\
	0
\end{smallmatrix}; z_2,\tau\right)}   h_1(0) g^{(1)}(-z_2)  \nnl
	=&  \int_0^{z_3}\dd z_2\,  e^{-s_{12}\tilde{\Gamma}\left(\begin{smallmatrix}
			1\\
			0
		\end{smallmatrix}; z_2,\tau\right)}  [ h_1(z_2) - h_1(0)] g^{(1)}(-z_2)  + \frac{1}{s_{12}}  \exp[-s_{12}\tilde{\Gamma}\left(\begin{smallmatrix}
	1\\
	0
\end{smallmatrix}; z_3,\tau\right)]  \, .
\end{align}

To obtain third line of \eqn{eq:V1tauComputation} we integrate a total $z_2-$ derivative, and assume we can analytically continue $s_{12}$ from a positive value so we discard the $z_2\rightarrow 0$ limit. More importantly, in the third line of  \eqn{eq:V1tauComputation}, the leftover integrand can now be  $\ap-$expanded. More explicitly, the formula for this $\eta^0_2-$component of  $V^\tau_1$ is:

\begin{align} \label{eq:V1tauExplicitFormula}
	V^\tau_1\big|_{\eta^0_2}= &-\int_0^{z_3}\dd z_2\, e^{-s_{12}\tilde{\Gamma}\left(\begin{smallmatrix}
			1\\
			0
		\end{smallmatrix}; z_2,\tau\right)}  \big\{ \exp[ -s_{23}\tilde{\Gamma}\left(\begin{smallmatrix}
		1\\
		z_3
	\end{smallmatrix}; z_2,\tau\right) -s_{24} \tilde{\Gamma}\left(\begin{smallmatrix}
	1\\
	z_4
\end{smallmatrix}; z_2,\tau\right)]-1 \big\}  g^{(1)}(z_2)  \nnl 
&+\frac{1}{s_{12}} \exp[ -s_{12}\tilde{\Gamma}\left(\begin{smallmatrix}
	1\\
	0
\end{smallmatrix}; z_3,\tau\right) ] \, .
\end{align}

 For  the $\eta^0_2$- component of   $V^\tau_2$ we can write:

\begin{align} \label{eq:V2tauComputation}
	V^\tau_2\big|_{\eta^0_2}= &\int_0^{z_3}\dd z_2\, e^{-s_{23}\tilde{\Gamma}\left(\begin{smallmatrix}
			1\\
			z_3
		\end{smallmatrix}; z_2,\tau\right)}  h_{2}(z_2) g^{(1)}(z_3-z_2) \nnl  
	=& \int_0^{z_3}\dd z_2\,  e^{-s_{23}\tilde{\Gamma}\left(\begin{smallmatrix}
			1\\
			z_3
		\end{smallmatrix}; z_2,\tau\right)} [ h_{2}(z_2) - h_{2}(z_3)] g^{(1)}(-z_2)  + \int_0^{z_3}\dd z_2\,  e^{-s_{23}\tilde{\Gamma}\left(\begin{smallmatrix}
		1\\
		z_3
	\end{smallmatrix}; z_2,\tau\right)}  h_{2}(z_3) g^{(1)}(-z_2)  \nnl
	=&  \int_0^{z_3}\dd z_2\,  e^{-s_{23}\tilde{\Gamma}\left(\begin{smallmatrix}
			1\\
			z_3
		\end{smallmatrix}; z_2,\tau\right)} [ h_{2}(z_2) - h_{2}(z_3)] g^{(1)}(-z_2)  \nnl 
	&\quad - \frac{1}{s_{23}}  \exp[-s_{12}\tilde{\Gamma}\left(\begin{smallmatrix}
		1\\
		0
	\end{smallmatrix}; z_3,\tau\right)-s_{24}\tilde{\Gamma}\left(\begin{smallmatrix}
	1\\
	z_4
\end{smallmatrix}; z_3,\tau\right)]  \, .
\end{align}
The integral in the last line of \eqn{eq:V2tauComputation} can be safely $\ap-$ expanded. More explicitly,  the $\eta^0_2$- component of   $V^\tau_2$ is given by:

\begin{align} \label{eq:V2tauExplicitFormula}
	V^\tau_2\big|_{\eta^0_2}= &-\int_0^{z_3}\dd z_2\, e^{-s_{23}\tilde{\Gamma}\left(\begin{smallmatrix}
			1\\
			z_3
		\end{smallmatrix}; z_2,\tau\right)}  \big\{ \exp[ -s_{12}\tilde{\Gamma}\left(\begin{smallmatrix}
		1\\
		0
	\end{smallmatrix}; z_2,\tau\right) -s_{24} \tilde{\Gamma}\left(\begin{smallmatrix}
		1\\
		z_4
	\end{smallmatrix}; z_2,\tau\right)]  \nnl
& \quad \quad \quad \quad\quad\quad\quad \quad\quad\quad\quad-
\exp[ -s_{12}\tilde{\Gamma}\left(\begin{smallmatrix}
	1\\
	0
\end{smallmatrix}; z_3,\tau\right) -s_{24} \tilde{\Gamma}\left(\begin{smallmatrix}
	1\\
	z_4
\end{smallmatrix}; z_3,\tau\right)]
 \big\}  g^{(1)}(z_2-z_3)  \nnl 
	&- \frac{1}{s_{23}}  \exp[-s_{12}\tilde{\Gamma}\left(\begin{smallmatrix}
		1\\
		0
	\end{smallmatrix}; z_3,\tau\right)-s_{24}\tilde{\Gamma}\left(\begin{smallmatrix}
		1\\
		z_4
	\end{smallmatrix}; z_3,\tau\right)]  \, .
\end{align}

\section{Solving for fibration basis changes via commutation relations}\label{app:fibrationBasisChange}

In this section, we outline a method to solve generating function equations such as \eqn{eqn:analyticContGbraid} relying solely on the commutation relations of   $\boldsymbol{x}^{(k)}_{ij}$ and $\boldsymbol{x}^{(0)}_{i}$ in \eqn{eqn:comRealij}. The idea is to treat $\boldsymbol{x}^{(k)}_{ij}$ and $\boldsymbol{x}^{(0)}_{j}$ as abstract non-commutative symbols subject to  \eqn{eqn:comRealij} without using any specific matrix representation. Throughout this section, we will treat $\boldsymbol{x}^{(k)}_{ij}$ and $\boldsymbol{x}^{(0)}_{i}$ as abstract noncommuting variables. Keeping in mind the dictionary set up in \secref{subsection:Dictionary}, this means that we are purely using relations of $\bar{\mathfrak{t}}_{1,N}$, the genus-one Drinfeld-Kohno algebra. However, in this appendix we will follow closely the notation $\boldsymbol{x}^{(k)}_{ij}$ and $\boldsymbol{x}^{(0)}_{i}$ of \secref{subsec:AnalyticContn4p1}, including the numbering notation for the punctures. 

\subsection{Solving for fibration basis changes in two variables: $\bar{\mathfrak{t}}_{1,3}$}

We will now proceed to try and solve \eqn{eqn:analyticContn4p1braid34}. As a first step, we isolate the generating function $\boldsymbol{\Gamma}_{4}(z_4,z_3)$:\footnote{One can write an inverse of all the generating functions $\boldsymbol{\Gamma}_m$, by virtue of them being path-ordered integrals.}
\begin{align} \label{eqn:Gammaz4ofz4z3Isolated}
	\boldsymbol{\Gamma}_{4}(z_4,z_3) = \boldsymbol{\Gamma}_{3}(z_3,z_4) \boldsymbol{\Gamma}_4(z_4)\, \mathbb{X}(\sigma_{34})   \left[\boldsymbol{\Gamma}_{3}(z_3)   \right]^{-1} \, .
\end{align}
According to \eqn{eqn:Gammaz4ofz4z3}, we  can obtain the analytic continuation of, $\Gamma\left(\begin{smallmatrix}
	k_1&k_2\\
	z_{j_1}&z_{j_2}
\end{smallmatrix}; z_4,\tau\right)$, by looking at the component of the RHS of \eqn{eqn:Gammaz4ofz4z3Isolated} that accompanies $\boldsymbol{x}_{4,j_1}^{(k_1)} \boldsymbol{x}_{4,j_2}^{(k_2)}$, according to the definition of the LHS of this equation. However,  the RHS of \eqn{eqn:Gammaz4ofz4z3Isolated} as written above contains matrices $\boldsymbol{x}_{ij}^{(k)}$ that do not appear in its LHS. In other words, this equation only holds after one takes into account the commutation relations among the $\boldsymbol{x}_{ij}^{(k)}$ and $\boldsymbol{x}_{i}^{(0)}$. We can try to overcome this obstacle by using a sufficiently large subset of the commutation relations of $\bar{\mathfrak{t}}_{1,3}$ to put the RHS of this equation into a form that allows comparison with the LHS, i.e. a \textit{canonical form}.

To obtain a canonical form, we will make use of the commutation relations that we have found in \secref{subsec:comRel}. In particular, we make use of the commutation relations to move $\boldsymbol{x}_{j,4}^{(k)}$ and $\boldsymbol{x}_{4}^{(0)}$ to the \textit{rightmost} position of every word $w\in \bar{\mathfrak{t}}_{1,3}$, to the right of any other letter except either of them.  That this process terminates is clear from looking at the commutation relations. After this process is done on the RHS of \eqn{eqn:Gammaz4ofz4z3Isolated}, we can look at the eMPLs of each word to obtain a change-of-fibration-basis identity for the corresponding eMPL. As an example, we can look at the component of the RHS multiplying the letter $\boldsymbol{x}_{34}^{(1)}$, to obtain an identity for $\tilde{\Gamma}\left(\begin{smallmatrix}
	1\\
	z_3
\end{smallmatrix}; z_4,\tau\right)$:

\begin{align}
	\tilde{\Gamma}\left(\begin{smallmatrix}
		1\\
		z_3
	\end{smallmatrix}; z_4,\tau\right) =  i \pi  - \tilde{\Gamma}\left(\begin{smallmatrix}
		1\\
		0
	\end{smallmatrix}; z_3,\tau\right) + \tilde{\Gamma}\left(\begin{smallmatrix}
		1\\
		0
	\end{smallmatrix}; z_4,\tau\right)+\tilde{\Gamma}\left(\begin{smallmatrix}
		1\\
		z_4
	\end{smallmatrix}; z_3,\tau\right) \; ,
\end{align}

where there is no actual  need to use the commutation relations of $\bar{\mathfrak{t}}_{1,3}$  because this is the component of a single letter, but notice that it coincides with the result of \eqn{eqn:firstEasyFibrationn4p1}.  Now, let's look at a less trivial example, which is the coefficient multiplying the word $\boldsymbol{x}_{14}^{(1)}\boldsymbol{x}_{34}^{(1)}$:

\begin{align} \label{eq:appFibrationLength2}
	\tilde{\Gamma}\left(\begin{smallmatrix}
		1 & 1\\
		0 & z_3
	\end{smallmatrix}; z_4,\tau\right) =& - 2 \zeta_2  - i \pi  \tilde{\Gamma}\left(\begin{smallmatrix}
		1\\
		0
	\end{smallmatrix}; z_3,\tau\right) + i \pi \tilde{\Gamma}\left(\begin{smallmatrix}
		1\\
		0
	\end{smallmatrix}; z_4,\tau\right)- \tilde{\Gamma}\left(\begin{smallmatrix}
		1\\
		0
	\end{smallmatrix}; z_3,\tau\right)\tilde{\Gamma}\left(\begin{smallmatrix}
	1\\
	0
\end{smallmatrix}; z_4,\tau\right)   \nnl
& -\tilde{\Gamma}\left(\begin{smallmatrix}
	0\\
	0
\end{smallmatrix}; z_4,\tau\right)\tilde{\Gamma}\left(\begin{smallmatrix}
	2\\
	0
\end{smallmatrix}; z_3,\tau\right)  
-2  \tilde{\Gamma}\left(\begin{smallmatrix}
	0\\
	0
\end{smallmatrix}; z_3,\tau\right)\tilde{\Gamma}\left(\begin{smallmatrix}
	2\\
	0
\end{smallmatrix}; z_4,\tau\right)  
-\tilde{\Gamma}\left(\begin{smallmatrix}
	0&2\\
	0&0
\end{smallmatrix}; z_3,\tau\right)  \nnl
&+\tilde{\Gamma}\left(\begin{smallmatrix}
	0&2\\
	0&z_4
\end{smallmatrix}; z_3,\tau\right)
+\tilde{\Gamma}\left(\begin{smallmatrix}
	1&1\\
	0&0
\end{smallmatrix}; z_3,\tau\right)
+\tilde{\Gamma}\left(\begin{smallmatrix}
	1&1\\
	0&0
\end{smallmatrix}; z_4,\tau\right) 
- \tilde{\Gamma}\left(\begin{smallmatrix}
	1&1\\
	0&z_4
\end{smallmatrix}; z_3,\tau\right) \, .
\end{align}
As a final example, let's look at the  coefficient multiplying the word $\boldsymbol{x}_{14}^{(1)}\boldsymbol{x}_{34}^{(1)}\boldsymbol{x}_{34}^{(1)}$, which contains a $\zeta_3$ :

\begin{align} \label{eq:appFibrationLength3}
	\tilde{\Gamma}\left(\begin{smallmatrix}
		1 & 1& 1\\
		0 & z_3& z_3
	\end{smallmatrix}; z_4,\tau\right) =& - \frac{i \pi^3}{6}  + 3 \zeta_2    \tilde{\Gamma}\left(\begin{smallmatrix}
		1\\
		0
	\end{smallmatrix}; z_3,\tau\right) - 3 \zeta_2  \tilde{\Gamma}\left(\begin{smallmatrix}
		1\\
		0
	\end{smallmatrix}; z_4,\tau\right)- i \pi  \tilde{\Gamma}\left(\begin{smallmatrix}
		1\\
		0
	\end{smallmatrix}; z_3,\tau\right)\tilde{\Gamma}\left(\begin{smallmatrix}
		1\\
		0
	\end{smallmatrix}; z_4,\tau\right)   \nnl
	&- i \pi  \tilde{\Gamma}\left(\begin{smallmatrix}
		0\\
		0
	\end{smallmatrix}; z_4,\tau\right)\tilde{\Gamma}\left(\begin{smallmatrix}
		2\\
		0
	\end{smallmatrix}; z_3,\tau\right)
	-2 i \pi   \tilde{\Gamma}\left(\begin{smallmatrix}
		0\\
		0
	\end{smallmatrix}; z_3,\tau\right)\tilde{\Gamma}\left(\begin{smallmatrix}
		2\\
		0
	\end{smallmatrix}; z_4,\tau\right)  \nnl
	&+3 \tilde{\Gamma}\left(\begin{smallmatrix}
		3\\
		0
	\end{smallmatrix}; z_4,\tau\right)\tilde{\Gamma}\left(\begin{smallmatrix}
		0&0\\
		0&0
	\end{smallmatrix}; z_3,\tau\right) 
+2 \tilde{\Gamma}\left(\begin{smallmatrix}
	2\\
	0
\end{smallmatrix}; z_4,\tau\right)\tilde{\Gamma}\left(\begin{smallmatrix}
	0&1\\
	0&0
\end{smallmatrix}; z_3,\tau\right) \nnl
& -  \tilde{\Gamma}\left(\begin{smallmatrix}
	2\\
	0
\end{smallmatrix}; z_3,\tau\right)\tilde{\Gamma}\left(\begin{smallmatrix}
	0&1\\
	0&0
\end{smallmatrix}; z_4,\tau\right)
+ i \pi  \tilde{\Gamma}\left(\begin{smallmatrix}
	0&2\\
	0&0
\end{smallmatrix}; z_3,\tau\right) \nnl 
& + \tilde{\Gamma}\left(\begin{smallmatrix}
	1\\
	0
\end{smallmatrix}; z_4,\tau\right)\tilde{\Gamma}\left(\begin{smallmatrix}
	0&2\\
	0&0
\end{smallmatrix}; z_3,\tau\right)
+ \tilde{\Gamma}\left(\begin{smallmatrix}
	0\\
	0
\end{smallmatrix}; z_4,\tau\right)\tilde{\Gamma}\left(\begin{smallmatrix}
	0&3\\
	0&0
\end{smallmatrix}; z_3,\tau\right) \nnl
&+ i \pi  \tilde{\Gamma}\left(\begin{smallmatrix}
	0&2\\
	0&0
\end{smallmatrix}; z_4,\tau\right)
+  \tilde{\Gamma}\left(\begin{smallmatrix}
	1\\
	0
\end{smallmatrix}; z_4,\tau\right)\tilde{\Gamma}\left(\begin{smallmatrix}
	0&2\\
	0&z_4
\end{smallmatrix}; z_3,\tau\right) \nnl
&+i \pi  \tilde{\Gamma}\left(\begin{smallmatrix}
	1&1\\
	0&0
\end{smallmatrix}; z_3,\tau\right)
+  \tilde{\Gamma}\left(\begin{smallmatrix}
	1\\
	0
\end{smallmatrix}; z_4,\tau\right)\tilde{\Gamma}\left(\begin{smallmatrix}
	1&1\\
	0&0
\end{smallmatrix}; z_3,\tau\right) \nnl
&+ i \pi  \tilde{\Gamma}\left(\begin{smallmatrix}
	1&1\\
	0&0
\end{smallmatrix}; z_4,\tau\right)
- \tilde{\Gamma}\left(\begin{smallmatrix}
	1\\
	0
\end{smallmatrix}; z_3,\tau\right)\tilde{\Gamma}\left(\begin{smallmatrix}
	1&1\\
	0&0
\end{smallmatrix}; z_4,\tau\right) \nnl
&- i \pi  \tilde{\Gamma}\left(\begin{smallmatrix}
	1&1\\
	0&z_4
\end{smallmatrix}; z_3,\tau\right) 
- \tilde{\Gamma}\left(\begin{smallmatrix}
	1\\
	0
\end{smallmatrix}; z_4,\tau\right)\tilde{\Gamma}\left(\begin{smallmatrix}
	1&1\\
	0&z_4
\end{smallmatrix}; z_3,\tau\right) \nnl
&+ i \pi  \tilde{\Gamma}\left(\begin{smallmatrix}
	2&0\\
	0&0
\end{smallmatrix}; z_3,\tau\right) 
+ \tilde{\Gamma}\left(\begin{smallmatrix}
	1\\
	0
\end{smallmatrix}; z_4,\tau\right)\tilde{\Gamma}\left(\begin{smallmatrix}
	2&0\\
	0&0
\end{smallmatrix}; z_3,\tau\right) \nnl
&+ \tilde{\Gamma}\left(\begin{smallmatrix}
	0\\
	0
\end{smallmatrix}; z_4,\tau\right)\tilde{\Gamma}\left(\begin{smallmatrix}
	2&1\\
	0&0
\end{smallmatrix}; z_3,\tau\right)
-2 \tilde{\Gamma}\left(\begin{smallmatrix}
	0\\
	0
\end{smallmatrix}; z_3,\tau\right)\tilde{\Gamma}\left(\begin{smallmatrix}
	2&1\\
	0&0
\end{smallmatrix}; z_4,\tau\right) \nnl
&- \tilde{\Gamma}\left(\begin{smallmatrix}
	0&0&3\\
	0&0&0
\end{smallmatrix}; z_3,\tau\right) 
+ \tilde{\Gamma}\left(\begin{smallmatrix}
	0&0&3\\
	0&0&z_4
\end{smallmatrix}; z_3,\tau\right) 
+2 \tilde{\Gamma}\left(\begin{smallmatrix}
	0&1&2\\
	0&0&0
\end{smallmatrix}; z_3,\tau\right) \nnl
&-2 \tilde{\Gamma}\left(\begin{smallmatrix}
	0&1&2\\
	0&0&z_4
\end{smallmatrix}; z_3,\tau\right) 
- \tilde{\Gamma}\left(\begin{smallmatrix}
	0&2&1\\
	0&0&0
\end{smallmatrix}; z_3,\tau\right) 
+ \tilde{\Gamma}\left(\begin{smallmatrix}
	0&2&1\\
	0&0&z_4
\end{smallmatrix}; z_3,\tau\right) \nnl
& - \tilde{\Gamma}\left(\begin{smallmatrix}
	0&2&1\\
	0&z_4&0
\end{smallmatrix}; z_3,\tau\right) 
+ \tilde{\Gamma}\left(\begin{smallmatrix}
	0&2&1\\
	0&z_4&z_4
\end{smallmatrix}; z_3,\tau\right) 
- \tilde{\Gamma}\left(\begin{smallmatrix}
	1&1&1\\
	0&0&0
\end{smallmatrix}; z_3,\tau\right) \nnl 
& + \tilde{\Gamma}\left(\begin{smallmatrix}
	1&1&1\\
	0&0&0
\end{smallmatrix}; z_4,\tau\right) 
+ \tilde{\Gamma}\left(\begin{smallmatrix}
	1&1&1\\
	0&0&z_4
\end{smallmatrix}; z_3,\tau\right) 
+ \tilde{\Gamma}\left(\begin{smallmatrix}
	1&1&1\\
	0&z_4&0
\end{smallmatrix}; z_3,\tau\right)  \nnl 
& - \tilde{\Gamma}\left(\begin{smallmatrix}
	1&1&1\\
	0&z_4&z_4
\end{smallmatrix}; z_3,\tau\right) 
- \tilde{\Gamma}\left(\begin{smallmatrix}
	2&0&1\\
	0&0&0
\end{smallmatrix}; z_3,\tau\right) 
+ \tilde{\Gamma}\left(\begin{smallmatrix}
	2&0&1\\
	0&0&z_4
\end{smallmatrix}; z_3,\tau\right) 
+ \zeta_3 
 \, .
\end{align}

We have performed numerical checks on these and several other identities that can be obtained via this method. These have been found to hold when  $\arg (z_3) > \arg(z_4)$ for  $z_3$ and $z_4$ in the fundamental parallelogram. We note that these  last two identities can also be derived from the explicit representation of the matrices $\boldsymbol{x}^{(0)}_i$ and $\boldsymbol{x}^{(k)}_{ij}$, as explained in \secref{subsec:TwoPuntureFibrationFromRep}. 

\subsection{Solving for fibration basis changes in three variables: $\bar{\mathfrak{t}}_{1,4}$}

In this subsection we will sketch how to use the commutation relations of $\bar{\mathfrak{t}}_{1,4}$ to solve for $\boldsymbol{\Gamma}_4(z_3,z_5,z_4)$ in \eqn{eqn:analyticContGbraidForN5P1sigma34sigma45}. However, we will do this in two steps, by studying separately the analytic continuations due to $\sigma_{34}$ and $\sigma_{45}$:

\begin{align}\label{eqn:analyticContGbraidForN5P1sigma34}
	\boldsymbol{\Gamma}_{3}(z_3,z_4,z_5) \boldsymbol{\Gamma}_{4}(z_4,z_5)\boldsymbol{\Gamma}_{5}(z_5) \mathbb{X}(\sigma_{34})   = 	\boldsymbol{\Gamma}_{4}(z_4,z_3,z_5) \boldsymbol{\Gamma}_{3}(z_3,z_5)\boldsymbol{\Gamma}_{5}(z_5) \, ,
\end{align}
and
\begin{align}\label{eqn:analyticContGbraidForN5P1sigma45}
	\boldsymbol{\Gamma}_{3}(z_3,z_4,z_5) \boldsymbol{\Gamma}_{4}(z_4,z_5)\boldsymbol{\Gamma}_{5}(z_5) \mathbb{X}(\sigma_{45})   = 		\boldsymbol{\Gamma}_{3}(z_3,z_5,z_4) \boldsymbol{\Gamma}_{5}(z_5,z_4)\boldsymbol{\Gamma}_{4}(z_4)  \, .
\end{align}
If we look carefully, it turns out that both $\boldsymbol{\Gamma}_{3}(z_3,z_4,z_5)$ and $\boldsymbol{\Gamma}_{3}(z_3,z_5,z_4) $ in \eqn{eqn:analyticContGbraidForN5P1sigma45} describe the same function -- in other words, the braiding $\sigma_{45}$ only acts nontrivially on the product of generating functions $ \boldsymbol{\Gamma}_{5}(z_5,z_4)\boldsymbol{\Gamma}_{4}(z_4) $. However, the leftover formula is an (albeit nontrivial) relabeling of the result of the previous section, with a  relabeling given by:
\begin{align} \label{eqn:relabelingAppendix}
(z_3,z_4,\boldsymbol{x}^{(0)}_3,\boldsymbol{x}^{(0)}_4,\boldsymbol{x}^{(k)}_{34},\boldsymbol{x}^{(k)}_{31},\boldsymbol{x}^{(k)}_{41}) \rightarrow 
(z_4,z_5,\boldsymbol{x}^{(0)}_4,\boldsymbol{x}^{(0)}_5,\boldsymbol{x}^{(k)}_{45},\boldsymbol{x}^{(k)}_{41}+\boldsymbol{x}^{(k)}_{43},\boldsymbol{x}^{(k)}_{51}+\boldsymbol{x}^{(k)}_{54})  \, .
\end{align}
Thus,  \eqn{eqn:analyticContGbraidForN5P1sigma45} has the same information as \eqn{eqn:Gammaz4ofz4z3Isolated} of the two-variable case. Because of this, we will focus on \eqn{eqn:analyticContGbraidForN5P1sigma34} for the rest of this subsection.

We proceed by  isolating the term $\boldsymbol{\Gamma}_{4}(z_4,z_3,z_5)$ in \eqn{eqn:analyticContGbraidForN5P1sigma34}:

\begin{align}\label{eqn:analyticContGbraidForN5P1sigma34isolatedZ4}
	\boldsymbol{\Gamma}_{4}(z_4,z_3,z_5) =    \boldsymbol{\Gamma}_{3}(z_3,z_4,z_5) \boldsymbol{\Gamma}_{4}(z_4,z_5)\boldsymbol{\Gamma}_{5}(z_5) \mathbb{X}(\sigma_{34}) [\boldsymbol{\Gamma}_{5}(z_5) ]^{-1} [	\boldsymbol{\Gamma}_{3}(z_3,z_5)]^{-1}  \, .
\end{align}
We now need to write the RHS of \eqn{eqn:analyticContGbraidForN5P1sigma34isolatedZ4} in a canonical form.  We follow the same logic as in the two-variable case, and try to use the commutation relations such that in every word in the RHS of  \eqn{eqn:analyticContGbraidForN5P1sigma34isolatedZ4} we move the letters $\boldsymbol{x}^{(k)}_{4,j}$ and $\boldsymbol{x}^{(0)}_{4}$ to the right of any letter $\boldsymbol{x}^{(m)}_{l,p}$ and $\boldsymbol{x}^{(0)}_{q}$ for $4\notin \{l,p,q\}$. However, as opposed to the two-puncture case, we need to make a choice for using the following commutation relation: we use $j=4$ when making use of the last commutation relation of \eqn{eqn:comRealij}. One has to additionally be careful in never using a commutation relation that would produce an element $\boldsymbol{x}^{(0)}_{1}$. 

It is not hard to convince oneself that the algorithm described above terminates. Moreover, we have experimentally checked that, as a pleasant property of the end result, every word left over in the RHS of \eqn{eqn:analyticContGbraidForN5P1sigma34isolatedZ4} is a word that appears in its LHS, when using eMPLs of weight and length up to 4. We do not currently have an argument for why this should be the case. Now, once the RHS of \eqn{eqn:analyticContGbraidForN5P1sigma34isolatedZ4} has been written in a canonical form, we can obtain identities by simply reading out the coefficients of the expected word of the LHS. For example, the coefficient of the word $\boldsymbol{x}^{(1)}_{45} \boldsymbol{x}^{(1)}_{34}$ of the RHS of \eqn{eqn:analyticContGbraidForN5P1sigma34isolatedZ4} is:

\begin{align} \label{eqn:threeVariableFibration}
	\tilde{\Gamma}\left(\begin{smallmatrix}
		1 & 1\\
		z_5 & z_3
	\end{smallmatrix}; z_4,\tau\right) =& -   \tilde{\Gamma}\left(\begin{smallmatrix}
	0\\
	0
\end{smallmatrix}; z_3,\tau\right) \tilde{\Gamma}\left(\begin{smallmatrix}
2\\
0
\end{smallmatrix}; z_4,\tau\right)   - i \pi  \tilde{\Gamma}\left(\begin{smallmatrix}
1\\
z_5
\end{smallmatrix}; z_3,\tau\right) -
 \tilde{\Gamma}\left(\begin{smallmatrix}
	1\\
	0
\end{smallmatrix}; z_4,\tau\right) 
 \tilde{\Gamma}\left(\begin{smallmatrix}
	1\\
	z_5
\end{smallmatrix}; z_3,\tau\right)   \nnl
	& + i \pi \tilde{\Gamma}\left(\begin{smallmatrix}
		1\\
		z_5
	\end{smallmatrix}; z_4,\tau\right)
	-  \tilde{\Gamma}\left(\begin{smallmatrix}
		1\\
		0
	\end{smallmatrix}; z_3,\tau\right)\tilde{\Gamma}\left(\begin{smallmatrix}
		1\\
		z_5
	\end{smallmatrix}; z_4,\tau\right)  
	+\tilde{\Gamma}\left(\begin{smallmatrix}
		1\\
		z_5
	\end{smallmatrix}; z_3,\tau\right)\tilde{\Gamma}\left(\begin{smallmatrix}
		1\\
		z_5
	\end{smallmatrix}; z_4,\tau\right)   \nnl
	& -  \tilde{\Gamma}\left(\begin{smallmatrix}
		0\\
		0
	\end{smallmatrix}; z_4,\tau\right)\tilde{\Gamma}\left(\begin{smallmatrix}
		2\\
		z_5
	\end{smallmatrix}; z_3,\tau\right)  
	-\tilde{\Gamma}\left(\begin{smallmatrix}
		0\\
		0
	\end{smallmatrix}; z_3,\tau\right)\tilde{\Gamma}\left(\begin{smallmatrix}
		2\\
		z_5
	\end{smallmatrix}; z_4,\tau\right) 
-\tilde{\Gamma}\left(\begin{smallmatrix}
	0&2\\
	0&0
\end{smallmatrix}; z_3,\tau\right)
	\nnl
	&+\tilde{\Gamma}\left(\begin{smallmatrix}
		0&2\\
		0&z_4
	\end{smallmatrix}; z_3,\tau\right)
	+\tilde{\Gamma}\left(\begin{smallmatrix}
		1&1\\
		z_5&0
	\end{smallmatrix}; z_3,\tau\right)
	+\tilde{\Gamma}\left(\begin{smallmatrix}
		1&1\\
		z_5&0
	\end{smallmatrix}; z_4,\tau\right) 
- \tilde{\Gamma}\left(\begin{smallmatrix}
	1&1\\
	z_5&z_4
\end{smallmatrix}; z_3,\tau\right) \, .
\end{align}

We have checked this last identity to be  valid numerically when $\arg (z_3) > \arg(z_4)> \arg(z_5)$ for  $z_3$, $z_4$, and $z_5$ in the fundamental parallelogram.

\section{Towards general integration domains along $A$-cycle} 

\label{app:LastAppendixGenerlaIntegrationDomain}

We now consider alternate integration domains for our genus-one integrals. The main goal of this appendix will be to outline the computation of the regularized initial values $\lim\limits_{n,p,n-p}  $ of these new types of integrals.  We start with a concrete example by looking at the case of $(n,p)=(3,1)$. We then outline the three possible outcomes of taking a single regularized limit $\lim\limits_{z_{p+2}\to 0} (- 2 \pi i z_{p+2})^{s_{12\dots p+2}}$.  One can thereafter iterate this process.

\subsection{An alternate integration domain for  $(n,p)=(3,1)$.}

Consider the $(n,p)=(3,1)$ integral with the integration contour $\gamma=\{z_3=0<z_2<1\equiv 0$, which we will denote as $	\boldsymbol{Z}_{\gamma}^{\tau}(z_3)$:

\begin{align}
	\label{eq:eqnZn3p1DifferentContour}
	\boldsymbol{Z}_{\gamma}^{\tau}(z_3)=\int_{z_3}^{1} \dd z_2\, \KN_{123}^\tau 
	\begin{pmatrix}
		F(z_{12},\eta_2,\tau )\\
		F(z_{32},\eta_2,\tau )
	\end{pmatrix}
	\,.
\end{align}

In order to use the methods of \secref{sec:alphaExpansion}, we need a regularized initial value as $z_3 \rightarrow 0$ of $	\boldsymbol{Z}_{\gamma}^{\tau}(z_3)$. At the integrand level, we can write

\begin{align} 
\KN_{123}^\tau  &=  \exp( -s_{12}\CG^\tau_{21} -s_{13}\CG^\tau_{31}-s_{23}\CG^\tau_{32}   ) \nnl
& =\exp (\pm \pi i s_{23} )\exp(  -s_{12}\CG^\tau_{21} -s_{13}\CG^\tau_{31}-s_{23}\CG^\tau_{23} ] \nnl
& =\exp (\pm \pi i s_{23} ) e^{s_{23} \omega(1,0|\tau)} (-2 \pi i z_3)^{-s_{23}} e^{-(s_{12}+s_{23})\CG^\tau_{21}}(1+O(z_3)) \, , \label{eqn:easiestPhaseFactor}
\end{align}
where in the first line we obtain a phase factor $\exp (\pm \pi i s_{23} )$ from writing $\CG^\tau_{23}$ in terms of $\CG^\tau_{32}$, and in the second line we use the local behavior of $	\tilde{\Gamma}\left(\begin{smallmatrix}
	1\\
	0
\end{smallmatrix}; z,\tau\right)  \sim \log(-2 \pi i z) + O(z)$ . Meanwhile, the $F(z_{32},\eta_2,\tau)$  simply degenerates to $F(z_{12},\eta_2,\tau)$ when taking the limit $z_3\rightarrow 0$ in the integrand. For the integration domain, however, there is a subtlety in taking the $z_3\rightarrow 0$ limit, and we cannot simply set $z_3$ to 0 in the integration domain\footnote{We will soon state conditions on the $\Re(s_{ij})$ that makes this simple approach work.}. This can be seen from writing e.g.:

\begin{align}
	\label{eq:eqnZn3p1SumOfDomains}
		\boldsymbol{Z}_{\gamma}^{\tau}(z_3) = 
	\int_{0}^{1} \dd z_2\, \KN_{123}^\tau 
	\begin{pmatrix}
		F(z_{12},\eta_2,\tau )\\
		F(z_{32},\eta_2,\tau )
	\end{pmatrix}
	\,  - 
 \int_{0}^{z_3} \dd z_2\, \KN_{123}^\tau 
	\begin{pmatrix}
		F(z_{12},\eta_2,\tau )\\
		F(z_{32},\eta_2,\tau )
	\end{pmatrix}
	\,         .
\end{align}

The RHS of \eqn{eq:eqnZn3p1SumOfDomains} has a first term leading order of $\mathcal{O}((-z_3)^{-s_{23}})$, which can be seen from the integrand behavior we just computed. The second term of the RHS, however, has a leading behavior of  $\mathcal{O}((-z_3)^{-s_{123}})$, see Appendix E.1 of \rcite{Broedel:2020tmd}. For this second term to be subleading, we additionally require that the Mandelstam variables satisfy $\Re(s_{123})<\Re(s_{23})<0$. Thus, by further constraining the Mandelstam variables, we can easily obtain the desired initial condition: 

\begin{align}
	\label{eq:eqnZn3p1InitialCondition}
	\lim \limits_{z_{3}\rightarrow 0} (- 2 \pi i z_{3})^{s_{13}} \boldsymbol{Z}_{\gamma}^{\tau}(z_3)=\exp (\pm \pi i s_{23} ) e^{s_{23} \omega(1,0|\tau)}  \int_{0}^{1} \dd z_2\, e^{-(s_{12}+s_{23})\CG^\tau_{21}} 
	\begin{pmatrix}
		F(z_{12},\eta_2,\tau )\\
		F(z_{12},\eta_2,\tau 
	\end{pmatrix}
	\,.
\end{align}
This result  coincides with the one of Appendix E.1 of \rcite{Broedel:2020tmd}, under the relabeling $\{z_3,z_2,s_{3j},\eta_2\}\rightarrow \{1-z_0,1-z_2,s_{0j},\eta\}$, except for the phase factor  $\exp (\pm \pi i s_{23} )$ which is due to the difference conventions for $\KN_{123}^\tau $.   The regularized  initial condition for $\boldsymbol{Z}_{\gamma}^{\tau}(z_3)$ has nontrivial $\tau$- and $\eta_2$-dependence, and precisely in the form of the $Z_2^{\tau}$ integrals introduced in \subsecref{subsec:ZnTauIntegrals}, under some relabeling of momenta.

\subsection{General integration domains along $A$-cycle for $(n,p)$ }

We will now consider more general integration contours $\gamma^{(n,p)}_{\vec{B}}$. These integration contours will have to describe how to integrate the punctures $z_2,z_3,\ldots,z_{p+1}$, given that any ordered partition of these  will be integrated between a pair of the  ordered, unintegrated punctures, $z_1=0<z_{p+2}<z_{p+3}<\ldots<z_{n}<1\equiv 0$. E.g. for $(n,p)=(5,2)$, we can integrate $\{z_2,z_3\}$ along the contour $\gamma=\{z_1=0<z_3<z_2<z_4<z_5<1\}$.

Generically, an integration contour  $\gamma^{(n,p)}_{\vec{B}}$ will have some punctures being integrated between two neighboring  unintegrated punctures $z_{j} < z_{k}$ , and we will label this (possibly empty) set of ordered punctures by the ordered partition $B^j$. For example, the contour $\gamma$ above involves integrates the punctures $z_3$ and $z_2$ integrated between the neighboring unintegrated punctures $z_1<z_4$, so it will have an ordered set $B^1=(3,2)$. Likewise, we notice that $\gamma$ has  $B^4=()$ because no punctures are being integrated between the pair $z_4$ and $z_5$. Likewise, $B^5=()$, and we can fully characterize the integration contour of  $\gamma=\{z_1=0<z_3<z_2<z_4<z_5<1\}$ by the ordered partition of $(2,3)$ given by $\vec{B}=(B^1,B^4,B^5)=\left((3,2),(),()\right)$.

We now give two  examples of this notation are given by:
\begin{align}
\gamma^{(6,2)}_{((),(),(2),(3))}=\{z_1=0<z_4<z_5<z_2<z_6<z_3<1\} \, ,
\end{align}
where $B^1=(0),B^4=(0),B^5=(2),B^6=(3)$ , and 
\begin{align}
	\gamma^{(6,3)}_{((4,2),(3),(0))}=\{z_1=0<z_4<z_2<z_5<z_3<z_6<1\} \, ,
\end{align}
where $B^1=(4,2),B^5=(3),B^6=(0)$ .

More explicitly,  any $B^k$ can be written as $B^k=(b^k_1,b^k_2,\ldots,b^k_{|B^k|})$, and any integration contour along the $A-$cycle can be written as: 

\begin{align}
\gamma^{(n,p)}_{\vec{B}
}&=\{z_1=0<z_{b_1^1}<\dots <z_{b_{|B^1|}^1}<z_{p+2}\nnl
&\ph{=\{ }<z_{b_1^{p+2}}<\dots <z_{b_{|B^{p+2}|}^{p+2}}<z_{p+3}<\dots\nnl
&\ph{=\{ }<z_{b_1^{n-1}}<\dots <z_{b_{|B^{n-1}|}^{n-1}}<z_{n}\nnl
&\ph{=\{ }<z_{b_1^{n}}<\dots <z_{b_{|B^{n}|}^{n}}<1\}\,,
\end{align}
and where  $\{B^1,B^{p+2},\dots,B^{n}\}$ is a partition of the labels of the $p$ integrated punctures, i.e. as an unordered set:
\begin{align}
\{B^1,B^{p+2},\dots,B^{n}\}=\{2,3,\dots,p+1\}\,.
\end{align}
Let us repeat the analysis from above to determine the degeneration of the type-$(n,p)$ integrals on the configurations $\gamma^{(n,p)}_{\vec{B}}$ for the single regularized limit 
\begin{align}
(z_{p+2},z_{p+3},z_{p+4}\dots,z_n)\to (0,z_{p+3},z_{p+4},\dots,z_n)\,.
\end{align}
We now focus on the  $z_{p+2}\to 0$  behavior of the Koba--Nielsen factor $\KN_{12\dots n}^{\tau}$, which can be seen understood with the change of variables $z_i=z_{p+2}x_i$ for $i\in \{1,B^1,p+2\}$, the Koba--Nielsen factor degenerates as follows\footnote{In these products, we  use the convention that $i\prec j$  for labels $i,j \in \{A)\}$ if $z_i<z_j$.  This rewriting of  the Koba--Nielsen factor generates a phase $\chi $, of the schematic form $\chi = \exp(\sum \limits_{ij} \pm  \pi i s_{ij})$, see e.g. \eqref{eqn:easiestPhaseFactor}.} 
\begin{align}
\KN_{12\dots n}^{\tau}
&= \chi \prod_{i \prec j\in \{1,B^1,p+2\}}e^{-s_{ij}\CG_{ji}^{\tau}}\prod_{\begin{smallmatrix}
	i\in \{1,B^1,p+2\}\\
	j\in \{B^{p+2},p+3,\dots,n,B^{n}\} 
	\end{smallmatrix}}e^{-s_{ij}\CG_{ji}^{\tau}}\prod_{i\prec j\in \{B^{p+2},p+3,\dots,n,B^{n}\}}e^{-s_{ij}\CG_{ji}^{\tau}}\nnl
&=\chi \,(-2 \pi i z_{p+2})^{-s_{(1,B^1,p+2)}}e^{s_{(1,B^1,p+2)}\omega(1,0)}\!\!\! \prod_{i\prec j\in \{1,B^1,p+2\}}x_{ji}^{-s_{ij}} \!\!\!\!\!\!\!\!\! \prod_{j\in \{B^{p+2},p+3,\dots,n,B^{n}\}}e^{-s_{(1,B^1,p+2),j}\CG_{j1}^{\tau}}\nnl
&\ph{=} \times \prod_{
	i\prec j\in \{B^{p+2},p+3,\dots,n,B^{n}\}}e^{-s_{ij}\CG_{ji}^{\tau}}(1+\mathcal{O}(z_{p+2}))\,,
\end{align}
such that
\begin{align}
\lim_{z_{p+2}\to 0}z_{p+2}^{s_{(1,B^1,p+2)}}\KN_{12\dots n}^{\tau}&=e^{s_{(1,B^1,p+2)}\omega(1,0)}\KN_{1,B^1,p+2}(x_i,\{s_{ij}\}) \KN^{\tau}_{1,B^{p+2},p+3,\dots,n,B^{n}}(z_i,\{s_{ij}^{p+2}\})\,,
\end{align}
where for $2\leq k\leq n-p$, we have relabeled Mandelstam variables
\begin{align}
s^{p+k}_{ij}&=\begin{cases}
s_{(1,B^1,p+2,B^{p+2},\dots,p+k),j}&\text{if }i=1\\
s_{ij}&\text{otherwise,}
\end{cases}
\end{align}
and where we have can identify tree-level and genus-one Koba--Nielsen factors\footnote{These Koba--Nielsen factors are now  written following the ordering  of punctures the punctures , i.e. using  $i \prec j $  in their definition. This  means that we  do not need to introduce more phase factors $\chi$ in iterating this limiting procedure. }. It is important to note that the $\textrm{SL}_2-$fixed tree-level Koba--Nielsen factor $\KN_{1,B^1,p+2}(x_i,\{s_{ij}\})$ is one corresponding to $(|B^1|+3)$ punctures (one of them $\textrm{SL}_2-$ fixed at infinity, so not explicitly seen). The genus-one Koba--Nielsen factor 
$\KN^{\tau}_{1,B^{p+2},p+3,\dots,n,B^{n}}(z_i,\{s_{ij}^{p+2}\})$ is one corresponding to $(n-|B^1|)$ punctures, with some Mandelstam variables relabeled.

The behavior of the Eisenstein--Kronecker chains under $z_{p+2}\to z_1=0$  is understood by first looking at the whole differential form:
\begin{align}
&\tilde{\varphi}^\tau(1,A^1)\prod_{k=p+2}^n\tilde{\varphi}^\tau(k,A^k)\prod_{i=2}^{p+1}\dd z_i\nnl
&=\tilde{\varphi}^\tau(1,A^1)\tilde{\varphi}^\tau(p+2,A^{p+2})\prod_{k=p+3}^n\tilde{\varphi}^\tau(k,A^k)\prod_{b\in B^1}\dd z_b\prod_{j=p+2}^n\prod_{b\in B^j}\dd z_b\nnl
&=\tilde{\varphi}^\tau(1,A^1)\tilde{\varphi}^\tau(p+2,A^{p+2})\prod_{k=p+3}^n\tilde{\varphi}^\tau(k,A^k)z_{p+2}^{|B^1|}\prod_{b\in B^1}\dd x_b\prod_{j=p+2}^n\prod_{b\in B^j}\dd z_b \, .
\end{align}
The factor $z_{p+2}^{|B^1|}$ from the change of variables $z_i=z_{p+2} x_i$ for $z_1\leq z_i\leq z_{p+2}$ can only be compensated if the sequences $A^1$ and $A^{p+2}$ \textit{begin} with the integration variables in $B^1$, i.e.\ if for an unordered partition 
\begin{align}
\{B^1_1,B^1_{p+2}\}&=B^1\,,
\end{align}
we have 
\begin{align}\label{eqn:condp2}
A^1&=(B^1_1,A^1_2)\,,\quad A^{p+2}=(B^1_{p+2},A^{p+2}_2)\,.
\end{align}
In this case, the above differential form degenerates to 
\begin{align}
&\tilde{\varphi}^\tau(1,A^1)\prod_{k=p+2}^n\tilde{\varphi}^\tau(k,A^k)\prod_{i=2}^{p+1}\dd z_i\nnl
&=\textrm{pt}(1,B^1_1)\textrm{pt}(p+2,B^1_{p+2})\prod_{b\in B^1}\dd x_b\nnl
&\ph{=}\times
\tilde{\varphi}^\tau(1,A^1_2\shuffle A^{p+2}_2)\prod_{k=p+3}^n\tilde{\varphi}^\tau(k,A^k)\prod_{j=p+2}^n\prod_{b\in B^j}\dd z_b + \mathcal{O}(z_{p+2})\,,
\end{align}
where the $\textrm{pt}$ are the genus-zero $\textrm{SL}_2-$fixed Parke-Taylor factors  with respect to the variables $x_i$ and for $2\leq k\leq n-p$, the punctures $x_{p+k}=1$ are additional punctures on genus-zero Riemann surface. Note that the products separates into factors $\textrm{pt}(1,B^1_1)\textrm{pt}(p+2,B^1_{p+2})$ containing the rescaled variables $x_i$ and factors $\tilde{\varphi}^\tau(1,A^1_2\shuffle A^{p+2}_2)\prod_{k=p+3}^n\tilde{\varphi}^\tau(k,A^k)\prod_{j=p+2}^n$ depending on the non-rescaled variables $z_i$. Chains which would mix the two are sub-leading.

Putting all together, in the limit $z_{p+2}\to 0$, the integral
\begin{align}
& Z^\tau_{n,p}(\gamma^{n,p}_{\vec{B}};(1,A^1),(p+2,A^{p+2}),(p+3,A^{p+3}),\dots,(n,A^n); z_{p+2},z_{p+3},\dots,z_n;\{s_{ij}\})\nonumber\\
&=\int_{\gamma^{(n,p)}_{\vec{B}
}}\prod_{i=2}^{p+1}\dd z_i\, \KN^{\tau}_{12\dots n}(z_i,\{s_{ij}\})\,\tilde{\varphi}^\tau(1,A^1)\prod_{k=p+2}^n\tilde{\varphi}^\tau(k,A^k)
\end{align}
vanishes unless the condition \eqref{eqn:condp2} is satisfied. When this condition is satisfied, the integral degenerates as follows
{\small %
\begin{align}
&\lim_{z_{p+2}\to 0}(-2 \pi i z_{p+2})^{s_{(1,B^1,p+2)}}Z^\tau_{n,p}(\gamma^{n,p}_{\vec{B}}, z_{p+2},z_{p+3},\dots,z_n;\{s_{ij}\})\\
&=e^{s_{(1,B^1,p+2)}\omega(1,0)}Z^\textrm{\tree}_{|B^1|+3,|B^1|}(\gamma^{|B^1|+3,|B^1|}_{B^1};(1,B^1_1),(p+2,B^{1}_{p+2});\{s_{ij}\})\nnl
&\ph{=}\lim_{z_{p+2}\to 0}\big [(-2 \pi i z_{p+2})^{s_{(1,B^1,p+2)}} \nnl
&\ph{=\lim_{z_{p+2}\to 0}}   Z^\tau_{n-|B^1|,p-|B^1|}(\gamma^{n-|B^1|,p-|B^1|}_{\vec{B'}};(1,A^1_2\shuffle A^{p+2}_2),(p+3,A^{p+3}),\dots,(n,A^n); z_{p+3},\dots,z_n;\{s^{p+2}_{ij}\})\big]\,,
\end{align}
}
where we have stripped off a factor of a genus-zero $Z^\textrm{\tree}$-integral with $n$ punctures, $n-3$ of which are integrated over and three of which are fixed to
\begin{align}
	(x_1,x_{n-1},x_n)&=(0,1,\infty)\,.
\end{align}
\begin{align}
&Z^\textrm{\tree}_{n,n+3}(\gamma^{n,p}_{B^1};(1,B^1_1),(p+2,B^{1}_{p+2});\{s_{ij}\})\nnl
&=\int_{x_1=0<x_{b^1_1}<x_{b^1_2}<\ldots<{x^1_{|B^1|}<1=x_{p+2}}}\prod_{i\in B^1}\dd x_i\, \KN_{1,B^1,p+2}(x_i,\{s_{ij}\})\,\textrm{pt}(1,B^1) 
\textrm{pt}(p+2,B^1_{p+2}) \, .
\end{align}

We have not yet finished evaluating the last limit because the  genus-1 integral now needs to be integrated over the $z_{p+2}-$dependent integration domain $\gamma^{n-|B^1|,p-|B^1|}_{\vec{B'}}$, where 
\begin{align}
	\gamma^{n-|B^1|,p-|B^1|}_{\vec{B'}}
	&=\{z_1=0<z_{p+2}\nnl
	&\ph{=\{ }<z_{b_1^{p+2}}<\dots <z_{b_{|B^{p+2}|}^{p+2}}<z_{p+3}<\dots\nnl
	&\ph{=\{ }<z_{b_1^{n-1}}<\dots <z_{b_{|B^{n-1}|}^{n-1}}<z_{n}\nnl
	&\ph{=\{ }<z_{b_1^{n}}<\dots <z_{b_{|B^{n}|}^{n}}<1\}\,,
\end{align}
i.e. the lower integration boundary is now $z_{p+2}$. It would be convenient to just set $z_{p+2}=0$ in the integration domain, but this is not correct, because other terms of similar leading behavior contribute, see e.g. the previous subsection. However, following the methods of Appendix E.2 of \rcite{Broedel:2020tmd}, we can make these other contributions subleading by requiring that\footnote{The key idea is to write the contour that we want, with $z_{p+2}=0$ as the lower integration boundary, as an union of contours in which the puncture $z_{p+2}$ is in all possible intermediate points between $z_1$ and $z_{p+3}$. The leading behavior of each of the new contours that we obtain this way is  $(-2\pi i z_{p+2})^{-(s^{p+2}_{1,p+2,b^2_1,b^2_2,\ldots,b^2_k})}$. To make these other contours subleading with respect to the contour we want, the condition stated follows.}, for any $1\leq l\leq |B^{p+22}|$:
\begin{align}
	\label{eqn:RequiredCondtionOnMandelstams}
	\Re(s^{p+2}_{1,p+2,b^{p+2}_1,b^{p+2}_2,\ldots,b^{p+2}_l})<\Re(s^{p+2}_{1,p+2})<0 \, .
\end{align}
Assuming that this last equation holds (or that we can analytically continue from such kinematic points), we write the regularized initial value of our integral:
{\small
\begin{align}
	\label{eqn:GeneralInitialValue}
	&\lim_{z_{p+2}\to 0}(-2 \pi i z_{p+2})^{s_{(1,B^1,p+2)}}Z^\tau_{n,p}(\gamma^{n,p}_{\vec{B}}, z_{p+2},z_{p+3},\dots,z_n;\{s_{ij}\})\\
	&=e^{s_{(1,B^1,p+2)}\omega(1,0)}Z^\textrm{\tree}_{|B^1|+3,|B^1|}(\gamma^{|B^1|+3,|B^1|}_{B^1};(1,B^1_1),(p+2,B^{1}_{p+2});\{s_{ij}\})\nnl
	&\ph{=}  Z^\tau_{n-|B^1|-1,p-|B^1|}(\gamma^{n-|B^1|-1,p-|B^1|}_{\vec{B''}};(1,A^1_2\shuffle A^{p+2}_2),(p+3,A^{p+3}),\dots,(n,A^n); z_{p+3},\dots,z_n;\{s^{p+2}_{ij}\})\big]\,,
\end{align}
}
where the integration domain $\gamma^{n-|B^1|-1,p-|B^1|}_{\vec{B''}}$ has no $z_{p+2}-$dependence:
\begin{align}
	\gamma^{n-|B^1|-1,p-|B^1|}_{\vec{B''}} 
	&=\{z_1=0<z_{b_1^{p+2}}<\dots <z_{b_{|B^{p+2}|}^{p+2}}<z_{p+3}<\dots\nnl
	&\ph{=\{ }<z_{b_1^{n-1}}<\dots <z_{b_{|B^{n-1}|}^{n-1}}<z_{n}\nnl
	&\ph{=\{ }<z_{b_1^{n}}<\dots <z_{b_{|B^{n}|}^{n}}<1\}\,.
\end{align}

We will now comment on the implications of the regularized-initial value \eqn{eqn:GeneralInitialValue}. When one takes a single limit, the $Z^\tau_{n,p}(\vec{B},\vec{A})$ integrals\footnote{We hope the notation here is intuitive: $\vec{A}$ describes the integrand and $\vec{B}$ describes the integration contour.} have a simple behavior: They either vanish at the leading order, depending on the compatibility of $\vec{A}$  and $\vec{B}$ by \eqn{eqn:condp2} (this happens e.g. with the third entry of \eqn{eqn:Z43Vector}). When the initial value does not vanish at leading order and if $B^1$ is nonempty, we obtain, as a factor, a genus-0 string integral of $(|B^1|+3)$ punctures\footnote{If $B^1$ is empty, this factor is 1. See e.g. \eqn{eq:eqnZn3p1InitialCondition}.}. Further, we are left with a genus-one integral with fewer total unintegrated and integrated punctures, and with some shift of the Mandelstam variables $s_{ij}\rightarrow s^{p+2}_{ij}$: $Z^\tau_{(n-|B^1|-1,p-|B^1|)}$. 

Note that if $p-|B^1|=0$, then the leftover $Z^\tau_{(n-|B^1|-1,p-|B^1|)}$ is not an integral, but simply a genus-one Koba--Nielsen factor. This  is precisely what happens with the integrals that are the focus of the main text,  see \eqn{eqn:simpleDependenceDroppingOut}. 

We note that one can iterate this process in a straightforward way to compute  $\lim\limits_{n,p,n-p} $.  From the rules above, we note that these regularized initial values will be given, if they do not vanish, by products of tree-level  integrals, $Z^\textrm{tree}$ and, for the cases where $B^{n}$ is nonempty, these products of tree-level integrals will multiply a genus-one integral $Z^\tau_{|B^{n}+1|}$. \footnote{We saw an example of such genus-one integral earlier in this appendix, in \eqn{eq:eqnZn3p1InitialCondition}.} 

Finally, we remark that in order to iterate this limiting procedure, we require the Mandelstam variables to satisfy, for $2 \leq k \leq n-p$ and $1 \leq l\leq B^{p+k}$:

\begin{align}
	\label{eqn:RequiredCondtionOnMandelstamsForIteration}
	\Re(s^{p+k}_{1,p+k,b^{k+2}_1,b^{k+2}_2,\ldots,b^{k+2}_l})<\Re(s^{p+k}_{1,p+k})<0 \, .
\end{align}

\vspace{-0.2cm}

\newpage

\bibliographystyle{JHEP}
\bibliography{references}

\end{document}